\documentstyle [12pt,eqsecnum,amsfonts,aps]{revtex}
\input epsf
\tighten
\draft
\widetext
\input epsf
\newcommand{\beq}{\begin{equation}}
\newcommand{\eeq}{\end{equation}}
\newcommand{\beqs}{\begin{eqnarray}}
\newcommand{\eeqs}{\end{eqnarray}}
\newcommand{\lsim}{\mathrel{\raisebox{-.6ex}{$\stackrel{\textstyle<}{\sim}$}}}

\begin{document}
\draft
\baselineskip 6.0mm

\bigskip
\bigskip

\title{Transfer Matrices for the Zero-Temperature Potts Antiferromagnet on 
Cyclic and M\"obius Lattice Strips}

\bigskip

\author{
Shu-Chiuan Chang$^{a,b}$ \thanks{email: scchang@phys.ntu.edu.tw} \and
Robert Shrock$^{c}$ \thanks{email: robert.shrock@sunysb.edu}}

\bigskip

\address{(a) \ Department of Applied Physics, Faculty of Science \\
Tokyo University of Science \\
Tokyo 162-8601, Japan}

\address{(b) \ Physics Division \\
National Center for Theoretical Sciences at Taipei \\
National Taiwan University  \\
Taipei 10617, Taiwan }

\address{(c) \ C. N. Yang Institute for Theoretical Physics \\
State University of New York \\
Stony Brook, N. Y. 11794}

\maketitle

\bigskip

\begin{abstract}

We present transfer matrices for the zero-temperature partition function of the
$q$-state Potts antiferromagnet (equivalently, the chromatic polynomial) on
cyclic and M\"obius strips of the square, triangular, and honeycomb lattices of
width $L_y$ and arbitrarily great length $L_x$.  We relate these results to our
earlier exact solutions for square-lattice strips with $L_y=3,4,5$,
triangular-lattice strips with $L_y=2,3,4$, and honeycomb-lattice strips with
$L_y=2,3$ and periodic or twisted periodic boundary conditions.  We give a
general expression for the chromatic polynomial of a M\"obius strip of a
lattice $\Lambda$ and exact results for a subset of honeycomb-lattice transfer
matrices, both of which are valid for arbitrary strip width $L_y$.  New results
are presented for the $L_y=5$ strip of the triangular lattice and the $L_y=4$
and $L_y=5$ strips of the honeycomb lattice.  Using these results and taking
the infinite-length limit $L_x \to \infty$, we determine the continuous
accumulation locus of the zeros of the above partition function in the complex
$q$ plane, including the maximal real point of nonanalyticity of the degeneracy
per site, $W$ as a function of $q$.

\end{abstract}

\bigskip
\bigskip

\newpage
\pagestyle{plain}
\pagenumbering{arabic}

\section{Introduction}

The $q$-state Potts antiferromagnet (AF) \cite{potts,wurev} exhibits nonzero
ground state entropy, $S_0 > 0$ (without frustration) for sufficiently large
$q$ on a given lattice $\Lambda$ or, more generally, on a graph $G$.  This is
equivalent to a ground state degeneracy per site $W > 1$, since $S_0 = k_B \ln
W$.  There is a close connection with graph theory
here, since the zero-temperature partition function of the above-mentioned
$q$-state Potts antiferromagnet on a graph $G=G(V,E)$ defined by vertex and
edge sets $V$ and $E$ satisfies
\beq
Z(G,q,T=0)_{PAF}=P(G,q)
\label{zp}
\eeq
where $P(G,q)$ is the chromatic polynomial expressing the number of ways of
coloring the vertices of the graph $G$ with $q$ colors such that no two
adjacent vertices have the same color \cite{wup}-\cite{bbook}.  Thus
\beq
W(\{G\},q) = \lim_{n \to \infty} P(G,q)^{1/n}
\label{w}
\eeq
where $n=|V|$ denotes the number of vertices of the graph $G$ and the symbol
$\{G\}$ formally denotes the set $\lim_{n \to \infty} G$. The minimum
number of colors that is necessary to color a graph $G$ subject to this
constraint is called the chromatic number of $G$, $\chi(G)$.

We represent a given strip as extending longitudinally in the $x$ direction and
transversely in the $y$ direction, with width $L_y$ vertices.  Each strip
involves a longitudinal repetition of $m$ copies of a particular subgraph.  For
the square-lattice strips, this is a column of squares.  It is convenient to
represent the strip of the triangular lattice as obtained from the
corresponding strip of the square lattice with additional diagonal edges
connecting, say, the upper-left to lower-right vertices in each square.  In
both these cases, the length is $L_x=m$ vertices.  We represent the strip of
the honeycomb lattice in the form of bricks oriented horizontally.  In this
case, since there are two vertices in 1-1 correspondence with each horizontal
side of a brick, $L_x=2m$ vertices.

The general structure of the Potts model partition function for cyclic strips
of the square lattice of width $L_y$, as a sum of powers of eigenvalues of a
formal transfer matrix multiplied by certain coefficients $c^{(d)}$, $0 \le d
\le L_y$, was given in Refs. \cite{saleur} (see also \cite{saleurcmp}).  The 
present authors (unaware of this finding in \cite{saleur}) rediscovered the 
result in Ref. \cite{cf} and showed that it applies also to cyclic strips of
the triangular and honeycomb lattices \cite{hca}.  The coefficients are 
polynomials of degree $d$ in the variable $q$ given by \cite{saleur,cf}
\beq
c^{(d)} = U_{2d}(q^{1/2}/2) = \sum_{j=0}^d (-1)^j {2d-j \choose j}
q^{d-j}
\label{cd}
\eeq
with $U_n(x)$ being the Chebyshev polynomial of the second kind.  The first few
of these coefficients are $c^{(0)}=1$, $c^{(1)}=q-1$, $c^{(2)}=q^2-3q+1$, and
$c^{(3)}=q^3-5q^2+6q-1$.  The $c^{(d)}$'s play a role analogous to
multiplicities of eigenvalues $\lambda_{\Lambda,L_y,d,j}$, although this
identification is formal, since $c^{(d)}$ may be zero or negative for the
physical values $q=1,2,3$.  For example, as shown in eqs. (2.19), (2.20) of 
Ref. \cite{cf}, if $q=2$, then $c^{(d)}=-1$ for $d=2$ or 3 mod 4, and if $q=3$,
then $c^{(d)}=-1$ for $d=3$ or 5 mod 6 and $c^{(d)}=-2$ for $d=4$ mod 6.  In
general, $c^{(d)}$ vanishes at $q=q_{d,k}$, where $q_{d,k}=4\cos^2(\pi
k/(2d+1))$, $k=1,2,...,d$, so that $c^{(d)}$ is positive for $q \ge 4$ for
arbitrary $d$.  

This structure also holds for the chromatic polynomial, with the difference
that, for $L_y \ge 2$, the number of eigenvalues for each $d$ is smaller than
the number for the full partition function.  We determined this number for
strips of the square and triangular lattice in Ref. \cite{cf}, for strips of
the honeycomb lattice in Ref. \cite{hca}, and for self-dual strips of the
square lattice in Ref. \cite{dg} (where we also noted how these numbers fill
out the entries in the relevant Bratelli diagrams). The chromatic polynomial
for a cyclic strip of the regular lattice $\Lambda$ has the form
\beq
P(\Lambda,L_y \times m,cyc.,q)=\sum_{d=0}^{L_y} c^{(d)}
\sum_{j=1}^{n_P(\Lambda,L_y,d)} (\lambda_{\Lambda,L_y,d,j})^m
\label{pgsum}
\eeq

Let $G^\prime=(V,E^\prime)$ be a spanning subgraph of $G$, i.e. a subgraph
having the same vertex set $V$ and an edge set $E^\prime \subseteq E$. Then
$P(G,q)$ can be written as the sum \cite{kf,fk}
\beq
P(G,q) = \sum_{G^\prime \subseteq G} q^{k(G^\prime)}(-1)^{|E^\prime|}
\label{cluster}
\eeq
where $k(G^\prime)$ denotes the number of connected components of $G^\prime$
and $|E^\prime|$ denotes the number of edges in the set $E^\prime$. 
Since we only consider connected graphs $G$, we have $k(G)=1$. The formula
(\ref{cluster}) enables one to generalize $q$ from ${\mathbb Z}_+$ to ${\mathbb
R}_+$.  The zeros of $P(G,q)$ in the complex $q$ plane are called chromatic
zeros. We denote the continuous accumulation set of these zeros in the $n \to
\infty$ limit as ${\cal B}$, which is the continuous locus of points
where $W(\{G\},q)$ is nonanalytic. The maximal value of $q$ where ${\cal B}$
intersects the (positive) real axis is labelled $q_c(\{G\})$.  This locus
occurs as a solution to the degeneracy in magnitude of $\lambda$'s of maximal
magnitude \cite{bkw1}-\cite{read91}.

In this paper we present transfer matrices $T_{\Lambda,L_y}$ for strips of the
square (sq), triangular (tri), and honeycomb (hc) lattices of fixed transverse
widths $L_y$ and arbitrarily great length with periodic (cyclic) longitudinal
boundary conditions.  We also give results for the corresponding strips with
twisted periodic (M\"obius) longitudinal boundary conditions.  We relate these
results to our earlier exact solutions for square-lattice strips with
$L_y=3,4,5$; triangular-lattice strips with $L_y=2,3,4$, and honeycomb-lattice
strips with $L_y=2,3$ and periodic or twisted periodic boundary conditions.  We
give a general expression for the chromatic polynomial of a M\"obius strip of a
lattice $\Lambda$ which is valid for arbitrary strip width $L_y$.  New results
are presented for the $L_y=5$ strip of the triangular lattice and the $L_y=4$
and $L_y=5$ strips of the honeycomb lattice.  Using these results and taking
the infinite-length limit $L_x \to \infty$, we determine certain properties of
the continuous accumulation locus of the zeros of the above partition function
in the complex $q$ plane, including the maximal point of nonanalyticity of the
degeneracy per site, $W$ as a function of $q$.  We also give a general trace
formula for the honeycomb strip which is valid for arbitrary $L_y$.  Chromatic
numbers for these lattice strips are given in the appendix.

There are several motivations for this work. One is that, as noted above, the
Potts antiferromagnet has the interesting property of nonzero ground state
entropy for sufficiently large $q$, which is an exception to the third law of
thermodynamics \cite{al,cw}.  Via eq. (\ref{w}), exact calculations of $P(G,q)$
yield calculations of $W\{ G \},q)$ and thus give insight into this property.
The constraint that no two adjacent vertices have the same value of $q$ becomes
less and less important as $q \to \infty$, and in this limit, $W(\{ G \},q) \to
q$.  As $q$ decreases, $W(\{ G \},q)$ remains a real analytic function of $q$
down to the point $q_c(\{G \})$, where it is nonanalytic.  In earlier work, we
studied how $q_c(\{G \})$ depends on the boundary conditions used for various
lattice strips \cite{w}-\cite{s5}.  It was found that a convenient property of
strips with periodic (or twisted periodic) longitudinal boundary conditions was
that they always defined a value of $q_c(\{ G \})$, and, furthermore, for a
given type of lattice, this point was observed to be a monotonically
nondecreasing function of the strip width for all cases calculated.  By
carrying out exact calculations of $P(G,q)$ and determining the analytic
structure of $W(\{ G \},q)$, one can study how $q_c(\{ G \})$ varies with
increasing strip width and how it approaches the values expected for the
corresponding two-dimensional lattices, namely $q_c(sq)=3$ \cite{lenard} and
$q_c(tri)=4$ \cite{baxter86,baxter87}.  This interpolation property is an
especially interesting use of these exact results since the resultant values of
$q_c(\{ G \})$ interpolate between the value $q_c=2$ for one-dimension and
values on two-dimensional lattices.  Similar results were found for the ground
state entropy itself \cite{w,ww,w3,wn,w2d}.  This should be contrasted with
other properties; for example, the critical temperature of the Potts
ferromagnet on an infinite-length lattice strip is zero for any width $L_y$
regardless of how great.  

For the honeycomb lattice, formal arguments give $q_c(hc)=(3+\sqrt{5})/2 \simeq
2.618$ \cite{ssbounds,p3afhc}.  The honeycomb value is only formal because if
$q$ is not a positive integer, then one cannot use the expression of the
partition function for the Potts antiferromagnet in terms of a sum of
positive-definite Boltzmann factors and instead must use the formula \cite{kf}
$Z(G,q,v)=\sum_{G^\prime \subseteq G} q^{k(G^\prime)} v^{|E^\prime|}$, where
$v=e^K-1$ with $K=\beta J$, where $\beta=(k_BT)^{-1}$ and $J$ is the spin-spin
coupling (which reduces to eq. (\ref{cluster}) for the $T=0$ antiferromagnet,
where $K=-\infty$).  However, for $-1 \le v < 0$ and non-integral $q$, this
does not, in general, define a Gibbs measure \cite{ssbounds,a}. This
non-integral property of $q_c(hc)$ for the two-dimensional honeycomb lattice
and the associated subtleties make it especially useful to have explicit
calculations of $q_c$ values for finite-width strips of this lattice, to check
that these are consistent with the expected behavior, according to which they
approach the above value of $q_c(hc)$ as $L_y$ gets large.  

A second motivation related to this is that when one generalizes $q$ from a
positive integer to a complex variable, one sees that the point $q_c(\{ G \})$
is the maximal point where a certain nonanalytic boundary ${\cal B}$ crosses
the real axis (discussed further below); thus, another purpose of these exact
calculations is to get further insight about this locus ${\cal B}$.  As is well
known, slight changes in coefficients of polynomials can have drastic changes
on the positions of the zeros of the polynomials, which means that one must
have the exact expression for $P(G,q)$ to study these zeros and their
accumulation set ${\cal B}$.  Since it was found that for (the $L_x \to \infty$
limit of) strips with free longitudinal boundary conditions, ${\cal B}$ does
not necessarily cross the real axis, so that no $q_c(\{ G \}, q)$ is defined
\cite{w,strip}, one cannot carry out this study in the same manner for these
strips (although for sufficiently wide strips with free longitudinal boundary
conditions, arcs on ${\cal B}$ have endpoints that are often close enough to
the real axis to allow one to extrapolate and define an effective $q_c$).

A third motivation is that certain structural features of the transfer matrices
can allow one to obtain general formulas that are applicable for arbitrarily
large strip widths.  We have already used this feature in Ref. \cite{s5}, where
we presented transfer matrices whose eigenvalues are $\lambda_{\Lambda,L_y,d}$
with degree $d=L_y-1$ for several lattice strips with periodic boundary
conditions, including those of the square and triangular lattices. Thus another
purpose of exhibiting explicit transfer matrices for $0 \le d \le L_y-2$ here
is to make them available for a wider community, leading, hopefully, to some
advances beyond those we have made with $d=L_y,L_y-1$ in the construction of
explicit formulas for $\lambda_{\Lambda,L_y,d}$ valid for arbitrary $L_y$.  All
of these degree-$d$ sectors contribute to the partition function (here, the
chromatic polynomial), so that they are of physical interest, although an
elementary result is that the degeneracy per site $W$ only depends on the
(maximal eigenvalue in the) $d=0$ sector.

The chromatic polynomial for the cyclic and M\"obius strips of the square
lattice were calculated (i) for $L_y=2$ in \cite{bds} (see also \cite{bm});
(ii) for $L_y=3$ cyclic in \cite{wcyl,wcy}, and M\"obius case in \cite{pm};
(iii) for $L_y=4$ cyclic/M\"obius in \cite{s4}, (iv) for $L_y=5$
cyclic/M\"obius in \cite{s5}.  The chromatic polynomials for the triangular
lattice strips were calculated for (i) $L_y=2$, cyclic/M\"obius in \cite{wcy},
(ii) $L_y=3,4$ in \cite{t}; and for the honeycomb lattice with (i) $L_y=2$,
cyclic/M\"obius in \cite{pg}, (ii) $L_y=3$ in \cite{hca}. (Refs. \cite{pg,nec}
showed that the structure (\ref{pgsum}) holds more generally for cyclic strip
graphs composed of iterated subgraphs that are not parts of regular lattices;
we shall not need this degree of generality here.)  We have proved that for
these strips the set of eigenvalues are the same for cyclic (cyc.) and M\"obius
(Mb.) boundary conditions and have given the transformation rules for how the
coefficients change for strips of the square lattice when one changes from
cyclic to M\"obius boundary conditions \cite{cf}.

In our previous works on chromatic polynomials we have sometimes given transfer
matrices (e.g. \cite{s5}) but often have expressed our results for the
$\lambda_{\Lambda,L_y,d,j}$ in terms of generating functions or solutions to
algebraic equations.  In one respect this is a maximally compact way of
presenting the results, since for a given set of $n_P(\Lambda,L_y,d)$
$\lambda_{\Lambda,L_y,d,j}$'s, it suffices to give the coefficients of the
algebraic equation of degree $n_P(\Lambda,L_y,d)$ and for the generating
function, it suffices to give $2n_P(\Lambda,L_y,d)$ coefficients, which are
polynomials in $q$.  Since the generating function is a rational function in
$q$, it has the appeal that one always deals with polynomials.  For the
expression of the chromatic polynomial in terms of transfer matrices, since
$T_{\Lambda,L_y,d}$ has dimension $n_P(\Lambda,L_y,d) \times
n_P(\Lambda,L_y,d)$, one must specify a greater number of polynomials, namely
$n_P(\Lambda,L_y,d)^2$ polynomials.  As $L_y$ increases, the number of
expressions that one must give thus grows quadratically rather than linearly,
as in the more compact approach using generating functions or algebraic
equations for the $\lambda_{\Lambda,L_y,d}$.  However, offsetting this
disadvantage, the transfer matrix method has the advantage that some of the
entries are rather simple, and, moreover, one can sometimes spot useful
patterns in the matrices that expedite or confirm the construction of general
formulas that are applicable for arbitrarily large strip widths $L_y$.

We mention some previous work.  Matrix methods and related recursive linear
algebraic techniques for calculating chromatic polynomials were used in early
work \cite{bds,bm,b} and more recent papers \cite{matmeth}-\cite{matmeth3}. The
coloring matrix technique of Ref. \cite{b} was applied in
Refs. \cite{ww}-\cite{wn} to obtain upper and lower bounds on $W$ for various
(infinite) two-dimensional lattices.  This application relied upon the fact
that these coloring matrices are non-negative, so that one could use the
Perron-Frobenius theorem in analyzing the eigenvalues.  Related recursive
linear algebraic methods were used in Refs. \cite{strip,strip2,hs} and in our
subsequent papers on chromatic polynomials.  Transfer matrices for chromatic
polynomials were used in Refs. \cite{baxter86,baxter87,klein}.  In Ref.
\cite{sqtran} transfer matrices for the Potts model were developed and were
applied there and in Refs. \cite{cyltran,tritran} to calculate chromatic
polynomials for strips of the square and triangular lattices with free
longitudinal boundary conditions and free or periodic transverse boundary
conditions.  These have been termed transfer matrices in the Fortuin-Kasteleyn
representation (see eq. (\ref{cluster})).  In Refs. \cite{ts,tt}, transfer
matrix methods were used to calculate full Potts model partition functions
with arbitrary $q$ and $v$ for strips of the square and triangular lattices
with free longitudinal boundary conditions.  Using transfer matrix methods
together with the sieve methods of Refs. \cite{matmeth}-\cite{matmeth3} (see
also \cite{dn,cprg,ka3}), we have proved theorems that determined
$T_{\Lambda,L_y,d}$ for $d=L_y-1$ and arbitrarily large $L_y$ for several
lattice strips with periodic boundary conditions \cite{s5}.

Some remarks are in order here concerning an important respect in which the
transfer matrices presented here differ from the conventional transfer matrices
in statistical mechanics and coloring matrices in mathematical graph theory.
Consider a strip graph, and let the spins on each transverse slice of this
strip be labeled $\sigma_{x,y}$ where the longitudinal position $x$ is fixed
and the transverse position $y$ varies.  As conventionally defined, the
transfer matrix of the (zero-field) Potts model at some temperature $T$ is
defined as the matrix $T_{x,x+1} = \langle \{ \sigma_{x,y} \} | e^{-\beta {\cal
H}} | \{ \sigma_{x+1,y} \} \rangle$, where ${\cal H} = -J \sum_{\langle i j
\rangle} \delta_{\sigma_i \sigma_j}$, $\sigma_i=1,...,q$ are the spin variables
on each vertex and $\langle i j \rangle$ denotes pairs of adjacent vertices.
Since there are $q^{L_y}$ spin configurations on each slice, this is a $q^{L_y}
\times q^{L_y}$ dimensional matrix. Although this dimension depends on $q$, the
individual elements themselves do not explicitly depend on $q$.  For physical
temperatures $0 \le T \le \infty$ this matrix is non-negative, and at nonzero
temperatures it is positive-definite.  For $T > 0$, on a strip of finite width,
for which this is a finite-dimensional matrix, one can then make powerful use
of the Perron-Frobenius theorem on positive-definite matrices to conclude that
there is a unique real positive eigenvalue of maximal magnitude.  Indeed, this
theorem provides a standard way to prove that the free energy of the Potts
model or other spin model with short-range interactions on a strip of infinite
length and finite width is analytic at all finite temperatures.  Similarly, the
coloring matrices used in Refs. \cite{b,ww,w3,wn} for studies of chromatic
polynomials and the degeneracy per site $W$ on various lattices are
non-negative matrices, and this property was necessary for the upper and lower
bounds derived in these papers.  In contrast, the transfer matrices in the
Fortuin-Kasteleyn representation are not positive-definite, and hence the
Perron-Frobenius does not apply to them.  The fact that it does not apply is
crucial for the property that the locus ${\cal B}$ crosses the positive real
axis and for the existence of $q_c(\{ G \})$, since this locus occurs where
there is a switch between at least two distinct eigenvalues of maximal
magnitude and hence non-uniqueness of such eigenvalues.

 From the exact expressions for $P(G,q)$, one can evaluate eq. (\ref{w}) to get
$W$ and analyze the accumulation locus ${\cal B}$ of zeros of $P(G,q)$ in the
complex $q$ plane.  We have done this for cyclic/M\"obius strips in a number of
previous works.  These loci are different for strips with different boundary
conditions, e.g., free, cylindrical, cyclic/M\"obius, and toroidal.  As we have
shown, since the $\lambda$'s are the same for the cyclic and M\"obius strips of
a given lattice, it follows that the locus ${\cal B}$ is the same for the
infinite-length limits of cyclic and M\"obius strips of a given lattice.  In
our previous work we have found a number of general properties of the locus
${\cal B}$, including the property that for strips of regular lattices with
periodic or twisted periodic longitudinal boundary conditions the locus ${\cal
B}$ is comprised of closed curves that enclose various regions and pass through
$q=0$ and at a maximal real point $q_c$ which depends on the lattice, as well
as possible other intermediate points.  The point $q_c(\{G\})$ is important
since it separates the interval $q > q_c(\{G\})$ on the positive real $q$ axis
where the Potts model (with $q$ extended from ${\mathbb Z}_+$ to ${\mathbb R}$)
exhibits nonzero ground state entropy (which increases with $q$, asymptotically
approaching $S_0 = k_B \ln q$ for large $q$) from the interval $0 \le q \le
q_c(\{G\})$ in which $S_0$ has a different analytic form.

\section{General Structure}

Our results for chromatic polynomials are obtained as special cases, for the
zero-temperature antiferromagnet, of general transfer matrices that we have
calculated for the full temperature-dependent Potts model partition function on
cyclic and M\"obius lattice strips \cite{js}.  The latter results, and the
general method, will be reported in a companion paper \cite{zt}. The
motivations for presenting the special cases of these results for the
zero-temperature Potts antiferromagnet (chromatic polynomial) have been given
in the introduction.  In addition to these physics motivations, a relevant
point is that the transfer matrices for the chromatic polynomials are
subtantially smaller in dimension and simpler in structure, depending only one
one variable instead of two, than the transfer matrices for the full Potts
model partition function.  Another reason for separating these analyses of the
full Potts model and the chromatic polynomial has to do with the determinants
$det(T_{Z,\Lambda,L_y,d})$ and $det(T_{P,\Lambda,L_y,d})$; in both the case of
the full Potts model and the chromatic polynomial, the individual eigenvalues
$\lambda_{X,\Lambda.L_y,d}$ for most strip widths are not expressible in
explicit algebraic form because the corresponding characteristic polynomials
are of fifth order or higher.  However, we have found a most interesting
property, that the products of the eigenvalues that define the determinant
$det(T_{Z,\Lambda,L_y,d})$ have a very simple form.  When one specializes to
$v=-1$, for all values of $L_y$ and $d=0,1,...L_y-1$, except the lowest case,
$L_y=1$, some of these eigenvalues vanish, so that $det(T_{Z,\Lambda,L_y,d})=0$
at $v=-1$, reflecting the fact that the transfer matrices $T_{P,\Lambda,L_y,d}$
are of smaller dimension than $T_{Z,\Lambda,L_y,d}$.  But in this $v=-1$ case,
when one removes the zero columns and corresponding rows of
$T_{Z,\Lambda,L_y,d}$ to form the transfer matrix $T_{P,\Lambda,L_y,d}$ for the
chromatic polynomial, the resultant determinant is, in general, nonzero.  And
again we find a very interesting feature, namely that although the eigenvalues
$\lambda_{P,\Lambda,L_y,d}$ themselves are roots of characteristic polynomials
that are often of fifth or higher order, precluding solutions in terms of
explicit algebraic expressions, the products that define
$det(T_{P,\Lambda,L_y,d})$ are often quite simple, especially for the strips of
the triangular and honeycomb lattices.  This is another motivation for our
presenting results for the chromatic polynomials separately from those for the
full Potts model partition function.

The chromatic polynomial for cyclic strips can be written in the form 
\cite{saleur,cf}
\beq 
P(\Lambda,L_y \times m,cyc.,q) = \sum_{d=0}^{L_y} c^{(d)}
Tr((T_{\Lambda,L_y,d})^m) 
\label{pgsumt} 
\eeq
where $T_{\Lambda,L_y,d}$ is the transfer matrix, with eigenvalues
$\lambda_{\Lambda,L_y,d}$. We denote the characteristic polynomial of
$T_{\Lambda,L_y,d}$ in the variable $z$ as $CP(T_{\Lambda,L_y,d},z)$.  We have
shown that the full transfer matrix $T_{\Lambda,L_y}$ has a block structure
formally specified by
\beq 
T_{\Lambda,L_y} = \bigoplus_{d=0}^{L_y} \prod
T_{\Lambda,L_y,d} 
\label{Tdirectsum} 
\eeq
where the product $\prod T_{\Lambda,L_y,d}$ means a set of square blocks, each
of dimension, $c^{(d)}$, of the form $\lambda_{\Lambda,L_y,d,j}$ times the
identity matrix. 
The dimension of the total transfer matrix, i.e., the total number
of eigenvalues $\lambda_{\Lambda,L_y,d,j}$ in eq. (\ref{pgsum}),
counting multiplicities, is thus
\beq 
dim(T_{\Lambda,L_y}) = \sum_{d=0}^{L_y}
dim(T_{\Lambda,L_y,d}) = \sum_{d=0}^{L_y} c^{(d)}
n_P(\Lambda,L_y,d) \ . 
\label{dimTL} 
\eeq

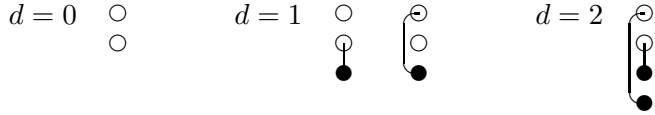
\begin{figure}
\unitlength 1mm \hspace*{5mm}
\begin{picture}(90,12)
\put(10,12){\makebox(0,0){{\small $d=0$}}}
\multiput(20,8)(0,4){2}{\circle{2}}
\put(40,12){\makebox(0,0){{\small $d=1$}}}
\multiput(50,4)(10,0){2}{\circle*{2}}
\multiput(50,8)(10,0){2}{\circle{2}}
\multiput(50,12)(10,0){2}{\circle{2}} \put(50,4){\line(0,1){4}}
\put(60,8){\oval(4,8)[l]} \put(80,12){\makebox(0,0){{\small
$d=2$}}} \multiput(90,0)(0,4){2}{\circle*{2}}
\multiput(90,8)(0,4){2}{\circle{2}} \put(90,4){\line(0,1){4}}
\put(90,6){\oval(4,12)[l]}
\end{picture}

\caption{\footnotesize{Partitions for the $L_y=2$ strip.}}
\label{L2partitions}
\end{figure}

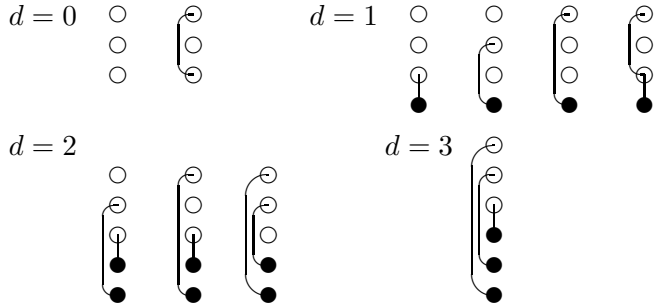
\begin{figure}
\unitlength 1mm \hspace*{5mm}
\begin{picture}(90,12)
\put(10,12){\makebox(0,0){{\small $d=0$}}}
\multiput(20,4)(0,4){3}{\circle{2}}
\multiput(30,4)(0,4){3}{\circle{2}} \put(30,8){\oval(4,8)[l]}
\put(50,12){\makebox(0,0){{\small $d=1$}}}
\multiput(60,0)(10,0){4}{\circle*{2}}
\multiput(60,4)(10,0){4}{\circle{2}}
\multiput(60,8)(10,0){4}{\circle{2}}
\multiput(60,12)(10,0){4}{\circle{2}} \put(60,0){\line(0,1){4}}
\put(70,4){\oval(4,8)[l]} \put(80,6){\oval(4,12)[l]}
\put(90,0){\line(0,1){4}} \put(90,8){\oval(4,8)[l]}
\end{picture}

\vspace*{5mm} \hspace*{5mm}
\begin{picture}(70,20)
\put(10,20){\makebox(0,0){{\small $d=2$}}}
\multiput(20,0)(10,0){3}{\circle*{2}}
\multiput(20,4)(10,0){3}{\circle*{2}}
\multiput(20,8)(10,0){3}{\circle{2}}
\multiput(20,12)(10,0){3}{\circle{2}}
\multiput(20,16)(10,0){3}{\circle{2}} \put(20,4){\line(0,1){4}}
\put(20,6){\oval(4,12)[l]} \put(30,4){\line(0,1){4}}
\put(30,8){\oval(4,16)[l]} \put(40,8){\oval(4,8)[l]}
\put(40,8){\oval(6,16)[l]} \put(60,20){\makebox(0,0){{\small
$d=3$}}} \multiput(70,0)(0,4){3}{\circle*{2}}
\multiput(70,12)(0,4){3}{\circle{2}} \put(70,8){\line(0,1){4}}
\put(70,10){\oval(4,12)[l]} \put(70,10){\oval(6,20)[l]}
\end{picture}

\caption{\footnotesize{Partitions for the $L_y=3$ strip.}}
\label{L3partitions}
\end{figure}

\begin{figure}
\unitlength 1mm \hspace*{5mm}
\begin{picture}(60,12)
\put(10,12){\makebox(0,0){{\small $d=0$}}}
\multiput(20,0)(10,0){4}{\circle{2}}
\multiput(20,4)(10,0){4}{\circle{2}}
\multiput(20,8)(10,0){4}{\circle{2}}
\multiput(20,12)(10,0){4}{\circle{2}} \put(30,8){\oval(4,8)[l]}
\put(40,6){\oval(4,12)[l]} \put(50,4){\oval(4,8)[l]}
\end{picture}

\vspace*{5mm} \hspace*{5mm}
\begin{picture}(100,16)
\put(10,16){\makebox(0,0){{\small $d=1$}}}
\multiput(20,0)(10,0){9}{\circle*{2}}
\multiput(20,4)(10,0){9}{\circle{2}}
\multiput(20,8)(10,0){9}{\circle{2}}
\multiput(20,12)(10,0){9}{\circle{2}}
\multiput(20,16)(10,0){9}{\circle{2}} \put(20,0){\line(0,1){4}}
\put(30,4){\oval(4,8)[l]} \put(40,6){\oval(4,12)[l]}
\put(50,8){\oval(4,16)[l]} \put(60,0){\line(0,1){4}}
\put(60,12){\oval(4,8)[l]} \put(70,4){\oval(4,8)[l]}
\put(70,12){\oval(4,8)[l]} \put(80,0){\line(0,1){4}}
\put(80,10){\oval(4,12)[l]} \put(90,8){\oval(4,8)[l]}
\put(90,0){\line(0,1){4}} \put(100,8){\oval(4,8)[l]}
\put(100,8){\oval(6,16)[l]}
\end{picture}

\vspace*{5mm} \hspace*{5mm}
\begin{picture}(90,20)
\put(10,20){\makebox(0,0){{\small $d=2$}}}
\multiput(20,0)(10,0){8}{\circle*{2}}
\multiput(20,4)(10,0){8}{\circle*{2}}
\multiput(20,8)(10,0){8}{\circle{2}}
\multiput(20,12)(10,0){8}{\circle{2}}
\multiput(20,16)(10,0){8}{\circle{2}}
\multiput(20,20)(10,0){8}{\circle{2}} \put(20,4){\line(0,1){4}}
\put(20,6){\oval(4,12)[l]} \put(30,4){\line(0,1){4}}
\put(30,8){\oval(4,16)[l]} \put(40,4){\line(0,1){4}}
\put(40,10){\oval(4,20)[l]} \put(50,8){\oval(4,8)[l]}
\put(50,8){\oval(6,16)[l]} \put(60,8){\oval(4,8)[l]}
\put(60,10){\oval(6,20)[l]} \put(70,10){\oval(4,12)[l]}
\put(70,10){\oval(6,20)[l]} \put(80,4){\line(0,1){4}}
\put(80,6){\oval(4,12)[l]} \put(80,16){\oval(4,8)[l]}
\put(90,12){\oval(4,8)[l]} \put(90,4){\line(0,1){4}}
\put(90,10){\oval(6,20)[l]}
\end{picture}

\vspace*{5mm} \hspace*{5mm}
\begin{picture}(50,24)
\put(10,24){\makebox(0,0){{\small $d=3$}}}
\multiput(20,0)(10,0){4}{\circle*{2}}
\multiput(20,4)(10,0){4}{\circle*{2}}
\multiput(20,8)(10,0){4}{\circle*{2}}
\multiput(20,12)(10,0){4}{\circle{2}}
\multiput(20,16)(10,0){4}{\circle{2}}
\multiput(20,20)(10,0){4}{\circle{2}}
\multiput(20,24)(10,0){4}{\circle{2}} \put(20,8){\line(0,1){4}}
\put(20,10){\oval(4,12)[l]} \put(20,10){\oval(6,20)[l]}
\put(30,8){\line(0,1){4}} \put(30,10){\oval(4,12)[l]}
\put(30,12){\oval(6,24)[l]} \put(40,8){\line(0,1){4}}
\put(40,12){\oval(4,16)[l]} \put(40,12){\oval(6,24)[l]}
\put(50,12){\oval(4,8)[l]} \put(50,12){\oval(6,16)[l]}
\put(50,12){\oval(8,24)[l]}
\end{picture}

\vspace*{5mm} \hspace*{5mm}
\begin{picture}(20,28)
\put(10,28){\makebox(0,0){{\small $d=4$}}}
\multiput(20,0)(0,4){4}{\circle*{2}}
\multiput(20,16)(0,4){4}{\circle{2}} \put(20,12){\line(0,1){4}}
\put(20,14){\oval(4,12)[l]} \put(20,14){\oval(6,20)[l]}
\put(20,14){\oval(8,28)[l]}
\end{picture}

\caption{\footnotesize{Partitions for the $L_y=4$ strip.}}
\label{L4partitions}
\end{figure}
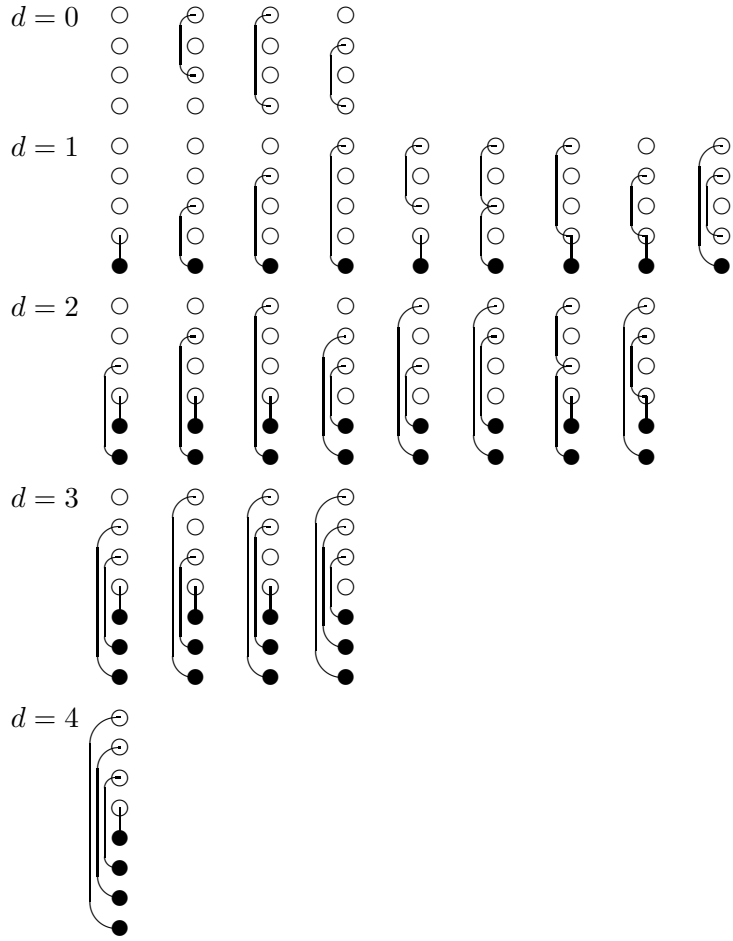

Each matrix $T_{\Lambda,L_y,d}$ is defined relative to a basis of $d$-color
assignments to a path graph with $L_y$ vertices corresponding to a transverse
slice across the strip.  Here one introduces partitions of the vertices in
these path graphs.  Using these as bases, we calculate the transfer matrix 
for each level (= degree) $d$, as discussed in our Ref. \cite{zt}. 
For the chromatic polynomials of the square and triangular
lattices, we have the additional requirement that adjacent vertices cannot
connect to each other. We list graphically all the possible partitions for
$L_y=2, 3, 4$ strips in Figs. \ref{L2partitions} to \ref{L4partitions}, where
white circles are the original $L_y$ vertices and each black circle corresponds
to a specific color assignment. We denote these partitions ${\cal P}_{L_y,d}$
for $2 \le L_y \le 5$ as follows:
\beq {\cal P}_{2,0} = \{ I \} \ , \qquad {\cal P}_{2,1} = \{ \bar
2; \bar 1 \} \ , \qquad {\cal P}_{2,2} = \{ \bar 1, \bar 2 \}
\label{L2partitionlist} \eeq
\beqs {\cal P}_{3,0} & = & \{ I; 13 \} \ , \qquad {\cal P}_{3,1} =
\{ \bar 3; \bar 2; \bar 1; \overline{13} \} \cr\cr {\cal P}_{3,2}
& = & \{ \bar 2, \bar 3; \bar 1, \bar 3; \bar 1, \bar 2 \} \ ,
\qquad {\cal P}_{3,3} = \{ \bar 1, \bar 2, \bar 3 \}
\label{L3partitionlist} \eeqs
\beqs {\cal P}_{4,0} & = & \{ I; 13; 14; 24 \} \ , \qquad {\cal
P}_{4,1} = \{ \bar 4; \bar 3; \bar 2; \bar 1; 13, \bar 4;
\overline{13}; \overline{14}; \overline{24}; 24, \bar 1 \} \cr\cr
{\cal P}_{4,2} & = & \{ \bar 3, \bar 4; \bar 2, \bar 4; \bar 1,
\bar 4; \bar 2, \bar 3; \bar 1, \bar 3; \bar 1, \bar 2;
\overline{13}, \bar 4; \bar 1, \overline{24} \} \cr\cr {\cal
P}_{4,3} & = & \{ \bar 2, \bar 3, \bar 4; \bar 1, \bar 3, \bar 4;
\bar 1, \bar 2, \bar 4; \bar 1, \bar 2, \bar 3 \} \ , \qquad {\cal
P}_{4,4} = \{ \bar 1, \bar 2, \bar 3, \bar 4 \}
\label{L4partitionlist} \eeqs
\beqs {\cal P}_{5,0} & = & \{ I; 13; 14; 15; 24; 25; 35; 24,15;
135 \} \cr\cr {\cal P}_{5,1} & = & \{ \bar 5; \bar 4; \bar 3; \bar
2; \bar 1; 13, \bar 5; 13, \bar 4; \overline{13}; 14, \bar 5;
\overline{14}; \overline{15}; 24, \bar 5; \overline{24}; 24, \bar
1; \overline{25}; 25, \bar 1; \overline{35}; 35, \bar 2; 35, \bar
1; \cr\cr & & 24, \overline{15}; \overline{135} \} \cr\cr {\cal
P}_{5,2} & = & \{ \bar 4, \bar 5; \bar 3, \bar 5; \bar 2, \bar 5;
\bar 1, \bar 5; \bar 3, \bar 4; \bar 2, \bar 4; \bar 1, \bar 4;
\bar 2, \bar 3; \bar 1, \bar 3; \bar 1, \bar 2; 13, \bar 4, \bar
5; \overline{13}, \bar 5; \overline{13}, \bar 4; \overline{14},
\bar 5; \overline{24}, \bar 5; 24, \bar 1, \bar 5; \cr\cr & & \bar
1, \overline{24}; \bar 1, \overline{25}; \bar 2, \overline{35};
\bar 1, \overline{35}; 35, \bar 1, \bar 2 \} \cr\cr {\cal P}_{5,3}
& = & \{ \bar 3, \bar 4, \bar 5; \bar 2, \bar 4, \bar 5; \bar 1,
\bar 4, \bar 5; \bar 2, \bar 3, \bar 5; \bar 1, \bar 3, \bar 5;
\bar 1, \bar 2, \bar 5; \bar 2, \bar 3, \bar 4; \bar 1, \bar 3,
\bar 4; \bar 1, \bar 2, \bar 4; \bar 1, \bar 2, \bar 3;
\overline{13}, \bar 4, \bar 5; \bar 1, \overline{24}, \bar 5;
\cr\cr & & \bar 1, \bar 2, \overline{35} \} \cr\cr {\cal P}_{5,4}
& = & \{ \bar 2, \bar 3, \bar 4, \bar 5; \bar 1, \bar 3, \bar 4,
\bar 5; \bar 1, \bar 2, \bar 4, \bar 5; \bar 1, \bar 2, \bar 3,
\bar 5; \bar 1, \bar 2, \bar 3, \bar 4 \} \cr\cr {\cal P}_{5,5} &
= & \{ \bar 1, \bar 2, \bar 3, \bar 4, \bar 5 \}
\label{L5partitionlist}  \eeqs
where partitions are separated by a colon, and each overline corresponds to a
color assignment.  That is, for $d=0$ there is no explicit color assignments;
for $d=1$, one vertex, or one connected set of vertices has its color
specified; for $d=2$, two vertices or two separate connected sets of vertices
have their colors specified; and so forth for higher values of $d$.  Therefore,
in eqs. (2.1) to (2.4), there is no overline for $d=0$, one overline for each
partition of $d=1$, two overlines for each partition of $d=2$, etc.

The number of partitions $n_P(\Lambda,L_y,d)$ is the dimension of the transfer
matrix $T_{\Lambda,L,d}$,
\beq 
dim(T_{\Lambda,L_y,d}) = n_P(\Lambda,L_y,d) \ . 
\label{dimTLd}
\eeq
In Ref. \cite{cf} we determined the number $n_P(\Lambda,L_y,d)$ (labelled
simply as $n_P(L_y,d)$) for the lattices $\Lambda = sq, tri$). 
Some special cases for $\Lambda=sq,tri$ are \cite{cf} as follows;
analogous results for the honeycomb lattice were given in Ref. \cite{hca}.
\beq
n_P(\Lambda,L_y,L_y)=1
\label{nplyly}
\eeq
\beq
n_P(\Lambda,L_y,L_y-1)=L_y
\label{nplylym1}
\eeq
\beq
n_P(\Lambda,L_y,0)=n_P(\Lambda,L_y-1,1)=M_{L_y-1} 
\label{nply0}
\eeq
where $M_n$ is the Motzkin number in combinatorics, given by 
\beq
M_n = \sum_{j=0}^n (-1)^j C_{n+1-j} {n \choose j}
\label{motzkin}
\eeq
where $C_n=(n+1)^{-1}{2n \choose n}$ is the Catalan number. (We use the same
symbol $C_n$ for the circuit graph; the meaning will be clear from context).
For reference, the first few Motzkin numbers are $M_n=1,2,4,9,21,51$ 
for $n=1,..., 6$, and we use the formal definition $M_0=1$.

Note that the number $n_P(tri,L_y,0)$ is the same as the dimension of the
corresponding transfer matrix for the triangular-lattice strip with free
boundary conditions obtained in Ref. \cite{sqtran}, while the transfer matrix
for the free strip of the square lattice has a dimension (for which a general
formula was given in Ref. \cite{ts}) which is smaller than $n_P(sq,L_y,0)$ for
$L_y \ge 4$.  Consistent with this, for $L_y \ge 4$, we find that the
characteristic polynomials for $T_{sq,4,0}$ and $T_{sq,5,0}$ factorize into
parts such that one factor is the characteristic polynomial for the transfer
matrix of the corresponding free strip, as will be seen below.

As in earlier work, we define $N_{P,\Lambda,L_y,\lambda}$ as the total number
of distinct eigenvalues of $T_{\Lambda,L_y}$, i.e.  the sum of the dimensions
of the submatrices $T_{\Lambda,L_y,d}$, modulo the multiplicity $c^{(d)}$:
\beq 
N_{P,\Lambda,L_y,\lambda} = \sum_{d=0}^{L_y}
n_P(\Lambda,L_y,d) \ . 
\label{nptot}
\eeq
Since we will use these results here, we include the relevant values in the
Appendix.  For cyclic or M\"obius strips of the square and triangular lattices,
\beq 
N_{P,\Lambda,L_y,\lambda}=2(L_y-1)! \
\sum_{j=0}^{[\frac{L_y}{2}]}\frac{(L_y-j)} {(j!)^2(L_y-2j)!} \quad
{\rm for} \ \ \Lambda=sq,tri 
\label{nptotform} 
\eeq
where $[\nu]$ denotes the integral part of $\nu$.  

For the corresponding strips of the honeycomb lattice, we calculated
$n_P(hc,L_y,d)$ and $N_{P,hc,L_y,\lambda}$ in Ref. \cite{hca}. The results that
are needed here are listed in the appendix. These correspond to the numbers of
partitions which allow the connections between vertices 2 and 3, between
vertices 4 and 5, etc. Therefore, in addition to the partitions given in eq.
(\ref{L3partitionlist}), the other partitions for the $L_y=3$ strip of the
honeycomb lattice are
\beq 
{\cal P}_{hc,3,0} = \{ 23 \} \ , \qquad {\cal P}_{hc,3,1} =
\{ \overline{23}; 23, \bar 1 \} \ , \qquad {\cal P}_{hc,3,2} = \{
\bar 1, \overline{23} \} \ . 
\eeq
In addition to the partitions given in eq.  (\ref{L4partitionlist}), the other
partitions for the $L_y=4$ strip of the honeycomb lattice are
\beqs 
{\cal P}_{hc,4,0} & = & \{ 23; 23, 14 \} \ , \qquad {\cal
P}_{hc,4,1} = \{ 23, \bar 4; \overline{23}; 23, \bar 1; 23,
\overline{14} \} \cr\cr {\cal P}_{hc,4,2} & = & \{ \overline{23},
\bar 4; 23, \bar 1, \bar 4; \bar 1, \overline{23} \} \ , \qquad
{\cal P}_{hc,4,3} = \{ \bar 1, \overline{23}, \bar 4 \} 
\eeqs
In addition to the partitions given in eq.  (\ref{L5partitionlist}), the other
partitions for the $L_y=5$ strip of the honeycomb lattice are
\beqs 
{\cal P}_{hc,5,0} & = & \{ 23; 45; 13, 45; 23, 14; 23, 15;
23, 45; 145; 235; 245; 23, 145 \} \cr\cr {\cal P}_{hc,5,1} & = &
\{ 23, \bar 5; 23, \bar 4; \overline{23}; 23, \bar 1;
\overline{45}; 45, \bar 3; 45, \bar 2; 45, \bar 1; 13,
\overline{45}; 45, \overline{13}; 14, 23, \bar 5; 23,
\overline{14}; 23, \overline{15}; \cr\cr & & 23, \overline{45};
45, \overline{23}; 23, 45, \bar 1; \overline{145}; \overline{235};
235, \bar 1; \overline{245}; 245, \bar 1; 23, \overline{145} \}
\cr\cr {\cal P}_{hc,5,2} & = & \{ 23, \bar 4, \bar 5;
\overline{23}, \bar 5; 23, \bar 1, \bar 5; \overline{23}, \bar 4;
23, \bar 1, \bar 4; \bar 1, \overline{23}; \bar 3, \overline{45};
\bar 2, \overline{45}; \bar 1, \overline{45}; 45, \bar 2, \bar 3;
45, \bar 1, \bar 3; 45, \bar 1, \bar 2; \cr\cr & & \overline{13},
\overline{45}; 23, \overline{14}, \bar 5; \overline{23},
\overline{45}; 23, \bar 1, \overline{45}; 45, \bar 1,
\overline{23}; \bar 1, \overline{235}; \bar 1, \overline{245} \}
\cr\cr {\cal P}_{hc,5,3} & = & \{ \overline{23}, \bar 4, \bar 5;
23, \bar 1, \bar 4, \bar 5; \bar 1, \overline{23}, \bar 5; \bar 1,
\overline{23}, \bar 4; \bar 2, \bar 3, \overline{45}; \bar 1, \bar
3, \overline{45}; \bar 1, \bar 2, \overline{45}; 45, \bar 1, \bar
2, \bar 3; \bar 1, \overline{23}, \overline{45} \} \cr\cr {\cal
P}_{hc,5,4} & = & \{ \bar 1, \overline{23}, \bar 4, \bar 5; \bar
1, \bar 2, \bar 3, \overline{45} \} \ . 
\eeqs

For cyclic strips of the square and triangular lattices, general
properties include first \cite{cf}
\beqs 
dim(T_{\Lambda,L_y}) & = & P(\Lambda,L_y \times m
,cyc.,q)_{m=0} =\sum_{d=0}^{L_y} c^{(d)}n_P(\Lambda,L_y,d) \cr\cr
& = & P({\rm Tree}_{L_y},q) = q(q-1)^{L_y-1} \ , \quad \Lambda=sq,tri 
\eeqs
where ${\rm Tree}_{L_y}$ denotes a tree graph (here a path graph) with
$L_y$ vertices; and
\beq 
Tr(T_{\Lambda,L_y}) = P(\Lambda,L_y \times m,cyc.,q)_{m=1} =
\sum_{d=0}^{L_y} c^{(d)}Tr(T_{\Lambda,L_y,d}) = 0 \ , \quad \Lambda=sq,tri 
\ . 
\label{trace_Ttotalzero} 
\eeq
For cyclic strips of the honeycomb lattice \cite{hca}
\beqs 
dim(T_{hc,L_y}) & = & P(hc,L_y \times m,cyc.,q)_{m=0}
=\sum_{d=0}^{L_y} c^{(d)}n_P(hc,L_y,d) \cr\cr & = &  \cases{
q\Bigl (q(q-1) \Bigr )^{\frac{L_y-1}{2}} & for \ $L_y$ \ odd \cr
          &   \cr
          \Bigl (q(q-1) \Bigr )^{\frac{L_y}{2}}  & for \ $L_y$ \ even }
\label{propfac} 
\eeqs
Here we give another general result for these strips of the honeycomb lattice,
governing the trace of the transfer matrix: 
\beqs 
Tr(T_{hc,L_y}) & = & P(hc, L_y \times m, cyc., q)_{m=1} =
\sum_{d=0}^{L_y} c^{(d)} Tr(T_{hc,L_y,d}) \cr\cr & = &
P({\rm Tree}_{2L_y},q) = q(q-1)^{2L_y-1} \ . 
\eeqs
This is proved by the same coloring matrix methods that we used in Ref.
\cite{cf} and \cite{hca}.  The reason for the difference relative to the
corresponding formula (\ref{trace_Ttotalzero}) for the cyclic strips of the
square and triangular lattices is that for $m=1$ on those strips the
longitudinal edges connect each vertex to itself, so the chromatic polynomial
vanishes, but, in contrast, for the strip of the honeycomb lattice, the
longitudinal edges have interior vertices of degree 2 associated with multiple
edges.  Because of this, and taking into account the elementary theorem that
the chromatic polynomial of a graph remains unchanged if one replaces any edge
by multiple copies of this edge, it follows that the resulting coloring is
described by the chromatic polynomial $P({\rm Tree}_{2L_y},q)$.  For the
reader's convenience in checking that our results satisfy these general trace
formulas, we shall list traces of our transfer matrices below.

In general,
\beq det(T_{\Lambda,L_y}) = \prod_{d=0}^{L_y}
[det(T_{\Lambda,L_y,d})]^{c^{(d)}} \ . 
\label{detTtotal} 
\eeq

We next discuss M\"obius strips.  One of the major results of the present paper
is a general formula, eq. (\ref{zgsum_transfermb}) together with
eqs. (\ref{cd0tran})-(\ref{cdoddtran}), for the partition function of the
zero-temperature $q$-state Potts antiferromagnet on the M\"obius strip of a
regular lattice $\Lambda$ of arbitrary width as well as arbitrary length.  This
formula involves a set of $\tilde T_{\Lambda,L_y,d}$ which occur once in each
$m$-fold product (with the $T_{\Lambda,L_y,d}$ thus raised to the power $m-1$)
and which differ from the $T_{\Lambda,L_y,d}$ in the interchange of certain
columns, encoding the M\"obius property of reversed-orientation periodic
longitudinal boundary conditions.  In our previous papers involving M\"obius
strips of the square and triangular lattices, we expressed the chromatic
polynomial in a form analogous to (\ref{pgsum}) but with different sets of
coefficients.  In Ref. \cite{cf} we derived a set of transformations that
governed how the coefficients for a certain $d$ change when one changes the
longitudinal boundary condition from cyclic to M\"obius for the square-lattice
strip.  However, as noted there, this did not apply to the triangular-lattice
strip; for that strip, if one expresses the chromatic polynomial in the same
manner for the cyclic and M\"obius cases, as a sum of $m$'th powers of
eigenvalues $\lambda_{tri,L_y,d,j}$, then some of the coefficients become
algebraic, rather than polynomial, functions of $q$.  This was evident in the
simplest nontrivial example, the $L_y=2$ strip \cite{wcy}.  The general
formalism that we introduce here for M\"obius strips has the advantage that the
coefficients stay in the set of $c^{(d)}$'s for triangular (and honeycomb) as
well as square lattice strips.  For $L_y \ge 3$, the $\tilde T_{\Lambda,L_y,d}$
matrices have eigenvalues that, in general, are not just related to those of
$\tilde T_{\Lambda,L_y,d}$ by possible sign changes.  In the limit of infinite
length, $m \to \infty$, because only the $T_{\Lambda,L_y,d}$ matrices are
raised to a power that goes to infinity, it is only the eigenvalues of these
matrices that determine locus ${\cal B}$.  Thus, the result of our previous
papers, that the locus ${\cal B}$ is the same for the infinite-length limit of
the strip of the lattice $\Lambda$ for cyclic or M\"obius boundary conditions,
is again evident in our present formulation.

We proceed to the details. For M\"obius strips one set of horizontal edges
reverses the order of connection. This corresponds to the exchange of the pair
of bases which switch to each other when the vertices reverse the order, i.e.,
the set of bases which do not have self-reflection symmetry with respect to the
center of the tree with $L_y$ vertices. For example, among the partitions for
the $L_y=2$ strip in Fig. \ref{L2partitions}, the first partition $\bar 2$ and
the second partition $\bar 1$ in ${\cal P}_{2,1}$ must be exchanged under this
reflection. The pairs of partitions for the $L_y=3$ strip in
Fig. \ref{L3partitions} are the first partition $\bar 3$ and the third
partition $\bar 1$ in ${\cal P}_{3,1}$, and the first partition $\bar 2, \bar
3$ and the third partition $\bar 1, \bar 2$ in ${\cal P}_{3,2}$. This
corresponds to the exchange of these pairs of columns of $T_{\Lambda,L_y,d}$,
and these matrices will be denoted as $\tilde T_{\Lambda,L_y,d}$
\cite{zt}. For general $L_y$ we have the following changes of coefficients for
the square, triangular and honeycomb lattices \cite{cf}:
\beq 
c^{(0)} \to c^{(0)} \label{cd0tran} 
\eeq
\beq 
c^{(2k)} \to -c^{(k-1)} \ , \qquad 1 \le k \le \Bigl
[\frac{L_y}{2} \Bigr ] 
\label{cdeventran} 
\eeq
\beq 
c^{(2k+1)} \to c^{(k+1)} \ , \qquad 0 \le k \le \Bigl
[\frac{L_y-1}{2} \Bigr ]  \ . 
\label{cdoddtran} 
\eeq
We thus obtain the important general formula, in terms of transfer matrices, 
for the chromatic polynomial for a M\"obius strip of the lattice $\Lambda$,
which is valid for arbitrary width $L_y$:
\beqs
P(\Lambda, L_y \times m, Mb., q) & = &
c^{(0)}Tr[(T_{\Lambda,L_y,0})^{m-1} \tilde T_{\Lambda,L_y,0}] \cr\cr & & +
\sum_{d=0}^{[(L_y-1)/2]} c^{(d+1)} Tr[(T_{\Lambda,L_y,2d+1})^{m-1}
\tilde T_{\Lambda,L_y,2d+1}] \cr\cr & & - \sum_{d=1}^{[L_y/2]}
c^{(d-1)} Tr[(T_{\Lambda,L_y,2d})^{m-1} \tilde T_{\Lambda,L_y,2d}] \ . 
\label{zgsum_transfermb}
\eeqs
(where the factor $c^{(0)}=1$ is displayed explicitly for uniformity). 

For the square lattice, the eigenvalues of $\tilde T_{sq,L_y,d}$
are the same as those of $T_{sq,L_y,d}$ except for possible changes
of signs. The number of eigenvalues with sign changes is the
number of column-exchanges from $T_{sq,L_y,d}$ to $\tilde
T_{sq,L_y,d}$.  Denote the number of eigenvalues that are the
same for $T_{sq,L_y,d}$ and $\tilde T_{sq,L_y,d}$ as
$n_P(sq,L_y,d,+)$, and the number of eigenvalues with different signs
as $n_P(sq,L_y,d,-)$. It is clear that
\beq n_P(sq,L_y,d,+) + n_P(sq,L_y,d,-) = n_P(sq,L_y,d) \ . \label{npld}
\eeq
Define
\beq \Delta n_P(sq,L_y,d) \equiv n_P(sq,L_y,d,+) - n_P(sq,L_y,d,-)
\label{deltanpld} \eeq
which gives the number of partitions having self-reflection
symmetry. For example, among the partitions for the $L_y=2$ strip
in Fig. \ref{L2partitions}, the partition $I$ in ${\cal P}_{2,0}$
and the partition $\bar 1, \bar 2$ in ${\cal P}_{2,2}$ have
self-reflection symmetry. Among the partitions for the $L_y=3$
strip in Fig. \ref{L3partitions}, it includes the two partitions
in ${\cal P}_{3,0}$, the second partition $\bar 2$ and the fourth
partition $\overline{13}$ in ${\cal P}_{3,1}$, the second
partition $\bar 1, \bar 3$ in ${\cal P}_{3,2}$, and the partition
in ${\cal P}_{3,3}$. We list $\Delta n_P(sq,L_y,d)$ for $1 \le L_y
\le 10$ in Table \ref{nppmtable}. The total number of these
partitions for each $L_y$ will be denoted as $\Delta N_{P,L_y}$,
and we have $\Delta N_{P,2n-1} = \Delta N_{P,2n}$ for $n>0$. The
relations between $\Delta n_P(sq,L_y,d)$ are
\beqs 
\Delta n_P(sq,2n-1,0) & = & \Delta n_P(sq,2n,0) \ , \qquad
\mbox{for} \ 0<n \cr\cr \Delta n_P(sq,2n-1,2m-1) & = & \Delta
n_P(sq,2n,2m)  \ , \qquad \mbox{for} \ 1 \le m \le n \cr\cr 
\Delta n_P(sq,2n-1,2m) & = & \Delta n_P(sq,2n,2m-1) \ , \qquad \mbox{for} \ 1
\le m \le n \cr\cr 
\Delta n_P(sq,2n+1,0) & = & 2\Delta n_P(sq,2n,0) +
\Delta n_P(sq,2n,1) \ , \qquad \mbox{for} \ 0<n \cr\cr 
\Delta
n_P(sq,2n+1,m) & = & \Delta n_P(sq,2n,m-1) + \Delta n_P(sq,2n,m) + \Delta
n_P(sq,2n,m+1) \ , \cr\cr
& & \mbox{for} \ 0<m \le 2n+1 \cr\cr 
\Delta n_P(sq,2n+1,2m) & = & \Delta n_P(sq,2n+1,2m+1) \ , \qquad \mbox{for} \ 0
\le m \le n \cr\cr 
\Delta n_P(sq,2n,0) & = & \Delta n_P(sq,2n,2) \cr\cr
\Delta n_P(sq,2n,2m-1) & = & \Delta n_P(sq,2n,2m+2) \ , \ \mbox{for} \
0<m \le n-1 \ .
\eeqs
We also list $n_P(sq,L_y,d,+)$ and $n_P(sq,L_y,d,-)$ for $2 \le L_y \le 10$ in
Table \ref{npctable}. Notice that $n_P(sq,L_y,0,+)$ is the number of
$\lambda_{P,sq,FF,L_y}$ proved in \cite{ts}. Recall the numbers of
$\lambda_{P,sq,L_y,j}$ for the M\"obius strips of the square lattice with
coefficients $\pm c^{(d)}$, defined as $n_{P,Mb}(L_y,d,\pm) \equiv
n_{P,Mb}(sq,L_y,d,\pm) $, have been given in \cite{cf}. With the eqs.
(\ref{cd0tran}) to (\ref{cdoddtran}), the relations between $n_P(sq,L_y,d,\pm)$
and $n_{P,Mb}(sq,L_y,d,\pm)$ are
\beqs 
n_{P,Mb}(sq,L_y,0,\pm) & = & n_P(sq,L_y,0,\pm) + n_P(sq,L_y,2,\mp)
\cr\cr n_{P,Mb}(sq,L_y,k,\pm) & = & n_P(sq,L_y,2k-1,\pm) +
n_P(sq,L_y,2k+2,\mp) \ , \qquad 1 \le k \le \Bigl [ \frac{L_y+1}{2}
\Bigr ] 
\label{npmbnp} 
\eeqs
We illustrate the application of this general formalism in the section where we
give results for specific transfer matrices.  For M\"obius strips of the square
lattice, the sum of the coefficients is given by
\beqs 
P(sq,L_y \times m ,Mb.,q)_{m=0} & = & c^{(0)} \Delta
n_P(sq,L_y,0) + \sum_{d=0}^{[(L_y-1)/2]} c^{(d+1)}\Delta
n_P(sq,L_y,2d+1) \cr\cr & & - \sum_{d=1}^{[L_y/2]} c^{(d-1)}\Delta
n_P(sq,L_y,2d) \cr\cr & = & c^{(0)} [\Delta n_P(sq,L_y,0) - \Delta
n_P(sq,L_y,2)] \cr\cr & & + \sum_{d=1}^{[(L_y+1)/2]} c^{(d)} [\Delta
n_P(sq,L_y,2d-1) - \Delta n_P(sq,L_y,2d+2)] \cr\cr & = & 
\cases{ 0 & for
\ $L_y$ \ even \cr & \cr P({\rm Tree}_{(L_y+1)/2},q)  & for \ $L_y$ \
odd } \ 
\eeqs
which agrees with Theorem 5 in \cite{cf} because of eqn.  (\ref{npmbnp}).  This
$m=0$ equation also holds for the triangular lattice strips. If
we take $m=1$ in eqn. (\ref{zgsum_transfermb}), the chromatic polynomial is
zero for the square lattice with odd $L_y$ or for the triangular lattice with
any $L_y$. This is what we expect since these M\"obius strips have at least one
edge connect a vertex to itself.

For the honeycomb lattice with $L_y$ even, the magnitudes of
eigenvalues of $T_{hc,L_y,d}$ and $\tilde T_{hc,L_y,d}$ are
again the same. We denote the number of eigenvalues that are the
same as $n_P(hc,L_y,d,+)$, and the number of eigenvalues with
different signs as $n_P(hc,L_y,d,-)$. Similarly to eqns.
(\ref{npld}) and (\ref{deltanpld}), we have
\beqs n_P(hc,L_y,d) & = & n_P(hc,L_y,d,+) + n_P(hc,L_y,d,-) \cr\cr
\Delta n_P(hc,L_y,d) & \equiv & n_P(hc,L_y,d,+) - n_P(hc,L_y,d,-) \ . 
\eeqs
The values of $\Delta n_P(hc,L_y,d)$ for even $L_y$ up to $L_y=10$ are listed
in Table \ref{nphcpmtable}. The values $\Delta n_P(hc,L_y,d)$ can be obtained
from $\Delta n_P(sq,L_y,d)$. For example,
\beqs \Delta n_P(hc,2,d) & = & \Delta n_P(sq,2,d) \cr\cr \Delta
n_P(hc,4,d) & = & \Delta n_P(sq,3,d)+\Delta n_P(sq,4,d) \cr\cr \Delta
n_P(hc,6,d) & = & \Delta n_P(sq,4,d)+\Delta n_P(sq,6,d) \cr\cr \Delta
n_P(hc,8,d) & = & \Delta n_P(sq,5,d)+\Delta n_P(sq,6,d)+\Delta
n_P(sq,7,d)+\Delta n_P(sq,8,d) \cr\cr \Delta n_P(hc,10,d) & = & \Delta
n_P(sq,6,d)+2\Delta n_P(sq,8,d)+\Delta n_P(sq,10,d) 
\eeqs
The relations between $\Delta n_P(hc,L_y,d)$ are
\beqs 
\Delta n_P(hc,4n,0) & = & 4\Delta n_P(hc,4n-2,0) + 2\Delta
n_P(hc,4n-2,1) \ , \qquad \mbox{for} \ 0<n \cr\cr 
\Delta n_P(hc,4n,2m-1) & = & n_P(hc,4n,2m) \cr\cr 
& = & 2(\Delta n_P(hc,4n-2,2m-1) + \Delta n_P(hc,4n-2,2m)) \cr\cr 
& & + \Delta n_P(hc,4n-2,2m-2) + \Delta n_P(hc,4n-2,2m+1) \ , \cr\cr
& & {\rm for} \ \ 1 \le m \le 2n \cr\cr 
\Delta n_P(hc,4n+2,0) & = & \Delta
n_P(hc,4n+2,2) \cr\cr 
& = & \Delta n_P(hc,4n,0) + \Delta n_P(hc,4n,2) \ , \qquad \mbox{for} 
\ 1 \le n \cr\cr \Delta
n_P(hc,4n+2,2m-1) & = & n_P(hc,4n+2,2m+2) \cr\cr & = & \Delta
n_P(hc,4n,2m-1) + \Delta n_P(hc,4n,2m+2) \ , \mbox{for} \ 1
\le m \le 2n \cr\cr
& & 
\eeqs
$n_P(hc,L_y,d,+)$ and $n_P(hc,L_y,d,-)$ for even $2 \le L_y \le
10$ are listed in Table \ref{npchctable}. For M\"obius strips of
the honeycomb lattice with even $L_y$, the sum of the coefficients
is given by
\beqs 
P(hc, L_y \times m ,Mb.,q)_{m=0} & = & c^{(0)} \Delta
n_P(hc,L_y,0) + \sum_{d=0}^{[(L_y-1)/2]} c^{(d+1)}\Delta
n_P(hc,L_y,2d+1) \cr\cr & & - \sum_{d=1}^{[L_y/2]} c^{(d-1)}\Delta
n_P(hc,L_y,2d) \cr\cr & = & c^{(0)} [\Delta n_P(hc,L_y,0) - \Delta
n_P(hc,L_y,2)] \cr\cr & & + \sum_{d=1}^{[(L_y+1)/2]} c^{(d)}
[\Delta n_P(hc,L_y,2d-1) - \Delta n_P(hc,L_y,2d+2)] \cr\cr & = &
\cases{ 0 & for \ $L_y/2$ \ odd \cr & \cr \Bigl ( q(q-1) \Bigr
)^{L_y/4} & for \ $L_y/2$ \ even } 
\eeqs

It will be convenient to introduce the following polynomial:
\beq
D_n = \frac{P(C_n,q)}{q(q-1)} = \sum_{j=0}^{n-2} (-1)^j {n-1 \choose j}
q^{n-2-j}
\label{dk}
\eeq
where $P(C_n,q)=(q-1)^n+(q-1)(-1)^n$ is the chromatic polynomial for the
circuit graph with $n$ vertices, $C_n$. Various properties of these polynomials
were given in Refs. \cite{hs,wa2}. The first few $D_n$'s are $D_1=0$, $D_2=1$,
$D_3=q-2$, $D_4=q^3-3q+3$. In addition to the identities given in Refs.
\cite{hs,wa2}, another identity is
\beq
D_n+D_{n-1}=(q-1)^{n-2} \ , n \ge 2 \ . 
\label{dnsum}
\eeq
Certain linear combinations of these polynomials also occur in the entries of
the transfer matrices.  These can be seen to arise as powers of basic linear
factors such as $q-a$ plus integer constants; e.g., $q^2-4q+5=(q-2)^2+1$,
$q^2-6q+10=(q-3)^2+1$, etc.

We shall write out the entries of the transfer matrices explicitly if this
can be done in the space available.  In cases where this is not feasible, we
shall express the entries in terms of the above polynomials, together with
the shorthand notation
\beq
q_a = q-a \ , \quad q_{a,b}=(q-a)(q-b)
\label{qab}
\eeq
and additional shorthand defined in the last appendix, including the
polynomials $F_{m,n}$, $G_{m,n}$, and others.

\section{Some Transfer Matrix Results for Arbitrary $L_{\lowercase{y}}$}

\subsection{$\lowercase{d}=L_{\lowercase{y}}$}

For arbitrary $L_y$, eq. (\ref{nplyly}) shows that for $d=L_y$ the transfer
matrix $T_{\Lambda,L_y,d=L}$ is one-dimensional, i.e., a scalar.  We have
previously proved that \cite{s5}
\beq
\lambda_{\Lambda,L_y,L_y}=(-1)^{L_y} \quad {\rm for} \quad \Lambda=sq,tri 
\ . 
\label{lamsqLL}
\eeq
Our results also led to the inference that \cite{pg,hca}
\beq
\lambda_{\Lambda,L_y,L_y}=1 \quad {\rm for} \quad \Lambda=hc \ . 
\label{lamhcLL}
\eeq
These results can also be seen to follow immediately from the transfer matrix
method.

\subsection{$\lowercase{d}=L_{\lowercase{y}}-1$}

\subsubsection{Square lattice} 

For arbitrary $L_y$, the dimension of the transfer matrix is specified by
eq. (\ref{nplylym1}) as $n_P(\Lambda,L_y,L_y-1)=L_y$ for $\Lambda=sq,tri$. For
this case we have given the transfer matrices $T_{\Lambda,L_y,d=L_y-1}$ for
cyclic strips of the square lattice in eqs. (7.1.1) and (7.1.4)-(7.1.6) of
Ref. \cite{s5} and for cyclic strips of the triangular lattice in
eqs. (7.3.1)-(7.3.4) of Ref. \cite{s5}.  General formulas for eigenvalues and
determinants were also given in Ref. \cite{s5}.  Since we will need these here,
we list them below:
\beq
\lambda_{sq,L_y,d=L_y-1,j} = (-1)^{L_y+1}(q-a_{sq,L_y,j}) \quad , \ 1 \le j
\le L_y
\label{lamdlym1}
\eeq
where
\beq
a_{sq,L_y,j} = 1 + 4\cos^2 \left ( \frac{(L_y+1-j) \pi}{2L_y} \right )
\quad , \ 1 \le j \le L_y \ .
\label{asqlyj}
\eeq
Hence, the determinant and trace of $T_{sq,L_y,L_y-1}$ are
\beq
det(T_{sq,L_y,L_y-1}) = \prod_{j=1}^{L_y}(q-a_{sq,L_y,j})
\label{detTsqLLminus1}
\eeq
\beq
Tr(T_{sq,L_y,L_y-1}) = (-1)^{L_y+1}[2+L_y(q-3)] \ . 
\label{traceTsqLLminus1}
\eeq
Although our general result (\ref{lamdlym1}) exhibits an explicit factorization
of the characteristic polynomial of $T_{sq,L_y,L_y-1}$ into linear factors,
this involves, in general, transcendental numbers (trigonometric functions) as
coefficients.  A few of these are of the form $\pm (q - k)$ where $k$ is an
integer; e.g., for $L_y=2, \ d=1$, one has $\lambda_{sq,2,1,j}=1-q,3-q$ and for
$L_y=3, \ d=2$, one has $\lambda_{sq,3,2,j}=q-1,q-2,q-4$.  Some others involve
terms of the form $\pm (q - a)$ where $a$ is an algebraic number; e.g., for
$L_y=4, \ d=3$, one has $\lambda_{4,3,j} =1-q,3-q,3 \pm \sqrt{2}-q$. In Table
\ref{sqfactors} we indicate the factorization properties of the various
characteristic polynomials of $T_{sq,L_y,d}$ in factors with integer
coefficients; for the illustrative case $L_y=4,d=3$, this is
$CP(T_{sq,4,3},z)=(z-1+q)(z-3+q)(z^2-2(3-q)z+q^2-6q+7)$, symbolized as
$(1^2,2)$ in the table.  We observe that the factorizations often correspond to
the numbers $n_P(sq,L_y,d,+)$ and $n_P(sq,L_y,d,-)$ in Table \ref{npctable},
although sometimes there are additional factorizations.

\subsubsection{Triangular lattice}

For $T_{tri,L_y,L_y-1}$ we proved that \cite{s5} 
\beq
det(T_{tri,L_y,L_y-1}) = (q-2)^2(q-3)^{L_y-2}
\label{detTtriLLminus1}
\eeq
\beq
Tr(T_{tri,L_y,L_y-1}) = (-1)^{L_y+1}[3+L_y(q-4)] \ . 
\label{traceTtriLLminus1}
\eeq
Since a matrix and its transpose have the same eigenvalues, it is a convention
which one is presented.  The transfer matrices obtained directly from our
calculation of the analogous matrices for the full Potts model are the
transposes of those given in \cite{s5}. This just amounts to a different
convention for the ordering of the basis of configurational states.

\subsubsection{Honeycomb lattice}

Here we give an exact determination of $T_{hc,L_y,d}$ for $d=L_y-1$ which is
valid for arbitrary $L_y$.  This goes beyond our results in Ref. \cite{s5},
which presented such formulas for the square and triangular lattices.  We first
observe that this matrix has dimension 
\beq
dim(T_{hc,L_y,L_y-1}) = n_P(hc,L_y,L_y-1) = \cases{ 
         \frac{3L_y-1}{2} & for \ $L_y$ \ odd \cr
          &   \cr
         \frac{3}{2}L_y-1  & for \ $L_y$ \ even }
\label{nphclylyminus1}
\eeq
This can be derived in a straightforward manner from results in \cite{hca}.  It
is convenient to work with two auxiliary sets of matrices, each of dimension
$n_Z(hc,L_y,L_y-1)=2L_y-1$, which are special cases of the transfer matrices
for the full Potts model on strips of the honeycomb lattice,
$S_{hc,L_y,L_y-1,j}$ for $j=1,2$.  The elements of these matrices will be given
below.  We then form the product
\beq
S_{hc,L_y,L_y-1}=S_{hc,L_y,L_y-1,2}S_{hc,L_y,L_y-1,1} \ .  
\label{smatrix}
\eeq
In this matrix certain columns are identically zero.  The number of
such zero columns is
\beq
n_Z(hc,L_y,L_y-1)-n_P(hc,L_y,L_y-1) =  \cases{
         \frac{L_y-1}{2} & for \ $L_y$ \ odd \cr
          &   \cr
         \frac{L_y}{2}  & for \ $L_y$ \ even }
\label{numberofzerocols}
\eeq
The transfer matrix $T_{hc,L_y,L_y-1}$ is then the submatrix of
$S_{hc,L_y,L_y-1}$ formed by removing these columns with vanishing entries and
the corresponding rows, which operation we denote with a subscript $red.$ for
``reduced'':

\beq
T_{hc,L_y,L_y-1}=S_{hc,L_y,L_y-1,red.} \ . 
\label{Thclylyminus1}
\eeq
To determine $T_{hc,L_y,L_y-1}$, it thus suffices to give the 
$S_{hc,L_y,L_y-1,j}$ matrices for $j=1,2$.  For the relevant range $L_y \ge 2$
where the strips of the honeycomb lattice are defined we calculate 
\beq 
(S_{hc,L_y,L_y-1,1})_{j,j}=2-q \quad {\rm for} \quad 1 \le j
\le L_y-1 
\label{TThc1x1} 
\eeq
\beq 
(S_{hc,L_y,L_y-1,1})_{L_y,L_y}= \cases{ 2-q & for $L_y$ even \cr
                                        1-q & for $L_y$ odd  }
\label{TThc1xl} 
\eeq
\beq 
(S_{hc,L_y,L_y-1,1})_{2j-1,2j}= (S_{hc,L_y,L_y-1,1})_{2j,2j-1}= 1
\quad {\rm for} \quad 1 \le j \le [L_y/2] 
\label{TThc1_1} 
\eeq
\beq 
(S_{hc,L_y,L_y-1,1})_{L_y+2j-1,L_y+2j-1}=0 \quad {\rm for} \quad 1
\le j \le [L_y/2] 
\label{TThc1ya} 
\eeq
\beq 
(S_{hc,L_y,L_y-1,1})_{L_y+2j,L_y+2j}=1 \quad {\rm for} \quad 1 \le
j \le [(L_y-1)/2] 
\label{TThc1yb} 
\eeq
\beq
(S_{hc,L_y,L_y-1,1})_{2j-1,L_y+2j-1}=(S_{hc,L_y,L_y-1,1})_{2j,L_y+2j-1}=0
\quad {\rm for} \quad 1 \le j \le [L_y/2]
\label{TThc1wa} 
\eeq
\beq 
(S_{hc,L_y,L_y-1,1})_{2j,L_y+2j}=(S_{hc,L_y,L_y-1,1})_{2j+1,L_y+2j}=-1
\quad {\rm for} \quad 1 \le j \le [(L_y-1)/2]
\label{TThc1wb} 
\eeq
\beq
(S_{hc,L_y,L_y-1,1})_{L_y+2j-1,2j-1}=(S_{hc,L_y,L_y-1,1})_{L_y+2j-1,2j}=-1 
\quad {\rm for} \quad 1 \le j \le [L_y/2] 
\label{TThc1v} 
\eeq
\beq (S_{hc,L_y,L_y-1,2})_{1,1}=1-q 
\label{TThc2x} 
\eeq
\beq 
(S_{hc,L_y,L_y-1,2})_{j,j}=2-q \quad {\rm for} \quad 2 \le j \le L_y-1 
\label{TThc2x1} 
\eeq
\beq 
(S_{hc,L_y,L_y-1,2})_{L_y,L_y}= \cases{ 2-q & for $L_y$ odd \cr
                                        1-q & for $L_y$ even  }
\label{TThc2xl} 
\eeq
\beq 
(S_{hc,L_y,L_y-1,2})_{2j,2j+1}= (S_{hc,L_y,L_y-1,1})_{2j+1,2j}= 1
\quad {\rm for} \quad 1 \le j \le [(L_y-1)/2] 
\label{TThc2_1} 
\eeq
\beq 
(S_{hc,L_y,L_y-1,2})_{L_y+2j-1,L_y+2j-1}=1 \quad {\rm for} \quad 1
\le j \le [L_y/2] 
\label{TThc2ya} 
\eeq
\beq 
(S_{hc,L_y,L_y-1,2})_{L_y+2j,L_y+2j}=0 \quad {\rm for} \quad 1 \le
j \le [(L_y-1)/2] 
\label{TThc2yb} 
\eeq
\beq
(S_{hc,L_y,L_y-1,2})_{2j-1,L_y+2j-1}=(S_{hc,L_y,L_y-1,2})_{2j,L_y+2j-1}=-1
 \quad {\rm for} \quad 1 \le j \le [L_y/2]
\label{TThc2wa} 
\eeq
\beq 
(S_{hc,L_y,L_y-1,2})_{2j,L_y+2j}=(S_{hc,L_y,L_y-1,2})_{2j+1,L_y+2j}=0
\quad {\rm for} \quad 1 \le j \le [(L_y-1)/2]
\label{TThc2wb} 
\eeq
\beq 
(S_{hc,L_y,L_y-1,2})_{L_y+2j,2j}=(S_{hc,L_y,L_y-1,1})_{L_y+2j,2j+1}=-1
\quad {\rm for} \quad 1 \le j \le [(L_y-1)/2] 
\label{TThc2v}
\eeq
with all other elements equal to zero. Using our general formulas, we find
that
\beq
det(T_{hc,L_y,L_y-1})=(q-1)^{L_y+1} \quad {\rm for \ odd} \ \ L_y \ge 3 \ . 
\label{detThclylyminus1}
\eeq
For even $L_y$, $det(T_{hc,L_y,L_y-1})$ is proportional to 
$(q-1)^{L_y}F_{4,3}=(q-1)^{L_y}(q^2-4q+5)$. 

We next proceed to list our explicit calculations of transfer matrices. Since
we have completely determined the transfer matrices in the cases $d=L_y$, where
it is a scalar, and for the case $d=L_y-1$, we do not list the former for each
specific $L_y$.

\section{Strips of the Square Lattice}

\subsection{$L_{\lowercase{y}}=2$}

We recall first that for $L_y=1$, an elementary calculation yields $P(sq,1
\times m, cyc.,q)=(q-1)^m+(q-1)(-1)^m$ so $T_{sq,1,0}=q-1$.  The chromatic
polynomials $P(sq,2 \times m,BC,q)$ for cyclic and M\"obius BC's were given in
Ref. \cite{bds} One has $n_P(\Lambda,2,0)=1$, $n_P(\Lambda,2,1)=2$, 
$n_P(\Lambda2,2)=1$ for $\Lambda=sq$ (or $tri$), $T_{sq,2,0}=q^2-3q+3$, and
\beq
T_{sq,2,1} = -\left( \begin{array}{cc}
             q-2 & -1  \\
             -1  & q-2 \end{array} \right )
\label{Tsq21}
\eeq
yielding the eigenvalues $\lambda_{sq,2,1,1}=1-q$ and
$\lambda_{sq,2,1,1}=3-q$.  The determinant $det(T_{sq,2,1})=(q-1)(q-3)$ and
trace $Tr(T_{sq,2,1})=-2(q-2)$ follow as the $L_y=2$ special case of our
results in \cite{s5}, viz., eqs. (\ref{detTsqLLminus1}) and
(\ref{traceTsqLLminus1}).  For the present strip, the general methods discussed
above yield, for the M\"obius strip, the matrix $\tilde T_{sq,2,1}$ which has
the first and second columns interchanged. 

\subsection{$L_{\lowercase{y}}=3$}

The chromatic polynomials $P(sq,3 \times m,BC,q)$ were given in \cite{wcyl,wcy}
for cyclic BC's and in \cite{pm} for M\"obius BC's.  We find 
\beq
T_{sq,3,0}=  \left( \begin{array}{cc}
    (q-2)(q^2-3q+4) & q^2-4q+5  \\
         1          & q-2        \end{array} \right )
\label{Tsq30}
\eeq

\bigskip

\beq
T_{sq,3,1}= \left( \begin{array}{cccc}
      -(q^2-4q+5) &  q-2      &  -1         &  2-q    \\
         q-2      & -(q-2)^2  &  q-2        &    1    \\
         -1       &  q-2      &-(q^2-4q+5)  &  2-q    \\
          1       &   1       &   1         &  q-2    \end{array} \right )
\label{Tsq31}
\eeq

\bigskip
\beq
T_{sq,3,2}= \left( \begin{array}{ccc}
         q-2   &   -1    &   0   \\
          -1   &   q-3   &  -1   \\
           0   &  -1     &  q-2  \end{array} \right )
\label{Tsq32}
\eeq
Although $T_{sq,3,0}$ is the same as the analogous transfer matrix for the free
strip of width $L_y=3$, $T_{sq,L_y,0}$ is larger than the free-strip transfer
matrix for $L_y \ge 4$.  Note that the upper left-hand $3 \times 3$ block of
$T_{sq,3,1}$ is symmetric.  The matrix $\tilde T_{sq,3,1}$ is obtained by
exchanging columns 1 and 3 of $T_{sq,3,1}$, and $\tilde T_{sq,3,2}$ is obtained
by exchanging columns 1 and 3 of $T_{sq,3,2}$. 

We have
\beq
det(T_{sq,3,0})=(q-1)(q^3-6q^2+13q-11)
\eeq
\beq
det(T_{sq,3,1})=-(q-1)(q-2)^2(q^4-9q^3+29q^2-40q+22)
\eeq
\beq
det(T_{sq,3,2})=(q-1)(q-2)(q-4)
\eeq
\beq
Tr(T_{sq,3,0})=(q-2)(q^2-3q+5)
\eeq
\beq
Tr(T_{sq,3,1})=-(3q^2-13q+16) \ . 
\eeq
$Tr(T_{sq,3,2})$ is given by the $L_y=3,d=2$ special case of the general
formula (7.1.28) of \cite{s5}, given above as eq. (\ref{traceTsqLLminus1}).
The characteristic polynomial $CP(T_{sq,3,1},z)$ factorizes into polynomials of
degree 1 and 3 in $z$.  This is indicated in Table \ref{sqfactors}. The
eigenvalue corresponding to the linear factor is $\lambda_{sq,3,1,1}=-(q-2)^2$
\cite{wcyl}.

\subsection{$L_{\lowercase{y}}=4$}

The chromatic polynomials $P(sq,4 \times m,BC,q)$ for BC=cyclic and M\"obius
were given in \cite{s4}.  We find
\beq
T_{sq,4,0}= \left( \begin{array}{cccc}
s_{21} & (q-2)p_6   & r_{20}    & (q-2)p_6  \\
  q-2  &  (q-2)^2   & 3-q       & 1        \\
  -1   &   2-q      & q^2-5q+7  & 2-q      \\
  q-2  &    1       & 3-q       & (q-2)^2  \end{array} \right )
\label{Tsq40}
\eeq
where the polynomials $s_{21}$, $p_6$, and $r_{20}$ are defined in the
appendix.  

We have
\beqs
& & det(T_{sq,4,0}) = (q-1)(q-3)(q^8-16q^7+112q^6-449q^5+1130q^4-1829q^3 \cr\cr
& & +1858q^2-1084q+279)
\eeqs
and
\beq
Tr(T_{sq,4,0}) = q^4-7q^3+24q^2-45q+36 \ . 
\eeq
As indicated in Table \ref{sqfactors}, the characteristic polynomial of
$T_{sq,4,0}$ consists of factors of degree 1 and 3 in $z$. The eigenvalue which
is the root of the linear factor is $\lambda_{sq,4,0,1}=(q-1)(q-3)$; the
eigenvalues of the cubic factor are the same as the eigenvalues for the free
strip of width $L_y=4$ determined in Ref. \cite{strip}.  The matrix 
$\tilde T_{sq,4,0}$ is obtained by exchanging columns 2 and 4 of $T_{sq,4,0}$.

\bigskip

\beq
T_{sq,4,1}= \left( \begin{array}{ccccccccc}
-r_{13} &F_{4,3}     & -q_2 & 1  &-G_{4,3}& q_2   &-G_{4,3} & -q_2^2 & q_2  \\
F_{4,3} &-q_2F_{4,3} &q_2^2 &-q_2& q_2    &-q_2^2 & q_3   & q_2 & -1   \\
-q_2  & q_2^2  & -q_2F_{4,3}  &F_{4,3}& -1  & q_2   & q_3 &-q_2^2 & q_2  \\
 1  & -q_2     &F_{4,3}&-r_{13}       & q_2 & -q_2^2&-G_{4,3}&q_2 &-G_{4,3} \\
 -1  & 0       & 0       & 0    &-q_2 & 0        & 0    & 0        & 0    \\
 -1  & q_2     & q_2     & q_2  &-q_2 & q_2^2    &-q_3  & 1        & 1    \\
 -1  & -1      & -1      & -1   &-q_2 &-q_2      & G_{4,3} &-q_2   &-q_2  \\
 q_2 & q_2     & q_2     & -1   & 1    & 1       &-q_3   &q_2^2    &-q_2  \\
 0   & 0       & 0       & -1   & 0    & 0        & 0    & 0       &-q_2
\end{array} \right )
\label{Tsq41}
\eeq
The shorthand notation used in eq. (\ref{Tsq41}) and below, such as $F_{m,n}$
and $G_{m,n}$, is defined in the last appendix.  Note that the upper
left-hand $4 \times 4$ block of $T_{sq,4,1}$ is symmetric.  We calculate
\beqs
& & det(T_{sq,4,1}) = (q-1)^3(q-3)^5(q^2-3q+3)^2 \cr\cr
& \times & (q^8-17q^7+125q^6-520q^5+1342q^4-2206q^3+2261q^2-1325q+341)
\label{detTsq41}
\eeqs
\beq
Tr(T_{sq,4,1}) = -4q^3+27q^2-69q+65 \ . 
\eeq
The matrix $\tilde T_{sq,4,1}$ is obtained by exchanging columns 1 and 4, 2 and
3, 5 and 9, and 6 and 8 of $T_{sq,4,1}$.  It is ironic that although the
chromatic polynomial is a special case of the full Potts model partition
function, the determinants of the transfer matrices $T_{\Lambda,L_y,d}$ for the
degree-$d$ subspaces for these square-lattice strips, such as
eq. (\ref{detTsq41}), are more complicated than the simple expression that we
found for the determinant of the transfer matrix $T_{Z,sq,L_y,d}$ in
Ref. \cite{zt},
\beq
det(T_{Z,sq,L_y,d}) = v^{L_y[ n_Z(L_y,d) - n_Z(L_y-1,d)]} 
\biggl [ (v+qr)^{L_y} (v+1)^{L_y-1} \biggr ]^{n_Z(L_y-1,d)}
\label{detTZsqld}
\eeq
where here $n_Z(L_y,d) \equiv n_Z(\Lambda,L_y,d)$ for $\Lambda=sq,tri$. 
Furthermore, the expression for $det(T_{sq,L_y,d})$ for these square-lattice
strips gets more complicated as $L_y$ increases, in contrast to
eq. (\ref{detTZsqld}) and also in contrast to the results that we find for 
$det(T_{\Lambda,L_y,d})$ for $\Lambda=tri,hc$ (see below).  

As indicated in Table \ref{sqfactors}, the characteristic polynomial of
$T_{sq,4,1}$ consists of factors of degree 4 and 5.

\bigskip

\beq
T_{sq,4,2}= \left( \begin{array}{cccccccc}
F_{4,3} & 2-q     & 1        &  0    &  0       &   0   & q- 2   &  0    \\
 2-q   & q^2-5q+6 & 3-q      & 2-q   &  1       &   0   & -1    &  0    \\
  1    & 3-q      & p_8      & 1     & 3-q      &   1   & q-2   & q-2   \\
  0    & 2-q      & 1        &(q-2)^2& 2-q      &   0   & 0     & 0     \\
  0    & 1        & 3-q      & 2-q   & q_{2,3}   & 2-q   & 0     & -1    \\
  0    & 0        & 1        & 0     & 2-q      &F_{4,3} & 0     & q-2   \\
 -1    & -1       & -1       & 0     & 0        & 0     & 2-q   & 0     \\
  0    & 0        & -1       & 0     & -1       & -1    & 0     & 2-q
\end{array} \right )
\label{Tsq42}
\eeq
where $F_{m,n}$ and $p_8$ are defined in the last appendix.  The
upper left-hand $6 \times 6$ block of $T_{sq,4,2}$ is symmetric.  The matrix
$\tilde T_{sq,4,2}$ is obtained by exchanging columns 1 and 6, 2 and 5, and 7
and 8 of $T_{sq,4,2}$. $CP(T_{sq,4,2},z)$ has factors of degree 3 and 5. We
have
\beqs
& & det(T_{sq,4,2})=(q-1)^2(q-3)^2(q^3-7q^2+15q-11) \cr\cr
& \times & (q^7-16q^6+106q^5-378q^4+788q^3-967q^2+653q-189)
\eeqs
\beq
Tr(T_{sq,4,2}) = 6q^2-29q+38
\eeq

\bigskip

\beq
T_{sq,4,3}= -\left( \begin{array}{cccc}
  q-2 & -1  &  0  &  0   \\
   -1 & q-3 & -1  &  0   \\
   0  & -1  & q-3 & -1   \\
   0  & 0   & -1  & q-2  \end{array} \right )
\label{Tsq43}
\eeq

 From the general formulas (\ref{detTsqLLminus1}) and (\ref{traceTsqLLminus1}),
we have $det(T_{sq,4,3})=(q-1)(q-3)(q^2-6q+7)$ and
$Tr(T_{sq,4,3})=-2(2q-5)$.

\subsection{$L_{\lowercase{y}}=5$}

The chromatic polynomials $P(sq,5 \times m,BC,q)$ for BC=cyclic and M\"obius
were given in \cite{s5}.  In addition to the general determination of
$T_{sq,L_y,L_y-1}$, the transfer matrix for $T_{sq,5,3}$ was given in
\cite{s5}. We display here our calculation of $T_{sq,5,0}$:
\beqs
& & T_{sq,5,0}= \cr\cr & & \left(
\begin{array}{ccccccccc} F_{4,3} r_{11} & s_{31} & s_{47} &
s_{61} & q_2^2p_7 & s_{47} & s_{31}& r_{19} & q_2p_8\\
F_{4,3}& q_2F_{4,3} & -q_{2,3} & 2q-5 & q_2 & q_3 & q_2 & 1 & q_2^2\\
-q_2 & -q_2^2 & q_2G_{4,3} & -p_8 & -q_2^2 & q_3 & -1 & -q_2 & -q_2\\
1 & q_2 & -G_{4,3} & r_{20} & q_2 & -G_{4,3} & q_2 & G_{4,3} & q_2^2\\
q_2^2 & q_2 & -q_{2,3} & 2q_3 & q_2^3 & -q_{2,3} & q_2 & q_{2,3}& 1\\
-q_2 & -1 & q_3 & -p_8 & -q_2^2 & q_2G_{4,3} & -q_2^2 & -q_2 & -q_2\\
F_{4,3} & q_2 & q_3 & 2q-5 & q_2 & -q_{2,3} & q_2F_{4,3} & 1 & q_2^2\\
0 & 0 & 0 & 1 & 0 & 0 & 0 & q_2 & 0\\
-1 & -q_2 & -q_3 & 5-2q & 1 & -q_3 & -q_2 & -1 & -q_2^2
\end{array} \right ) \cr\cr
& & \label{Tsq50}
\eeqs
where the various $p, r, s$ polynomials are defined in the appendix. 
The characteristic polynomial of $T_{sq,5,0}$ consists of factors of 
degree 2 and 7 in $z$. The eigenvalues that are the roots of the 
degree-7 factor are the same as those found in Ref. \cite{strip} for the 
corresponding $L_y=5$ free strip of the square lattice.  

The transfer matrices $T_{sq,5,d}$ for $d=1,2$ are both of dimension $21 \times
21$ and are too lengthy to present here; they are available from the
authors. We find
\beqs
& & det(T_{sq,5,0})=-(q-1)^3(q-2)^2(q^6-13q^5+69q^4-191q^3+292q^2-236q+79)
\cr\cr
& & \times (q^{15}-34q^{14}+538q^{13}-5259q^{12}+35541q^{11}-176036q^{10}
\cr\cr
& & +660682q^9-1914798q^8+4324155q^7-7615130q^6+10381339q^5 \cr\cr
& & -10768339q^4+8235159q^3-4388527q^2+1459163q-228580) \ . 
\eeqs
The determinant $det(T_{sq,5,1})$ factorizes as $(q-1)^6(q-2)^2P_{20}P_{29}$,
where $P_{20}$ and $P_{29}$ are polynomials in $q$ of degree 20 and 29 in
$q$. $det(T_{sq,5,2})$ factorizes as $(q-1)^6(q-2)P_{19}P_{23}$, where $P_{19}$
and $P_{23}$ are polynomials in $q$ of degree 19 and 23. For $d=3$ we have
\beqs
& & det(T_{sq,5,3})=(q-1)^3(q-2)(q^8-20q^7+171q^6-817q^5+2389q^4
\cr\cr
& & -4387q^3+4954q^2-3155q+869)(q^{11}-29q^{10}+375q^9-2853q^8+14188q^7
\cr\cr
& & -48439q^6+115934q^5-194762q^4+225461q^3-171683q^2+77617q-15835)
\eeqs
and, from eq. (\ref{detTsqLLminus1}),
\beq
det(T_{sq,5,4})=(q-1)(q^2-5q+5)(q^2-7q+11)
\eeq
We obtain the traces
\beq
Tr(T_{sq,5,0})=q^5-9q^4+42q^3-120q^2+195q-137
\eeq
\beq
Tr(T_{sq,5,1})=-5q^4+46q^3-180q^2+346q-269
\eeq
\beq
Tr(T_{sq,5,2})=10q^3-74q^2+198q-189
\eeq
\beq
Tr(T_{sq,5,3})=-10q^2+51q-69
\eeq
with $Tr(T_{sq,5,4})$ being given by eq. (\ref{traceTsqLLminus1}).

\section{Strips of the Triangular Lattice}

\subsection{$L_{\lowercase{y}}=2$}

The chromatic polynomials $P(tri,2 \times m,BC,q)$ for BC=cyclic and M\"obius
were given in \cite{wcy}.  One has $\lambda_{tri,2,0}=(q-2)^2$, and, as a
special case of our general result in eqs. (7.1.1) and (7.3.1)-(7.3.4) of
\cite{s5},
\beq
T_{tri,2,1} = -\left( \begin{array}{cc}
              q-3 & q-2  \\
              -1  & q-2 \end{array} \right )
\label{Ttri21} \eeq
with eigenvalues
\beq
\lambda_{tri,2,1,j} = \frac{1}{2}\biggl [ 5-2q \pm \sqrt{9-4q} \ \biggr ]
\label{lamtri21}
\eeq
Note that we use a different ordering convention for the basis configurations
here than the one we used in Ref. \cite{s5}, so that $T_{tri,L_y,L_y-1}$ in
terms of our present basis is the transpose of the corresponding matrix given
in \cite{s5}. This has no effect on the eigenvalues, since the eigenvalues of a
matrix $A$ and its transpose $A^T$ are the same.  We have
$det(T_{tri,2,1})=(q-2)^2$ and $Tr(T_{tri,2,1})=5-2q$, in accordance with the
general formulas (\ref{detTtriLLminus1}) and (\ref{traceTtriLLminus1}).  For
the M\"obius strip, the $\tilde T_{tri,2,1}$ matrix is obtained from
$T_{tri,2,1}$ by interchanging the two columns, as was true for the square
lattice strip.  However, in contrast to the case of the square-lattice strip,
where this just reverses the sign of one of the eigenvalues, here the $\tilde
T_{tri,2,1}$ matrix has different eigenvalues than $T_{tri,2,1}$ (namely,
$(1/2)(3-q \pm \sqrt{5q^2-22q+25} \ )$).  It is instructive to display how our
general formula operates.  For comparison, we first display the explicit
chromatic polynomial for the cyclic strip \cite{wcy}:
\beq
P(tri,2 \times m,cyc.,q)=c^{(0)}(\lambda_{tri,2,0})^m +
c^{(1)}\biggl [ (\lambda_{tri,2,1,1})^m + (\lambda_{tri,2,1,2})^m \biggr ]
 + c^{(2)}
\label{tpxy2}
\eeq
If one expresses the chromatic polynomial for the
M\"obius strip as a sum of powers of the same eigenvalues as for the cyclic 
strip, the result is \cite{wcy}
\beq
P(tri,2 \times m,Mb.,q)=c^{(0)}(\lambda_{tri,2,0})^m 
-\frac{(q-1)(q-3)}{\sqrt{9-4q}}\biggl [ (\lambda_{tri,2,1,1})^m - 
(\lambda_{tri,2,1,2})^m \biggr ] - c^{(0)}
\label{tpxy2mb}
\eeq
One sees that the coefficient of the second and third terms is no longer in the
set of $c^{(d)}$'s and, indeed, is not a polynomial function of $q$.  Of
course, the square root in the denominator of this coefficient cancels so that,
as must be true, the chromatic polynomial is a polynomial in $q$.  As discussed
in \cite{pm}, this can be seen as a consequence of the theorem on symmetric
functions of roots of a polynomial equation whose coefficients are polynomials
in $q$; although these roots are not, in general polynomial functions of $q$,
the symmetric functions can be expressed in terms of polynomials in $q$.  In
the present case, the second and third terms can be written in a manifestly
symmetric manner by observing that the denominator of the coefficient is equal
to $\lambda_{tri,2,1,1}-\lambda_{tri,2,1,2}$.  With our general formula
(\ref{zgsum_transfermb}) we express the chromatic polynomial in a form that
keeps the coefficients in the set of $c^{(d)}$ polynomials. 

\subsection{$L_{\lowercase{y}}=3$}

The chromatic polynomials $P(tri,3 \times m,BC,q)$ for BC=cyclic and M\"obius
were given in \cite{t}.  We have
\beq
T_{tri,3,0} = \left( \begin{array}{cc}
     (q-2)(q^2-5q+7) & (q-3)^2  \\
          2-q        & q-3 \end{array} \right )
\label{Ttri30}
\eeq

\bigskip
This yields the same eigenvalues that were earlier calculated for the free
$L_y=3$ strip in Ref. \cite{strip} and can be taken to be the transfer matrix 
for that strip. 
\beq
T_{tri,3,1}= \left( \begin{array}{cccc}
 -(q^2-6q+10)& -(q-2)(q-3)  & q-2         &  3-q  \\
     q-3     & -(q-2)(q-3)  & -(q-2)(q-3) &  3-q  \\
     -1      & q-2          & -(q-2)(q-3) &  3-q  \\
     2       & 2-q          & 2-q     &  q-3  \end{array} \right )
\label{Ttri31}
\eeq

\bigskip

The characteristic polynomial $CP(T_{tri,3,0},z)$ is a quadratic, while
$CP(T_{tri,3,1},z)$ has a linear and cubic factor, as indicated in Table
\ref{trifactors}.  We note that for odd $L_y$ the factorizations given in Table
\ref{trifactors} correspond to the numbers $n_P(sq,L_y,d,+)$ and
$n_P(sq,L_y,d,-)$ given in Table \ref{npctable}, but there are no nontrivial
factorizations for even $L_y$.

\bigskip
As a special case of our general formulas given in eqs. (7.3.1)-(7.3.4) of
\cite{s5}, we have
\beq
T_{tri,3,2}= \left( \begin{array}{ccc}
         q-3   &  q-4     &  q-2   \\
         -1    &  q-4     &  q-2   \\
         0     &  -1      &  q-2  \end{array} \right )
\label{Ttri32} \eeq
We calculate
\beq
det(T_{tri,3,0})=(q-2)^3(q-3)
\eeq
\beq
det(T_{tri,3,1}) = -(q-2)^5(q-3)^2
\eeq
\beq
det(T_{tri,3,2})=(q-2)^2(q-3)
\eeq
\beq
Tr(T_{tri,3,0})=q^3-7q^2+18q-17
\eeq
\beq
Tr(T_{tri,3,1})=-3q^2+17q-25
\eeq
\beq
Tr(T_{tri,3,2})=3(q-3) \ . 
\eeq

\subsection{$L_{\lowercase{y}}=4$}

The chromatic polynomials $P(tri,4 \times m,cyc.,q)$ were given in \cite{t}.
The transfer matrices that we give below, together with our general procedure
for obtaining $\tilde T_{\Lambda,L,d}$ from $T_{\Lambda,L,d}$, yields a
new solution for $P(tri,4 \times m,Mb.,q)$. We find
\beq
T_{tri,4,0}= \left( \begin{array}{cccc}
 (q-2)(q-3)p_8   & (q-2)(q-3)(q-4) & (q-3)p_{15}  & (q-3)p_{10}   \\
-(q-2)(q-3)      & (q-2)(q-3)      & 2(3-q)       & 3-q           \\
  q-2            & (q-2)(q-3)      & q^2-7q+13    & 3-q           \\
-(q-2)(q-3)      & (q-2)(q-3)      & (q-3)(q-5)   & (q-3)^2
\end{array} \right )
\label{Ttri40}
\eeq
where $p_8$, etc. are defined in the appendix.  This yields
the same characteristic polynomial and eigenvalues as those obtained for 
the free $L_y=4$ strip in Ref. \cite{strip} and can be taken as the transfer
matrix for that strip.  We have 
\beq
det(T_{tri,4,0})=(q-2)^6 (q-3)^4
\eeq
\beq
Tr(T_{tri,4,0})=q^4-10q^3+42q^2-88q+76 \ . 
\eeq
(We recall that the actual form of the matrix is basis-dependent; for example,
for the free $L_y=4$ strip, a different matrix which, however, also has the
same characteristic polynomial and eigenvalues originally calculated in
Ref. \cite{strip}, was given in Ref. \cite{tritran}.)
\beq
T_{tri,4,1}=\left( \begin{array}{ccccccccc}
-r_{34}&-q_2p_{10} &q_{2,3}  &-q_2 &-q_{3,5} &-q_{2,3} &-p_{13} &-q_3^2 &q_3 \\
p_{10} &-q_2p_{10} &-q_2q_3^2&q_{2,3}&q_3 &-q_{2,3} &2q_3 &-q_3^2 &q_3  \\
-q_3   &q_{2,3}&-q_2q_3^2&-q_2p_{10} &q_3 &-q_{2,3} &-p_{13}&-q_3^2 &-q_{3,4} \\
1    &-q_2  &q_{2,3} &-q_2p_{10} &q_3 &-q_{2,3} &-p_{13} &q_3  &-q_{3,4} \\
q_4  &q_2   & 0      & 0         &-q_3& 0       & 0      & 0   & 0       \\
-2   &2q_2  & -q_{2,3}&-q_{2,3}  &-q_3& q_{2,3} & -2q_3  &-q_3 & -q_3    \\
-2   &q_2   & q_2     & q_2      &-2q_3&q_{2,3} &p_{13}  &-q_3 & -q_3    \\
2q_3 &-q_{2,3}&-q_{2,3}&2q_2     &-2q_3&q_{2,3} &p_{14}  &q_3^2&-2q_3    \\
0    & 0    & 0        & q_2     & 0   & 0      & 1      & 0   & -q_3
\end{array} \right )
\label{Ttri41}
\eeq
We calculate
\beq
det(T_{tri,4,1})=(q-2)^{12}(q-3)^8
\eeq
\beq
Tr(T_{tri,4,1})=-2(q-3)(2q^2-12q+21)
\eeq
\beq
T_{tri,4,2}=\left( \begin{array}{cccccccc}
 p_{10} & q_{3,4} & -q_4   & q_{2,3} & -q_2    & 0       & q_3 & 0    \\
 -q_3   & q_{3,4} & p_{14} & q_{2,3} & q_{2,4} & -q_2    & q_3 & q_3  \\
 1      & -q_4    & p_{14} & -q_2    & q_{2,4} & -q_2    & q_3 & q_3  \\
 0      & -q_3    & -q_4   & q_{2,3} & q_{2,4} & q_{2,3} & 0   & q_3  \\
 0      & 1       & -q_4   & -q_2    & q_{2,4} & q_{2,3} & 0   & q_3  \\
 0      & 0       & 1      & 0       & -q_2    & q_{2,3} & 0   & q_3  \\
-2      & q_4     & q_4    & q_2     & q_2     & 0       & -q_3& 0    \\
0       & 0       & -2     & 0       & q_2     & q_2     & 0   & -q_3
\end{array} \right )
\label{Ttri42}
\eeq
\beq
det(T_{tri,4,2})=(q-2)^8(q-3)^6
\eeq
\beq
Tr(T_{tri,4,2})=6q^2-38q+62
\eeq

As a special case of our general result in \cite{s5}, we have
\beq
T_{tri,4,3}= -\left( \begin{array}{cccc}
  q-3  &  q-4  &  q-4 &  q-2   \\
  -1   &  q-4  &  q-4 &  q-2   \\
  0    &  -1   &  q-4 &  q-2   \\
  0    &  0    &  -1  &  q-2
\end{array} \right )
\label{Ttri43}
\eeq
The determinant and trace are the $L_y=4$ special case of our general formulas
(\ref{detTtriLLminus1}) and (\ref{traceTtriLLminus1}), viz.,
$det(T_{tri,4,3})=(q-2)^2(q-3)^2$ and $Tr(T_{tri,4,3})=13-4q$.

\subsection{$L_{\lowercase{y}}=5$}

We present here new results for $P(tri,5 \times m,BC,q)$ for BC=cyclic and
M\"obius.  For $d=0$ we calculate
\beqs
& & T_{tri,5,0}= \cr\cr
& & \left( \begin{array}{ccccccccc}
q_2 s_{82} & q_{2,3}p_{13} & q_2 r_{58} & s_{173} & q_{2,3}p_{13} & s_{148}
  &q_3 r_{34}&q_3 p_{17}&q_5q_3^2\\
-q_2 p_{10}&q_2q_3^2 &-2q_{2,3} & -q_4^2& -q_{2,3}&-p_{13} &-q_{3,4}&-q_3
  &q_3^2\\
 q_{2,3}& q_2q_3^2 &q_2p_{13} & -q_4^2& -q_{2,3}  & 2q_3  & q_3   &-q_3
  &q_3^2\\
-q_2   & -q_{2,3}  &q_2 p_{13}  &r_{47} & -q_{2,3} &-p_{13} & q_3 &q_{3,5}
  &q_3^2\\
-q_2q_3^2& q_2q_3^2&q_2q_4^2 &-q_4(2q-7)&q_2q_3^2&-2q_3^2&-q_3^2&q_{3,4}
  &q_3^2\\
q_{2,3}&-q_{2,3}&q_2p_{13}& r_{49} & q_2q_3^2 &q_3p_{13}&-q_3^2&q_3(2q-7)
  &q_3^2 \\
-q_2 p_{10} &-q_{2,3} &-2q_{2,3} &-q_4(2q-7) &q_2q_3^2 &r_{46} &q_3p_{10}
  &q_{3,4}&q_3^2 \\
  0    & 0        & -q_2   & -q_4   & 0        & 0    & 0     & q_3  & 0    \\
2q_2 & q_{2,3}   & 2q_{2,3} & q_{4,5} & q_{2,3}  &p_{14} &-2q_3  &2q_3
  & -q_3^2
\end{array} \right ) \cr\cr
& &
\label{Ttri50}
\eeqs
where the various $p$, $r$, and $s$ polynomials are defined in the appendix.
We find
\beq
det(T_{tri,5,0})=-(q-2)^{14}(q-3)^{12}
\eeq
\beq
Tr(T_{tri,5,0})=(q-3)(q^4-10q^3+46q^2-112q+118) \ . 
\eeq
The characteristic polynomial of $T_{tri,5,0}$ consists of factors of degree 2
and 7.

The transfer matrices $T_{tri,5,d}$ for $d=1,2$ are both of dimension $21
\times 21$ and are too lengthy to present here; they are available from the
authors.  However, we do note the simple results
\beq
det(T_{tri,5,1})=(q-2)^{30}(q-3)^{27}
\eeq
\beq
Tr(T_{tri,5,1})=-5q^4+62q^3-306q^2+708q-642
\eeq
\beq
det(T_{tri,5,2})=(q-2)^{25}(q-3)^{24}
\eeq
\beq
Tr(T_{tri,5,2})=10q^3-98q^2+332q-387 \ . 
\eeq
Factorizations of $CP(T_{tri,5,j},z)$, $j=1,2$, are given in Table
\ref{trifactors}.

For $d=3$ we calculate
\beqs
& & T_{tri,5,3}= \cr\cr
& & \left( \begin{array}{ccccccccccccc}
-p_{10}&-q_{3,4}&q_4    &-q_{3,4}&q_4 &0  &-q_{2,3} &q_2 &0 &0 &-q_3 &0 & 0 \\
q_3&-q_{3,4}&-p_{14}&-q_{3,4}&-q_4^2 &q_4 &-q_{2,3}&-q_{2,4}&q_2&0&-q_3&-q_3&0\\
-1 & q_4 &-p_{14} &q_4 &-q_4^2 &q_4 &q_2 &-q_{2,4} &q_2 &0 &-q_3 &-q_3 & 0  \\
0&q_3&q_4&-q_{3,4}&-q_4^2&-p_{14} &-q_{2,3} &-q_{2,4}&-q_{2,4}&q_2&0&-q_3&-q_3\\
0&-1 &q_4 &q_4 &-q_4^2 &-p_{14} &q_2 &-q_{2,4}&-q_{2,4}&q_2 &0 &-q_3 &-q_3 \\
0 &0 &-1 &0 &q_4 &-p_{14} &0 &q_2 &-q_{2,4} &q_2 &0 &-q_3 &-q_3 \\
0 &0 &0 &q_3 &q_4 &q_4 &-q_{2,3} &-q_{2,4} &-q_{2,4} &-q_{2,3} &0 &0 &-q_3 \\
0 & 0 & 0 &-1 &q_4 &q_4  &q_2&-q_{2,4} &-q_{2,4} &-q_{2,3}& 0 & 0 &-q_3 \\
0 & 0 & 0 & 0 & -1 & q_4 & 0 & q_2 & -q_{2,4} & -q_{2,3} & 0 & 0 & -q_3 \\
0 & 0 & 0 & 0 & 0  & -1  & 0 & 0 & q_2 & -q_{2,3} & 0 & 0 & -q_3\\
2 & -q_4 & -q_4 & -q_4 & -q_4 & 0 & -q_2 & -q_2 & 0 & 0 & q_3 & 0 & 0 \\
0 & 0 & 2 & 0 & -q_4 & -q_4 & 0 & -q_2 & -q_2 & 0 & 0 & q_3 & 0 \\
0 & 0 & 0 & 0 & 0 & 2 & 0 & 0 & -q_2 & -q_2 & 0 & 0 & q_3
\end{array} \right ) \cr\cr
& &
\label{Ttri53}
\eeqs
We calculate
\beq
det(T_{tri,5,3})=(q-2)^{11}(q-3)^{12}
\eeq
\beq
Tr(T_{tri,5,3})=-10q^2+67q-115 \ . 
\eeq
The characteristic polynomial of $T_{tri,5,3}$ has factors of degree 5 and 8.

For $d=4$, our general theorem in \cite{s5} yields
\beq
T_{tri,5,4}= \left( \begin{array}{ccccc}
  q-3 & q-4 & q-4 & q-4 & q-2  \\
  -1  & q-4 & q-4 & q-4 & q-2  \\
  0   & -1  & q-4 & q-4 & q-2  \\
  0   & 0   & -1  & q-4 & q-2  \\
  0   & 0   & 0   & -1  & q-2   \end{array} \right )
\label{Ttri54} \eeq
In accordance with our general results (\ref{detTtriLLminus1}) and
(\ref{traceTtriLLminus1}), we have
\beq
det(T_{tri,5,4})=(q-2)^2(q-3)^3
\eeq

\beq
Tr(T_{tri,5,4})=5q-17 \ . 
\eeq
The characteristic polynomial of $T_{tri,5,4}$ has factors of degree 2 and 3.

\section{Strips of the Honeycomb Lattice}

\subsection{$L_{\lowercase{y}}=2$}

The chromatic polynomials $P(hc,2 \times m,BC,q)$ for BC=cyclic and M\"obius
were given in \cite{pg}.  For $d=0$, $T_{hc,2,0}$ is a scalar,
$T_{hc,2,0}=\lambda_{hc,2,0}=D_6=q^4-5q^3+10q^2-10q+5$. For $d=1$, from our
general formulas above, we obtain 
\beq
S_{hc,2,1,1}= \left( \begin{array}{ccc}
           2-q  & 1   &  0  \\
           1    & 2-q &  0  \\
           -1   & -1  &  0  \end{array} \right )
\label{Shc211}
\eeq
\beq
S_{hc,2,1,2}= \left( \begin{array}{ccc}
           1-q  & 0   & -1  \\
           0    & 1-q & -1  \\
           0    & 0   &  1  \end{array} \right )
\label{Shc212}
\eeq
whence 
\beq
T_{hc,2,1} = \left( \begin{array}{cc}
           D_4      & -D_3  \\
           -D_3     &  D_4   \end{array} \right ) =
 \left( \begin{array}{cc}
           q^2-3q+3      & 2-q  \\
           2-q     &  q^2-3q+3   \end{array} \right )
\label{Thc21}
\eeq
yielding the relevant special case of the eigenvalues given in eq. (6), (7)
of \cite{pg}, $\lambda_{hc,2,1,1}=D_4-D_3=F_{4,3}=q^2-4q+5$ and
$\lambda_{hc,2,1,2}=D_4+D_3=(q-1)^2$.  We note that
\beq
det(T_{hc,2,1})=(q-1)^2(q^2-4q+5)
\label{detT_hc21}
\eeq
and $Tr(T_{hc,2,1})=2D_4$

\subsection{$L_{\lowercase{y}}=3$}

Further, 
\beq
T_{hc,3,0}= \left( \begin{array}{ccc}
   q_{1,2}s_7 & q_2s_{3,5}   &  t_{5,5}   \\
   q_1        & F_{4,3}       &  -q_3      \\
-F_{6,2}      & -q_2p_4      & -F_{5,4}   \end{array} \right )
\label{Thc30}
\eeq
so that 
\beq
det(T_{hc,3,0})=(q-1)^4(q-2)^2
\eeq
\beq
Tr(T_{hc,3,0})=q^6-8q^5+28q^4-56q^3+71q^2-58q+26 \ . 
\eeq

For $d=1$, 
\beq
T_{hc,3,1}= \left( \begin{array}{cccccc}
q_{1,2}p_4  &-q_1 F_{4,3} & 1       & r_{13}  & -G_{4,3} & q_2 p_6    \\
-q_1 q_2^2  &q_{1,2} D_4  &-D_5     &-q_{2,3} & q_3 D_4  & -2q_2^2    \\
q_{1,2}     &-q_1D_4      & F_{6,5} &-q_{2,3} & q_3 D_4  &q_2 F_{4,3} \\
-q_{1,2}    & q_1         & 0       &-q_3     & 1        & -q_2       \\
-q_{1,2}    & q_1 D_4     &D_5      &-2q_2    & 2D_4     & q_{2,3}    \\
  q_1       & q_1         & -1      &-q_3     & -q_3     & F_{4,3}
\end{array} \right )
\label{Thc31}
\eeq
 from which it follows that 
\beq
det(T_{hc,3,1})=(q-1)^8(q-2)^2
\eeq
\beq
Tr(T_{hc,3,1})=3q^4-18q^3+46q^2-60q+37 \ . 
\eeq

For $d=2$, applying our general formulas above, we calculate
\beq
S_{hc,3,2,1}= \left( \begin{array}{ccccc}
  2-q  &   1   &   0   &   0   &   0   \\
  1    &  2-q  &   0   &   0   &  -1   \\
  0    &  0    &  1-q  &   0   &   -1  \\
  -1   &  -1   &   0   &   0   &   0   \\
  0    &   0   &   0   &   0   &   1  
\end{array} \right )
\label{Thc321}
\eeq
\beq
S_{hc,3,2,2}= \left( \begin{array}{ccccc}
  1-q  &   0   &   0   &  -1   &   0   \\
   0   &  2-q  &   1   &  -1   &   0   \\
   0   &   1   &  2-q  &   0   &   0   \\
   0   &   0   &   0   &   1   &   0   \\
   0   &  -1   &  -1   &   0   &   0
\end{array} \right )
\label{Th322}
\eeq
The factorizations for $\Lambda=hc$ are given in Table \ref{hcfactors}.  We
note that for even $L_y$, the factorizations in this table correspond to the
numbers $n_P(hc,L_y,d,+)$ and $n_P(hc,L_y,d,-)$ given in Table \ref{nphctable},
when these numbers can be defined, but there are no nontrivial factorizations
for odd $L_y$. We calculate
\beq
det(T_{hc,3,2})=(q-1)^4
\eeq
\beq
Tr(T_{hc,3,2})=3q^2-10q+12
\eeq

\subsection{$L_{\lowercase{y}}=4$}

For the $6 \times 6$ dimensional matrix $T_{hc,4,0}$ we calculate 
\beq
det(T_{hc,4,0})=2(q-1)^8(q-2)^2(q^2-5q+7)
(2q^6-20q^5+83q^4-185q^3+239q^2-175q+60)
\eeq
\beq
Tr(T_{hc,4,0})=q^8-11q^7+55q^6-165q^5+333q^4-480q^3+503q^2-362q+142 \ . 
\eeq
The characteristic polynomial has factors of degree 1 and 5, as indicated in
Table \ref{hcfactors}.  The eigenvalues are
\beq
\lambda_{hc,4,0,1}=(q-1)^2(q^2-5q+7)
\eeq
and the roots of the factor of degree 5.

For the $13 \times 13$ dimensional matrix $T_{hc,4,1}$ we calculate 
\beqs
& & det(T_{hc,4,1})=4(q-1)^{16}(q-2)^2
(q^8-15q^7+100q^6-389q^5+974q^4-1623q^3 \cr\cr
& & +1774q^2-1171q+360)(q^{12}-20q^{11}+185q^{10}-1048q^9+4061q^8
\cr\cr
& & -11385q^7+23784q^6-37468q^5+44352q^4-38630q^3+23631q^2-9200q+1750)
\cr\cr
& &
\eeqs
\beq
Tr(T_{hc,4,1})=4q^6-36q^5+147q^4-351q^3+529q^2-489q+229 \ . 
\eeq
The characteristic polynomial factorizes into factors of degree 5 and 8.

For the $11 \times 11$ dimensional matrix $T_{hc,4,2}$ we find
\beqs
& & det(T_{hc,4,2})=(q-1)^{12}(q^6-10q^5+44q^4-108q^3+160q^2-141q+60) \cr\cr
& & \times (2q^8-29q^7+186q^6-690q^5+1634q^4-2564q^3+2647q^2-1670q+500)
\eeqs
\beq
Tr(T_{hc,4,2})=6q^4-39q^3+110q^2-157q+104
\eeq

For $T_{hc,4,3}$, applying our general formula (\ref{Thclylyminus1}), we
calculate 
\beq
T_{hc,4,3}= \left( \begin{array}{ccccc}
D_4     & 2-q     & 0        & 0      & 0         \\
3-q     & F_{4,3} & 2-q      & 1      & q-3       \\
1       & 2-q     & F_{4,3}  & 3-q    & q-3       \\
0       & 0       & 2-q      & D_4    & 0         \\
-1      & q-2     & q-2      & -1     & 2    \end{array} \right )
\label{Thc43}
\eeq
whence 
\beq
det(T_{hc,4,3})=2(q-1)^4(q^2-4q+5)
\eeq
\beq
Tr(T_{hc,4,3})=2(2q^2-7q+9)
\eeq
The eigenvalues are
\beq
\lambda_{hc,4,3,1}=(q-1)^2
\eeq
\beq
\lambda_{hc,4,3,2}=F_{4,3}=q^2-4q+5
\eeq
and the three roots of
\beq
z^3-2(q^2-4q+6)z^2+(q^4-8q^3+26q^2-36q+21)z-2(q-1)^2=0 \ . 
\eeq

\subsection{$L_{\lowercase{y}}=5$}

We have calculated the $T_{hc,5,d}$ and find 
\beq
det(T_{hc,5,0}) =(q-1)^{36}(q-2)^{12}
\label{detThc50}
\eeq
\beqs
Tr(T_{hc,5,0}) & = & q^{10}-14q^9+91q^8-364q^7+1007q^6-2058q^5+3232q^4 \cr\cr
& & -3950q^3+3662q^2-2350q+808
\eeqs
\beq
det(T_{hc,5,1}) =(q-1)^{78}(q-2)^{24}
\label{detThc51}
\eeq
\beqs
Tr(T_{hc,5,1}) & = & 5q^8-60q^7+336q^6-1152q^5+2676q^4-4374q^3 \cr\cr
& & +4999q^2-3720q+1429
\eeqs
\beq
det(T_{hc,5,2})=(q-1)^{66}(q-2)^{16}
\label{detThc52}
\eeq
\beq
Tr((T_{hc,5,2})=10q^6-96q^5+422q^4-1084q^3+1742q^2-1686q+805
\eeq
\beq
det(T_{hc,5,3})=(q-1)^{30}(q-2)^4
\label{detThc53}
\eeq
\beq
Tr(T_{hc,5,3})=10q^4-68q^3+202q^2-302q+207
\eeq
\beq
det(T_{hc,5,4})=(q-1)^6
\label{detThc54}
\eeq
\beq
Tr(T_{hc,5,4})=5q^2-18q+24
\eeq
The matrix $T_{hc,5,4}$ is sufficiently small that we can list it here:
\beq
T_{hc,5,4}= \left( \begin{array}{ccccccc}
 D_4   &  2-q    &   0    &   0    &  0     &  0     &  0      \\
 3-q   &F_{4,3}  & 2-q    &   1    &  0     & q-3    &  0      \\
  1    & 2-q     &F_{4,3} & 3-q    &  0     & q-3    &  0      \\
  0    & 0       & 3-q    &F_{4,3} &  1-q   & 0      &  q-3    \\
  0    & 0       &  1     & 2-q    & q_{1,2}& 0      &  q-3    \\
 -1    & q-2     &  q-2   & -1     & 0      & 2      &  0      \\
  0    & 0       &  -1    & q-2    & q-1    & 0      &  2
    \end{array} \right )
\label{Thc54}
\eeq
Factorizations of characteristic polynomials are given in Table
\ref{hcfactors}.

\section{Some Algebraic Properties of Evaluations of Transfer Matrices}

We have studied algebraic properties of evaluations of transfer matrices
$T_{\Lambda,L_y,d}$ at special values of $q$.  We find many interesting
results, and mention only a few here.  

First, from our exact results for the $T_{\Lambda,L_y,d}$ for $d=L_y-1$, we can
calculate properties of powers of these matrices, in particular,
when evaluated at special values of $q$. For example,
\beq
[T_{sq,2,1}]^m = 2^{m-1}\left( \begin{array}{cc}
   1 & 1  \\
   1 & 1 \end{array} \right ) \quad {\rm for} \quad q=1, \quad n \ge 1 
\label{Tsq21_q1_power}
\eeq
and, for odd $m=2n+1$ and even $m=2n$, 
\beq
[T_{sq,2,1}]^{2n+1} = \left( \begin{array}{cc}
    0 & 1  \\
    1 & 0 \end{array} \right ) \quad {\rm for} \quad q=2, \quad n \ge 0 
\label{Tsq21_q2_oddpower}
\eeq
\beq
[T_{sq,2,1}]^{2n} = \left( \begin{array}{cc}
     1 & 0  \\
     0 & 1 \end{array} \right ) \quad {\rm for} \quad q=2, \quad n \ge 1 
\label{Tsq21_q2_evenpower}
\eeq
and
\beq
[T_{sq,2,1}]^m = (-1)^m 2^{m-1}\left( \begin{array}{cc}
      1 & -1  \\
     -1  & 1 \end{array} \right ) \quad {\rm for} \quad q=3, \quad m \ge 1 \ .
\label{Tsq21_q3_power}
\eeq
Two other identities are given below (where $n \ge 1$ and there is no sum on
$j$):
\beq
[(T_{sq,4,3})^{2n}]_{jj} = 2^{n-1}(1+2^{n-1})  \quad {\rm for} \quad q=3
\eeq
\beq
[(T_{sq,6,5})^{2n}]_{jj} = \frac{1}{3}\Bigl [ 2^{2n-1} + 3^n + 1 \Bigr ]
\quad {\rm for} \quad q=3 \ . 
\eeq

We next consider transfer matrices for strips of the triangular lattice. 
In general, from eq. (\ref{detTtriLLminus1}), it follows that
the evaluation of $T_{tri,L_y,L_y-1}$ at $q=2$ is noninvertible and the
evaluation of this matrix at $q=3$ is noninvertible if $L_y \ge
3$. First, we observe that 
\beq
[T_{tri,2,1}]^m = \left( \begin{array}{cc}
    1 & 0  \\
    1 & 0 \end{array} \right ) \quad {\rm for} \quad q=2 \ . 
\label{Ttri21_q2}
\eeq
The matrix $T_{tri,2,1}$ at $q=3$ and its powers form the cyclic group of order
3, \ ${\mathbb Z}_3$, with $[T_{tri,2,1}]^{m+3}=[T_{tri,2,1}]^m$ and 
\beqs
& & 
T_{tri,2,1} = \left( \begin{array}{cc}
    0 & -1  \\
    1 & -1 \end{array} \right ) \ ,  \quad 
[T_{tri,2,1}]^2 = \left( \begin{array}{cc}
   -1 & 1  \\
   -1 & 0 \end{array} \right ) \ , \quad
[T_{tri,2,1}]^3 = \left( \begin{array}{cc}
    1 & 0  \\
    0 & 1 \end{array} \right ) \cr\cr
& & \quad {\rm for} \quad q=3
\label{Ttri21_q3}
\eeqs
Given that the transfer matrices are real, if such a matrix satisfies
$(T_{\Lambda,L_y,d})^p=I_k$, where $I_k$ denotes the $k \times k$ identity
matrix, this implies that $det(T_{\Lambda,L_y,d})= \pm 1$, and if $p$ is odd,
as is the case here, with $p=3$, then $det(T_{\Lambda,L_y,d})=1$. Referring to
our general result (\ref{detTtriLLminus1}), we see that this condition is,
indeed, met with $k=2$ if $L_y=2$, $d=1$, and $q=3$.

The matrix $T_{tri,3,0}$ evaluated at $q=2$ satisfies the relation (denoting
this by $T$ for short) $T^3=T$, so, together with the $2 \times 2$ identity
matrix $I_2$, $(T_{tri,3,0})_{q=2}$, $[(T_{tri,3,0})_{q=2}]^2$ form a
multiplicative semigroup of order 3.  (Here, we recall that a semigroup
satisfies the same axioms as a group but without the axiom that inverses of
elements must exist; this is also called a semigroup with unit or monoid
\cite{jacobson}.  The same matrix evaluated at $q=3$ is idempotent, i.e.,
satisfies $T^2=T$, so that $I_2$ and $(T_{tri,3,0})_{q=3}$ form a semigroup of
order 2.  One surmises that connections can be made between these cycle lengths
and the dependence of the chromatic number of the cyclic strip of the 
triangular lattice on its length (see appendix), but we have not pursued this. 

The matrix $T_{tri,3,1}$ evaluated at $q=3$ satisfies the relation (again
denoting this by $T$ for short) $T^4=T$ so that $I_4$, $(T_{tri,3,1})_{q=3}$,
$[(T_{tri,3,1})_{q=3}]^2$, and $[(T_{tri,3,1})_{q=3}]^3$ form a semigroup of
order 4.  The matrix $T_{tri,3,2}$ evaluated at $q=3$ satisfies the relation
$T^3=T$, and hence the elements $I_3$, $(T_{tri,3,2})_{q=3}$, and
$[(T_{tri,3,2})_{q=3}]^2$ form a semigroup of order 3.  (These are not groups
since $(T_{tri,3,1})_{q=3}$ and $(T_{tri,3,2})_{q=3}$ are not invertible.)  We
have found a number of similar algebraic properties of powers of the
$T_{\Lambda,L_y,d}$ matrices evaluated for special values of $q$.

\section{Discussion of Determinants}

\subsection{Square-Lattice Strips} 

\begin{figure}[hbtp]
\centering
\leavevmode
\epsfxsize=4.0in
\begin{center}
\leavevmode
\epsffile{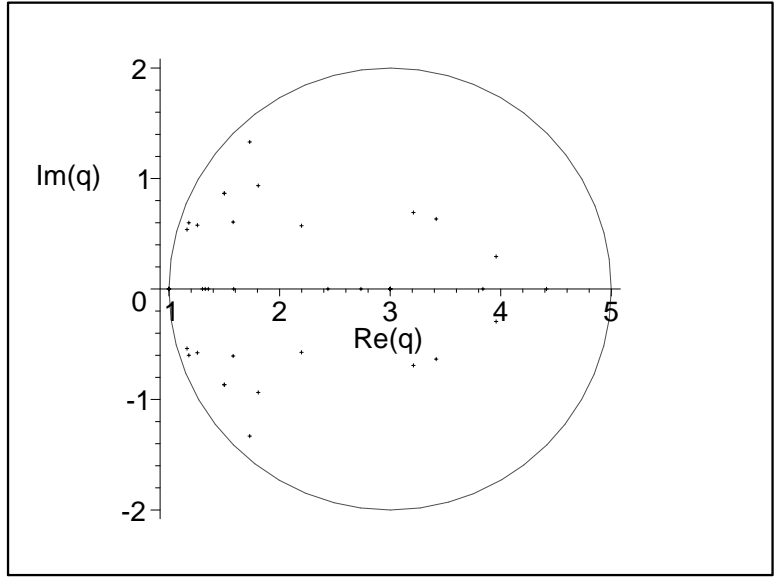}
\end{center}
\caption{\footnotesize{Zeros of $det(T_{sq,L})$ for $L=4$ in the complex $q$ 
plane.}}
\label{zerodetsqpxy4}
\end{figure}

\begin{figure}[hbtp]
\centering
\leavevmode
\epsfxsize=4.0in
\begin{center}
\leavevmode
\epsffile{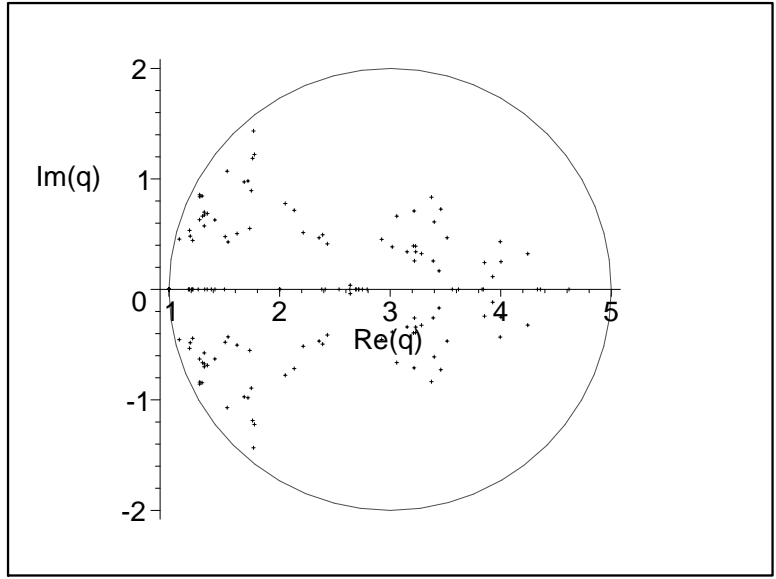}
\end{center}
\caption{\footnotesize{Zeros of $det(T_{sq,L})$ for $L=5$ in the complex $q$ 
plane.}}
\label{zerodetsqpxy5}
\end{figure}

Here we discuss further the determinants $det(T_{\Lambda,L_y,d})$ for
$\Lambda=sq,tri,hc$, starting with the square lattice .  We first consider the
zeros of the determinants, restricting to the range $0 \le d \le L_y-1$ since
$T_{sq,L_y,L_y}=(-1)^{L_y}$ has no zeros. From our theorems in Ref. \cite{s5},
it follows that $det(T_{sq,L_y,L_y-1})$ has (i) the factor $(q-1)$ always; (ii)
the factor $(q-3)$ if $L_y=0$ mod 2; (iii) the factors $(q-2)$ and $(q-4)$ if
$L_y=0$ mod 3; further, all of the zeros of $det(T_{sq,L_y,L_y-1})$ lie in the
interval $1 \le q \le 5$ and, as $L_y \to \infty$, they become dense in this
interval.  One may also investigate the properties of the zeros of
$T_{sq,L_y,d}$ for other values of $d$ than $L_y-1$. Below we list some zeros;
for multiple zeros, we indicate the multiplicity in parentheses, $(m=2)$, etc.
\beq
det(T_{sq,2,0})=0 \quad {\rm at} \quad q=\frac{1}{2}(3 \pm i \sqrt{3} \ ) =
1.5 \pm 0.866i
\label{detTsq20z}
\eeq
\beq
det(T_{sq,3,0})=0 \quad {\rm at} \quad q=1, \quad 1.659 \pm 1.1615i, \quad
2.682
\label{detTsq30}
\eeq
\beq
det(T_{sq,3,1})=0 \quad {\rm at} \quad q=1, \quad 1.229 \pm 0.712i, \quad
2\ (m=2), \quad 3.271 \pm 0.449i
\label{detTsq31z}
\eeq
\beq
det(T_{sq,3,2})=0 \quad {\rm at} \quad q=1, \ 2, \ 4
\label{detTsq32z}
\eeq
\beq
det(T_{sq,4,0})=0 \ \ {\rm at} \ \  q=1, \ 1.163 \pm 0.539i, \ 1.355, \
1.730 \pm 1.331i, \ 2.437, 3, 3.210 \pm 0.6918i
\label{detTsq40z}
\eeq
\beqs
& & det(T_{sq,4,1})=0 \quad {\rm at} \quad q=1 \ (m=3), \ 1.257 \pm 0.578i, \
1.302, \ 1.5 \pm 0.866i \ (m=2), \cr\cr
& & 1.807 \pm 0.936i,  2.736, \ 3 (m=5), \ 3.4169 \pm 0.6339i
\label{detTsq41z}
\eeqs
\beqs
& & det(T_{sq,4,2})=0 \quad {\rm at} \quad q=1 \ (m=2), \ 1.180 \pm 0.600i, \
1.327, \ 1.580 \pm 0.606i, \cr\cr
& &  2.198 \pm 0.573i, \ 3 \ (m=2), \ 3.839, \ 3.959 \pm 0.294i
\label{detTsq42z}
\eeqs
\beq
det(T_{sq,4,3})=0 \quad {\rm at} \quad q=1, \ 1.586, \ 3, \ 4.414
\label{detTsq43z}
\eeq
and so forth for higher $L_y$ values.  We plot these zeros in
Figs. \ref{zerodetsqpxy4}-\ref{zerodetsqpxy5}. (The figures are formatted such
that, to save space, the vertical axis is shown intersecting the horizontal
axis slightly to the left of $q=1$.) 

We observe that all of the zeros of $det(T_{sq,L_y,d})$ that we have calculated
satisfy the condition $|q-3| \le 2$.  As is evident in eq. (\ref{detThc20z}),
some zeros of $det(T_{hc,L_y,d})$ have real parts that lie outside the interval
$1 \le Re(q) \le 5$.  It would be of interest to prove a theorem bounding the
complex zeros of the determinants analogous to results that have been obtained
for chromatic zeros, such as the theorem in Ref. \cite{sokalzeros}.  We also
note that for $d \ne 0$ (as well as $d \ne L$), for the cases that we have
calculated, $det(T_{sq,L_y,d})$ has the factor $(q-1)$. Although
$det(T_{sq,2,0})=T_{sq,2,0}$ does not have this factor, we observe that for the
cases we have calculated with $L_y \ge 3$, $det(T_{sq,L_y,0})$ contains the
factor $(q-1)$.

\subsection{Triangular-Lattice Strips}

An interesting finding is that the determinants of
$T_{tri,L_y,d}$ are the simplest among the regular homopolygonal lattices,
square, triangular, and honeycomb.  Restricting to the range $0 \le d \le
L_y-1$ to exclude $T_{tri,L_y,L_y}=(-1)^{L_y}$, which has no zeros, we observe
that all of the results that we have obtained are consistent with the following
formula, which we conjecture to hold for general $L_y$ (over the range 
$L_y \ge 2$ where the triangular lattice strip is defined):
\beq
det(T_{tri,L_y,d})= \eta (q-2)^a(q-3)^b
\label{detTtrilyd}
\eeq
where $\eta=\pm 1$ and 
\beq
a= 2 n_P(tri,,L_y-1,d) + (L_y-2) n_P(tri,L_y-2,d)
\label{aform}
\eeq
and
\beq
b=(L_y-2) n_P(tri,L_y-1,d) \ . 
\label{bform}
\eeq
Here the sign $\eta$ depends on $L_y$ and $d$, and we take $n_P(tri,1,d) \equiv
n_P(sq,1,d)$, in accord with the general equality
$n_P(tri,L_y,d)=n_P(sq,L_y,d)$ \cite{cf}.  (It is not necessary to give a
formal definition of $n_P(tri,0,d)$ because the only time it appears in the
equation for $a$, for the lowest nontrivial case $L_y=2$, it is multiplied by
zero.)  Some structural properties of this conjecture are indirectly motivated
by the exact formula that we have found in Ref. \cite{zt} for
$det(T_{Z,tri,L_y,d})$, where $T_{Z,tri,L_y,d}$ is the transfer matrix, in the
degree-$d$ subspace, for the full Potts model partition function on the given
strip.  However, aside from the factors of $(v+1)$ in $T_{Z,tri,L_y,d}$ that
cause it to vanish in the special case $v=-1$ corresponding to the chromatic
polynomial, the rest of the expression that we have established for
$det(T_{Z,tri,L_y,d})$ does not involve factors that reduce to powers of
$(q-2)$ or $(q-3)$ when $v=-1$; instead, it involves powers of $(q+v)$ and $v$,
which reduce to $(q-1)$ and $\pm 1$.  The main structural feature of our known
result for $det(T_{Z,tri,L_y,d})$ that influenced our present conjecture is
that the powers of these factors involve quantities such as $n_Z(L_y,d)$ and
$n_Z(L_y-1,d)$, where $n_Z(L_y,d)$ is the dimension of $T_{Z,\Lambda,L_y,d}$
(which is the same for $\Lambda=sq,tri,hc$) \cite{cf,hca}.  This led us to
construct and test conjectured forms for the chromatic polynomial
$det(T_{tri,L_y,d})$ that involve powers of the analogous dimensions
$n_P(tri,L_y,d)$.

 The conjecture (\ref{detTtrilyd}) implies that the zeros of these determinants
$det(T_{tri,L_y,d})$ occur at the two points $q=3$ and $q=2$ (for $0 \le d \le
L_y-1$), in accordance with all of our explicit calculations.  It is
interesting to observe that these are, respectively, the chromatic number
$\chi(tri)$ and $\chi(tri)-1$ of the two-dimensional lattice.  For the special
case $d=L_y-1$, eq. (\ref{detTtrilyd}) (with the plus sign) reduces to our
previously proved result (\ref{detTtriLLminus1}).  For $L_y \ge 2$ and
$d=L_y-2$, eq. (\ref{detTtrilyd}) can be written explicitly as
\beq
det(T_{tri,L_y,L_y-2}) = \eta (q-2)^{3L_y-4}(q-3)^{(L_y-1)(L_y-2)}
\label{detTtrilylyminus2}
\eeq
Starting from the other end of the range of $d$, for $d=0,1$ (and $L_y \ge 2$),
eq. (\ref{detTtrilyd}) can be written explicitly in terms of the Motzkin
numbers $M_n$ defined in eq. (\ref{motzkin}) as 
\beq
det(T_{tri,L_y,0})= \eta (q-2)^{2M_{L_y-2}+(L_y-2)M_{L_y-3}}
(q-3)^{(L_y-2)M_{L_y-2}}
\label{detTtrily0}
\eeq
\beq
det(T_{tri,L_y,1})= \eta (q-2)^{2M_{L_y-1}+(L_y-2)M_{L_y-2}}
(q-3)^{(L_y-2)M_{L_y-1}}
\label{detTtrily1}
\eeq
Similar explicit formulas can be obtained from (\ref{detTtrilyd}) for other
values of $d$.

\subsection{Honeycomb-Lattice Strips}

In the case of the honeycomb lattice, we have
\beq
det(T_{hc,2,0})=0 \quad {\rm at} \quad 0.691 \pm 0.951i, \ 1.809 \pm 0.588i
\label{detThc20z}
\eeq
\beq
det(T_{hc,2,1})=0 \quad {\rm at} \quad 1 \ (m=2), \ 2 \pm i
\label{detThc21z}
\eeq
As is evident in our results above, for $0 \le d \le 2$, the zeros of 
$det(T_{hc,3,d})$ occur at the two points $q=1$ and $q=2$. 
Restricting to $0 \le d \le L_y-1$ to exclude the constant $T_{hc,L_y,L_y}=1$,
and to odd widths $L_y$, we observe that all of our results are consistent 
with the following formula, which we conjecture to hold in general: 
\beq
det(T_{hc,L_y,d})=(q-1)^s(q-2)^t  \quad {\rm for \ odd} \quad L_y 
\label{detThclyd}
\eeq
where
\beq
s=(L_y+1) n_P(hc,L_y-1,d)
\label{cform}
\eeq
and
\beq
t=(L_y-1) n_P(hc,L_y-2,d)
\label{dform}
\eeq
where we take $n_P(hc,1,d) \equiv n_P(sq,1,d)$.  The motivations for this
conjecture are based in part on the exact expression that we have found 
elsewhere \cite{zt} for $det(T_{Z,hc,L_y,d})$, where $T_{Z,hc,L_y,d}$ is the
transfer matrix, in the degree-$d$ subspace, for the partition function of 
the full Potts model on the given strip. However, we note that 
our exact expression for $det(T_{Z,hc,L_y,d})$ does not have a factor that
reduces to a power of $(q-2)$ for the special case $v=-1$ that defines the
chromatic polynomial.  Thus, as in the case of the other lattices, the detailed
structure of the determinants of $T_{Z,\Lambda,L_y,d}$ for the full partition
function and $T_{P,\Lambda,L_y,d}$, denoted here simply as $T_{\Lambda,L_y,d}$,
for the special case $v=-1$ are different.  This is not surprising, since 
the matrix $T_{P,\Lambda,L_y,d}$ does not arise, in general, from a block
decomposition of $T_{Z,\Lambda,L_y,d}$ but instead by removing zero columns 
and corresponding rows of the latter matrix. It is interesting that the
zeros of $det(T_{hc,L_y,d})$ occur at $q=\chi(hc)$ and $q=\chi(hc)-1$, where 
$\chi(hc)=2$ is the chromatic number for the honeycomb lattice.  For the
special case $d=L_y-1$, eq. (\ref{detThclyd}) agrees with the result
(\ref{detThclylyminus1}) that we have proved from our general calculation of
$T_{hc,L_y,L_y-1}$.  For the special case $d=L_y-2$, eq. (\ref{detThclyd}) can
be written explicitly as 
\beq
det(T_{hc,L_y,L_y-2})=(q-1)^{(L_y+1)(3L_y-5)/2} (q-2)^{L_y-1}
\quad {\rm for \ odd} \quad L_y \ . 
\label{detThclylyminus2}
\eeq
Similar explicit formulas can be obtained from (\ref{detThclyd}) for other
values of $d$.  Both of our conjectured general formulas for
$det(T_{\Lambda,L_y,d})$ have the form
\beq
det(T_{\Lambda,L_y,d}) = \eta \biggl [q-(\chi(\Lambda)-1) \biggr ]^{p_1}
\biggl [q-\chi(\Lambda) \biggr ]^{p_2}
 \quad {\rm for} \ \ \Lambda=tri,hc
\label{detform}
\eeq
where $\chi(\Lambda)$ is the chromatic number of the respective two-dimensional
lattices $\Lambda=tri,hc$; $\eta=\pm 1$ for $\Lambda=tri$ and $\eta=1$ for
$\Lambda=hc$; $L_y \ge 3$ is odd for the honeycomb-lattice strips; and the
powers $p_1$ and $p_2$ have been given in (\ref{aform}), (\ref{bform}),
(\ref{cform}), and (\ref{dform}).

\section{Accumulation Locus of Chromatic Zeros}

Although the locus ${\cal B}$ for a strip with a given width $L_y$ and free
longitudinal boundary conditions (and free or periodic transverse boundary
conditions) is different from that for the corresponding strip with periodic or
twisted periodic longitudinal boundary conditions, one notices that in some
ways these loci become more similar as $L_y$ increases.  One expects, for
example, that in the limit $L_y \to \infty$, with any of these boundary
conditions, one will obtain the same value of $q_c$ for the ${\cal B}$ for the
Potts model on the resultant infinite 2D lattice.  From our new exact
calculations, we find the following results. First, for the $5 \times \infty$
cyclic/M\"obius strip of the triangular lattice, ${\cal B}$ crosses the real
$q$ axis at $q=0,2,3$, and the maximal point
\beq
q_c=3.33245.. \ , \quad \Lambda=tri, \quad L_y \times L_x = 5 \times \infty,
\quad {\rm BC} \ = \ {\rm cyclic/Mobius} \ . 
\label{qc_tpxy5}
\eeq
 From our new calculations for the $L_y=4$ and $L_y=5$ cyclic/M\"obius strips
of the honeycomb lattice and our analysis of the locus ${\cal B}$ in the $L_x
\to \infty$ limit, we find that in both cases
${\cal B}$ crosses the real $q$ axis at the points $q=0$ and $q=2$, and at
the maximal crossing points
\beq
q_c=2.15476..  \ , \quad \Lambda=hc, \quad L_y \times L_x = 4 \times \infty,
\quad {\rm BC} \ = \ {\rm cyclic/Mobius}
\label{qc_hpxy4}
\eeq
\beq
q_c=2.26407...\ ,  \quad \Lambda=hc, \quad L_y \times L_x = 5 \times \infty,
\quad {\rm BC} \ = \ {\rm cyclic/Mobius} \ . 
\label{qc_hpxy5}
\eeq
In Table \ref{qctable} we present a summary of these new results, together with
the results that we have obtained for smaller values of $L_y$ in earlier work.
Our present calculations are in agreement with our previous conjecture, based
on all of the exact results that we had obtained, that for the infinite-length
limit of a cyclic or M\"obius strip graph of a given lattice $\Lambda$, with a
given width $L_y$, $q_c$ is a nondecreasing function of $L_y$.  For each type
of lattice, the values of $q_c$ increase toward the values for the infinite 2D
lattices, $q_c(sq)=3$ \cite{lenard}, $q_c(tri)=4$ \cite{baxter87}, and,
formally, $q_c(hc)=(3+\sqrt{5})/2 \simeq 2.618$ \cite{ssbounds,p3afhc}.
Although the values of $q_c$ for each of these types of lattice strips are less
than the value of $q_c$ for the respective two-dimensional lattices, we have
shown previously that this need not necessarily be the case.  For example, for
(the $L_x \to \infty$ limit of ) the $L_y=3$ strip of the square lattice with
toroidal boundary conditions \cite{tk}, $q_c=3$, equal to the value for the
square lattice, and, indeed, for the cyclic self-dual strip of the square
lattice, $q_c=3$ for each of the widths for which we obtained exact
calculations \cite{jz,dg,sdg}.  Nevertheless, the observed monotonically
nondecreasing behavior of the $q_c$ values provides a nice interpolation
between the value $q_c=2$ for the one-dimensional Potts antiferromagnet and the
values on the respective two-dimensional lattices.  As noted before, in
general, if one uses strips with free longitudinal boundary conditions, the
locus ${\cal B}$ does not necessarily cross the real $q$ axis \cite{strip}, and
in cases where it does not, there is, strictly speaking, no $q_c(\{G \})$. In
these cases, the way that one gains information is different; one examines the
endpoints of the rightmost arcs on ${\cal B}$ that are closest to the real
axis.  These are found to exhibit the general tendency to move gradually to the
right with endpoints that move closer to the real axis as the strip width is
increased \cite{strip,strip2,bcc,sqtran,cyltran,tritran,s4,s5,t}.  The results
are consistent with the inference that as $L_y \to \infty$, the arcs will merge
to form closed curves on ${\cal B}$ and the right-most part of ${\cal B}$ will
cross the real axis at the value of $q_c$ for the corresponding two-dimensional
lattice. (One can attempt to carry out a similar type of study for
three-dimensional lattices, and exploratory work has been done in \cite{hd}.)

\begin{figure}[hbtp]
\centering
\leavevmode
\epsfxsize=4.0in
\begin{center}
\leavevmode
\epsffile{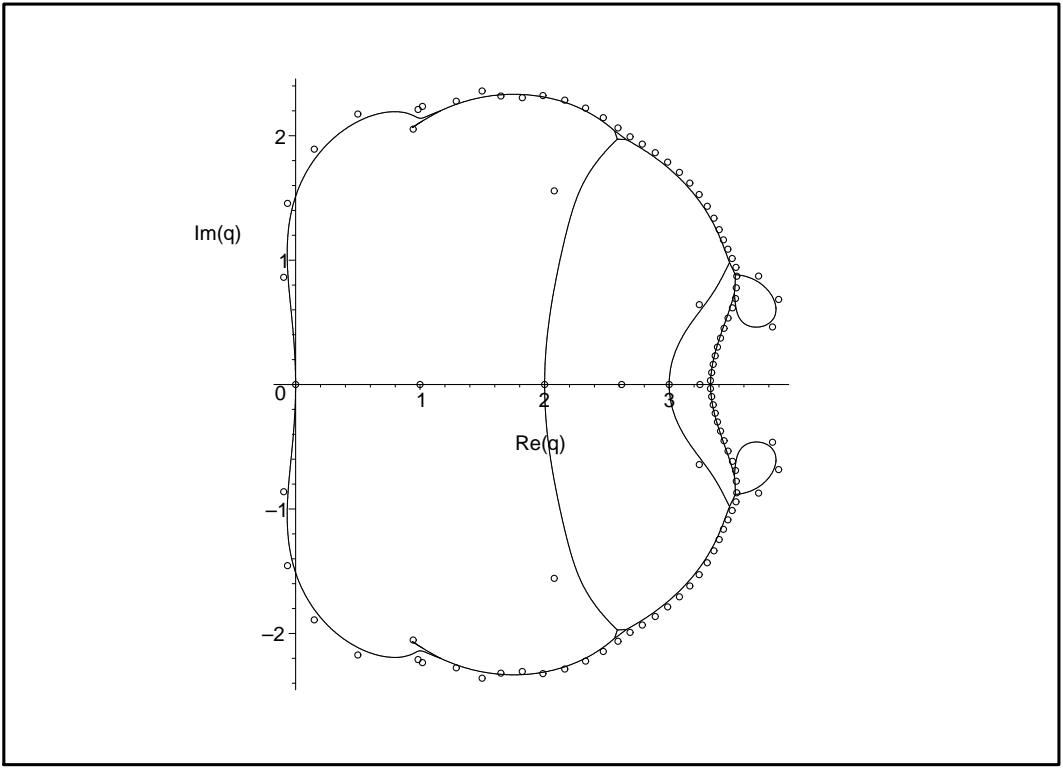}
\end{center}
\caption{\footnotesize{Singular locus ${\cal B}$ in the complex $q$ plane for
the $L_x \to \infty$ limit of the cyclic strip of the triangular lattice with
width $L_y=5$.  For comparison, the plot also shows zeros of $P(tri,L_y \times
L_x,cyc.,q)$ for $L_y=5$ and a typical large value of the length, $L_x=20$.}}
\label{tpxy5zeros}
\end{figure}

\begin{figure}[hbtp]
\centering
\leavevmode
\epsfxsize=4.0in
\begin{center}
\leavevmode
\epsffile{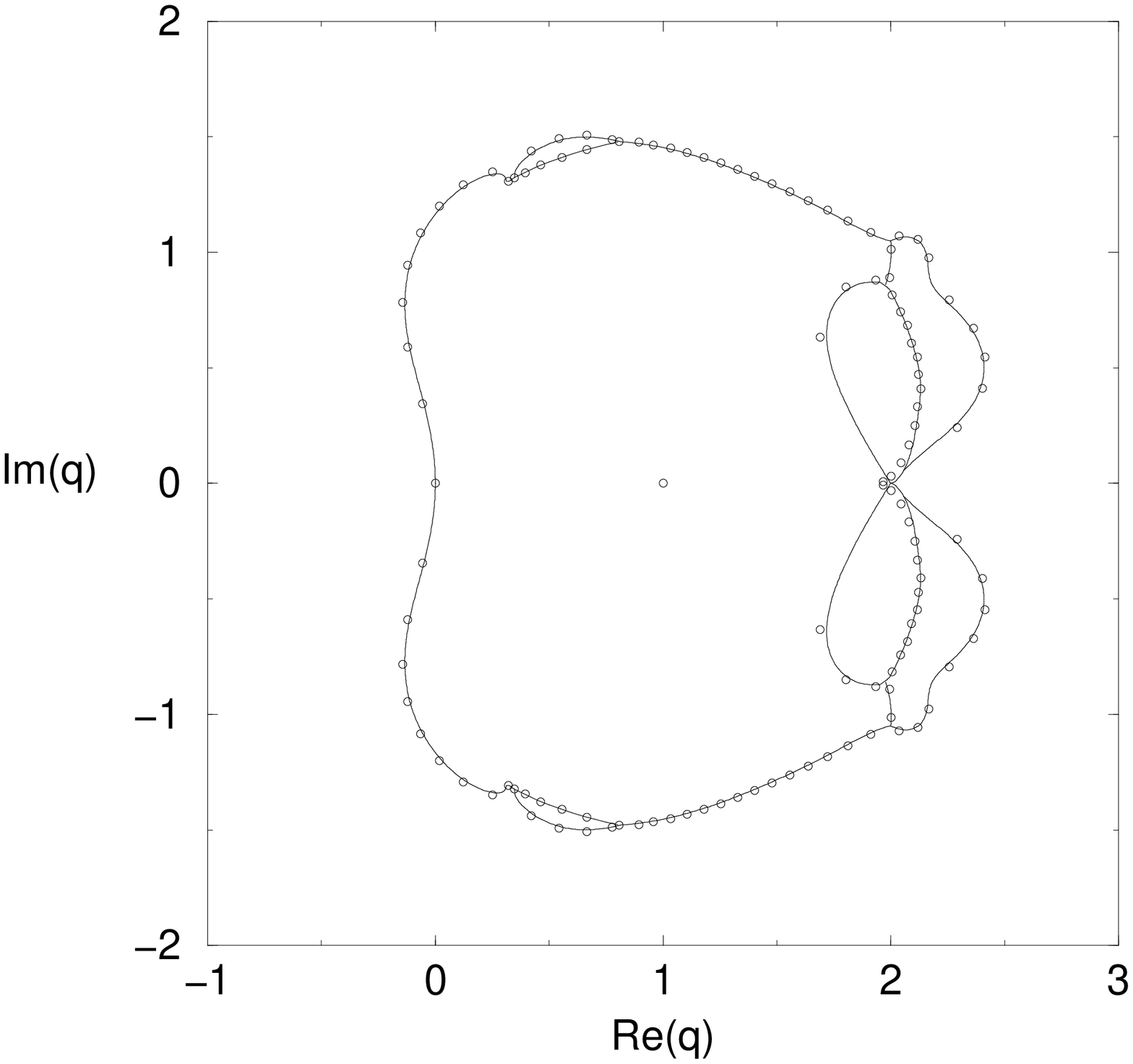}
\end{center}
\caption{\footnotesize{Singular locus ${\cal B}$ in the complex $q$ plane for
the $L_x \to \infty$ limit of the cyclic strip of the honeycomb lattice with
width $L_y=3$.  For comparison, the plot also shows zeros of $P(tri,L_y \times
L_x,cyc.,q)$ with $L_y=3$ and a typical large value of the length, $L_x=40$.}}
\label{hcpxy3zeros}
\end{figure}

\begin{figure}[hbtp]
\centering
\leavevmode
\epsfxsize=4.0in
\begin{center}
\leavevmode
\epsffile{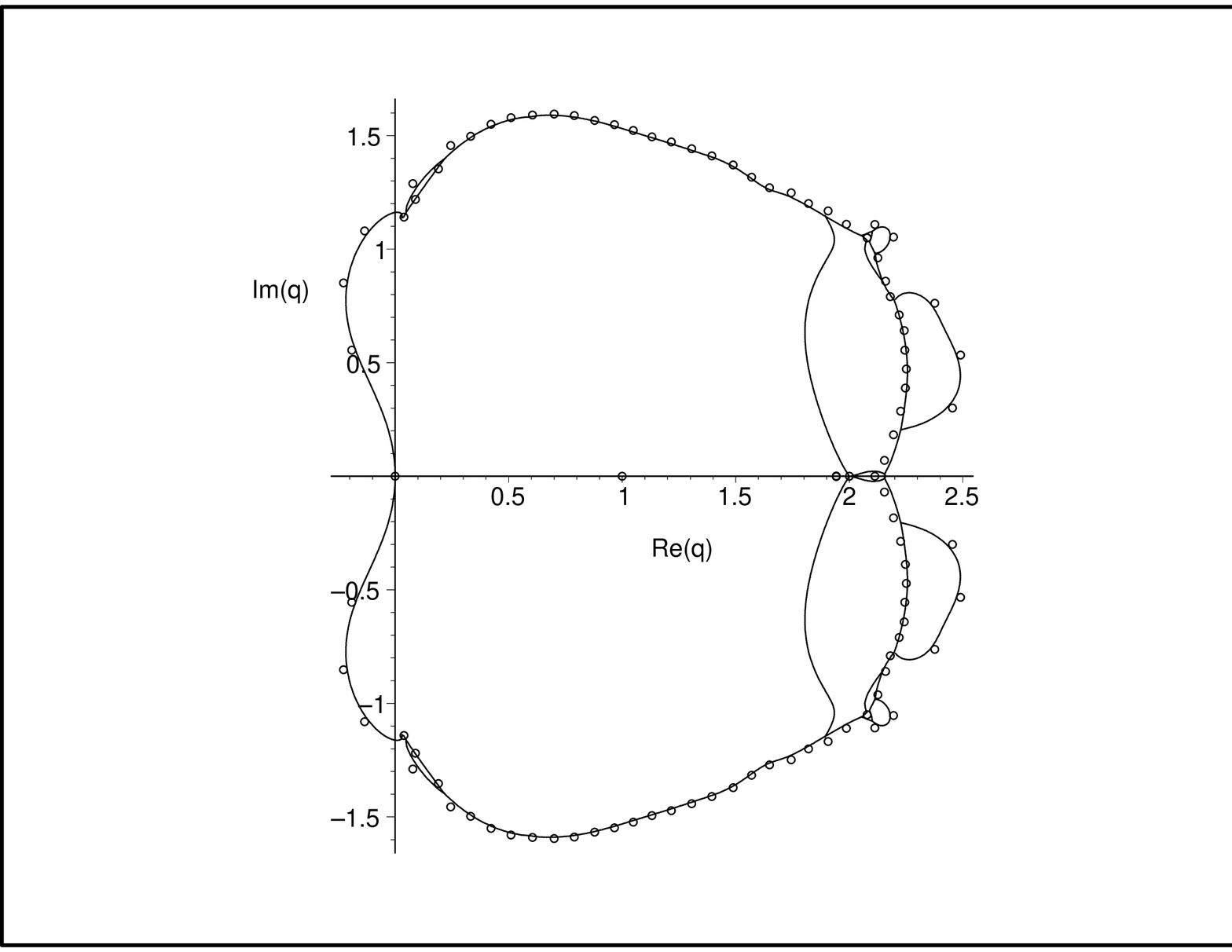}
\end{center}
\caption{\footnotesize{Singular locus ${\cal B}$ in the complex $q$ plane for
the $L_x \to \infty$ limit of the cyclic strip of the honeycomb lattice with
width $L_y=4$.  For comparison, the plot also shows zeros of $P(tri,L_y \times
L_x,cyc.,q)$ with $L_y=4$ and a typical large value of the length, $L_x=24$.}}
\label{hcpxy4zeros}
\end{figure}

\begin{figure}[hbtp]
\centering
\leavevmode
\epsfxsize=4.0in
\begin{center}
\leavevmode
\epsffile{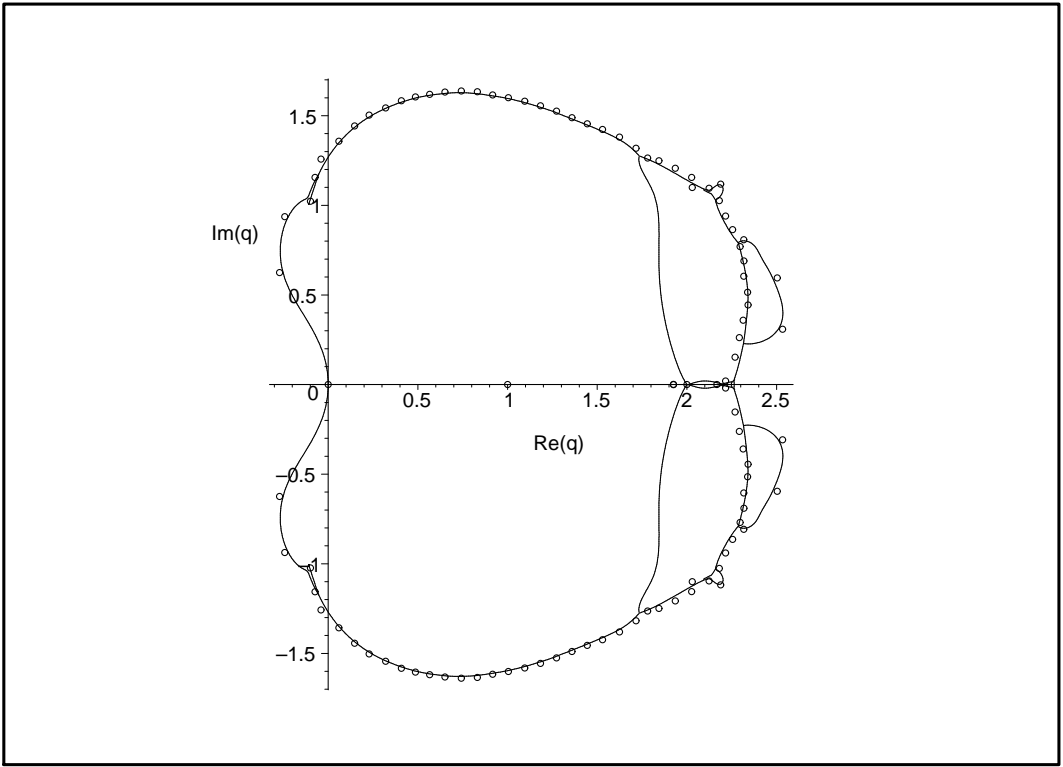}
\end{center}
\caption{\footnotesize{Singular locus ${\cal B}$ in the complex $q$ plane for
the $L_x \to \infty$ limit of the cyclic strip of the honeycomb lattice with
width $L_y=5$.  For comparison, the plot also shows zeros of $P(tri,L_y \times
L_x,cyc.,q)$ with $L_y=5$ and a typical large value of the length, $L_x=20$.}}
\label{hcpxy5zeros}
\end{figure}

In Figs. \ref{tpxy5zeros}-\ref{hcpxy5zeros} we show the singular loci ${\cal
B}$ for infinite-length strips of the triangular lattice with width $L_y=5$ and
of the honeycomb lattice with widths $L_y=3,4,5$, together with zeros of the
chromatic polynomials (chromatic zeros) for long finite-length strips of each
type.  We first discuss the triangular-lattice strip (Fig. \ref{tpxy5zeros}).
This should be compared with our earlier results for the $L_y=2$ strip in
Fig. 3 in Ref. \cite{wcy} and for the $L_y=3$ and $L_y=4$ strips in Figs. 2 and
3 of Ref. \cite{t}.  As expected, the comparison is particularly close with the
$L_y=4$ case.  For $L_y=3,4$ we found that there are two inner curves that run
between points on the outer envelope and cross the real $q$ axis at $q=2$ and
$q=3$, separating the interior of the envelope curve into regions that include
the real segments (i) $0 \le q \le 2$, (ii) $2 \le q \le 3$, and (iii) $3 \le q
\le q_c$.  We also found that the densities of zeros on these curves passing
through $q=2$ and $q=3$ were slightly less than the densities of the zeros on
the outer envelope curves.  Another feature that we found was the existence of
two ``bubble'' regions protruding to the right from the main curve that passes
through $q_c$. Finally, we observed that as $L_y$ increases from $L_y=3$ to
$L_y=4$, there are more zeros, and more support for a part of ${\cal B}$ in the
half-plane with $Re(q) < 0$. Our current results for $L_y=5$ exhibit these same
features, and show an increase, relative to our $L_y=4$ results, in the portion
of ${\cal B}$ extending into the half-plane with $Re(q) < 0$.  This is
indicated in Table \ref{qctable}.  In this table we also characterize the form
of the curve ${\cal B}$ as it passes through the point $q=0$; given that it
passes vertically through this point, and is invariant under complex
conjugation, there are two possibilities, viz., that it extends into the
half-plane with (i) $Re(q) > 0$ or (ii) $Re(q) < 0$.  We denote these two
possibilities as (i) convex and (ii) concave to the left.  Note that, {\it a
priori}, the conditions that ${\cal B}$ has support in the half-plane with
$Re(q) < 0$ and that it is concave to the left are not equivalent since it
might be convex to the left, i.e., curve into the $Re(q) > 0$ half-plane in the
neighborhood of the origin, but then curve back to the left at larger values of
$|Im(q)|$ and cross over into the $Re(q) < 0$ half-plane.  However, we find
that this does not happen for any of the strips for which we have done
calculations, as is evident in Table \ref{qctable}.  Thus, for these strips,
the properties that ${\cal B}$ has support for $Re(q) < 0$ and that it is
concave to the left at $q=0$ are observed to occur together.  We also note that
the zeros on the right-hand side of the locus ${\cal B}$ cross the real axis at
a point consistent with the asymptotic result for $q_c$ given in
eq. (\ref{qc_tpxy5}).  We observe small complex-conjugate bubble regions around
$q = 2.6 \pm 2i$, where the curve passing through $q=2$ intersects the outer
curve on ${\cal B}$.  We should also remark on the parts of the locus ${\cal
B}$ near the points $q=1 \pm 2.1i$: without magnification, these appear to be
arc endpoints, but in fact the end in small, very narrow, bubble regions.  This
is consistent with our finding for the $L_x \to \infty$ limit of all strips of
regular lattices with periodic longitudinal boundary conditions, that the
respective loci ${\cal B}$ do not have arc endpoints, in contrast with the
situation for strips with free longitudinal boundary conditions, for which the
loci ${\cal B}$ generically do exhibit such arc endpoints.  In our earlier work
we have found that loci ${\cal B}$ can exhibit tiny sliver regions, and we
mention that for this case of the $L_y=5$ cyclic strip of the triangular
lattice and the for the loci shown below for the strips of the honeycomb
lattice, such tiny sliver regions, if small enough, would elude our analysis
because of the finite size of the grid that we use for testing for equimodular
dominant $\lambda$'s.  

In addition to the continuous accumulation set of chromatic zeros that forms
${\cal B}$, one may also investigate discrete chromatic zeros.  In general, for
any graph with at least one edge, there is always a chromatic zero at $q=1$,
since there is no proper coloring of such a graph with just one color.  For
strips of the triangular lattice with width the $L_y=5$ and reasonably great
length, we also find a chromatic zero near to the point $q=(3+\sqrt{5})/2
\simeq 2.618$, which is a zero of $c^{(2)}$, where a
degree $d=2$ $\lambda$ is dominant, and a chromatic zero near to the point 
$q \simeq 3.2470$, a zero of $c^{(3)}$, where a degree $d=3$ $\lambda$ is
dominant.

We next show the singular loci ${\cal B}$ for the infinite-length strips of the
honeycomb lattice with $L_y=3,4,5$ in
Figs. \ref{hcpxy3zeros}-\ref{hcpxy5zeros}, together with chromatic zeros for
long finite-length strips of each respective width.  As background, we recall
that in the $L_x \to \infty$ limit for the smaller widths $L_y=2$ (Ref.
\cite{pg}) and $L_y=3$ (Ref. \cite{hca}), we found that six curves on ${\cal
B}$, forming three branches, intersect at the point $q_c$ (which is equal to 2
for those widths).  Hence, it was found that the region diagram included at
least six regions (which comprised the totality of regions for $L_y=2$ and were
augmented by two additional very small regions centered approximately at $q=0.5
\pm 0.45i$ for $L_y=3$): (i) the outermost region $R_1$, extending infinitely
far away from the origin and including the intervals $q > 2$ and $q < 0$ on the
real axis; (ii) the innermost region $R_2$, which includes the interval $0 \le
q \le 2$; (iii) a complex-conjugate (c.c.) pair of regions $R_3, \ R_3^*$
forming upper and lower outer crescent-shaped areas adjacent to $q_c$, and (iv)
the c.c. pair $R_4, \ R_4^*$ forming upper and lower inner crescent-shaped
areas adjacent to $q_c$.  It was also found that the density of zeros on the
inner two of the six curves passing through $q_c$ was somewhat lower than the
density of the zeros on other parts of ${\cal B}$.  The locus ${\cal B}$ in the
vicinity of the origin $q=0$ is concave to the left, which implies that it has
support for values of $q$ in the half-plane with $Re(q) < 0$.  This concavity
increases as the width increases.  In the loci ${\cal B}$ presented in
Fig. \ref{hcpxy4zeros} and \ref{hcpxy5zeros}, we see that the point $q=2$ is an
intersection point of four rather than six curves, and this feature is
correlated with the fact that for these widths, $q_c$ exceeds 2.  Thus, in the
plot for $L_y=4$ we see a small self-conjugate bubble region that includes the
interval $2 \le q \lsim q_{hc4c}$, where $q_{hc4c} \simeq 2.155$ is the value
of $q_c$ for this strip, given in eq. (\ref{qc_hpxy4}).  For $L_y=5$ we observe
three small self-conjugate regions in ${\cal B}$ (see Fig. \ref{hcpxy5zeros})
that include the three segments $2 \le q \le q_{hc5a}$, $q_{hc5a} \le q \le
q_{hc5b}$, and $q_{hc5b} \le q \le q_{hc5c}$, where $q_{hc5a} \simeq 2.1997$,
$q_{hc5b} \simeq 2.2468$, and $q_{hc5c} \simeq 2.26407$ is the value of $q_c$
for this strip, given in eq. (\ref{qc_hpxy5}).  Thus, the locus ${\cal B}$ for
the cyclic $L_y=5$ strip of the honeycomb lattice crosses the real $q$ axis
$q=0, \ 2, \ q_{hc5a}, \ q_{hc5b}$, and $q_{hc5c}$.  These values are listed,
together with those for other strips, in Table \ref{qctable}.  For the $L_y=3$
strip, one sees in Fig. \ref{hcpxy3zeros} a complex-conjugate pair of narrow
regions around $q \simeq 0.5 \pm 1.4i$; similarly, for the $L_y=4$ strip, one
sees in Fig. \ref{hcpxy4zeros} a complex-conjugate pair of very narrow sliver
regions around $q \simeq 0.1 \pm 1.3i$.  For the $L_y=5$ strip of the honeycomb
lattice, we find arcs ending in tiny bubble regions at approximately $q = -0.1
\pm i$. A general feature that we observe is increasing complexity of the loci
${\cal B}$ with increasing strip width.  As before, we note that regions that
are extremely small could be missed by our grid used in testing for
degeneracies between dominant eigenvalues that yield components of the locus
${\cal B}$.

\bigskip

Acknowledgment

We thank J. Salas for discussions on transfer matrix methods during the work
for Refs. \cite{ts,tt}, A. Sokal for related discussions, and N. Biggs for
discussion on sieve methods.  The research of R.S. was partially supported by
the NSF grant PHY-00-98527.  The research of S.C.C. was partially supported by
the Nishina and Inoue Foundations, and he thanks Prof. M. Suzuki for further
support.  The NCTS Taipei address for S.C.C. applies after April 12, 2004.

\newpage

\section{Appendix 1: Chromatic Numbers for Lattice Strips}

The following chromatic numbers apply for lengths that are greater than the
lowest few where strips sometimes degenerate.  
For cyclic strips of the square lattice, the chromatic number is given by 
\beq
\chi(sq, L_y \times L_x,cyc.) = \cases{ 2 & for even $L_x$ \cr
               3 & for odd $L_x$ \cr }
\label{chi_sq_cyc}
\eeq
independent of $L_y$.  For the M\"obius strips of the square lattice (e.g.,
\cite{s4}) 
\beq
\chi(sq, L_y \times L_x,Mb.) = \cases{ 2 & for $(L_x,L_y)=(e,o), \ (o,e)$ \cr
               3 & for $(L_x,L_y)=(e,e), \ (o,o)$ \cr }
\label{chi_sq_mb}
\eeq
where $e$ and $o$ denote even and odd.

The cyclic and M\"obius strips of the triangular lattice have chromatic numbers
(e.g. \cite{t}) 

\beq
\chi(tri, L_y \times L_x, cyc.) = \cases{ 3 & if $L_x=0$ \
mod 3 \cr 4 & if $L_x=1$ \ or $L_x =2$ mod 3 \cr }
\label{chi_tri_cyc}
\eeq
\beq
\chi(tri, L_y \times L_x, Mb.) = 4
\label{chi_tri_mb}
\eeq

Again for $L_x$ above the first few values ($L_x=1,2$ where strips can
degenerate), the cyclic and M\"obius strips of the honeycomb lattice have
chromatic numbers (e.g. \cite{hca})
\beq 
\chi(hc, L_y \times L_x,cyc.) = 2
\label{chi_hc_cyc} 
\eeq 
\beq 
\chi(hc, L_y \times L_x,Mb.) = 3 
\label{chi_hc_mb}
\eeq 

\newpage

\section{Appendix 2: Tables}

\begin{table}
\caption{\footnotesize{Table of numbers $n_P(\Lambda,L_y,d)$ and their sums,
$N_{P,\Lambda,L_y,\lambda}$ for cyclic strips of the square and triangular
lattices.  Blank entries are zero.}}
\begin{center}
\begin{tabular}{|c|c|c|c|c|c|c|c|c|}
$L_y \ \downarrow$ \ \ $d \ \rightarrow$
   & 0 & 1   & 2   & 3   & 4   & 5  & 6  & $N_{P,\Lambda,L_y,\lambda}$
\\ \hline\hline
1  & 1   & 1   &     &     &     &    &    & 2    \\ \hline
2  & 1   & 2   & 1   &     &     &    &    & 4    \\ \hline
3  & 2   & 4   & 3   & 1   &     &    &    & 10   \\ \hline
4  & 4   & 9   & 8   & 4   & 1   &    &    & 26   \\ \hline
5  & 9   & 21  & 21  & 13  & 5   & 1  &    & 70   \\
\end{tabular}
\end{center}
\label{nptable}
\end{table}

\begin{table}[hbtp]
\caption{\footnotesize{Table of numbers $n_P(hc,L_y,d)$ and their sums,
$N_{P,hc,L_y,\lambda}$ for cyclic strips of the honeycomb lattice.
Blank entries are zero.}}
\begin{center}
\begin{tabular}{|c|c|c|c|c|c|c|c|c|}
$L_y \ \downarrow$ \ \ $d \ \rightarrow$
   & 0 & 1   & 2   & 3   & 4   & 5  & 6  & $N_{P,hc,L_y,\lambda}$
\\ \hline\hline
2  & 1   & 2   & 1   &     &     &    &    & 4    \\ \hline
3  & 3   & 6   & 4   & 1   &     &    &    & 14   \\ \hline
4  & 6   & 13  & 11  & 5   & 1   &    &    & 36   \\ \hline
5  & 19  & 43  & 40  & 22  & 7   & 1  &    & 132   \\
\end{tabular}
\end{center}
\label{nphctable}
\end{table}

\begin{table}
\caption{\footnotesize{Table of $\Delta n_P(sq,L_y,d)$ for strips of the square
lattice. Blank entries are zero. The last entry for each value of $L_y$ is the
total number of partitions with self-reflection symmetry.}}
\begin{center}
\begin{tabular}{|c|c|c|c|c|c|c|c|c|c|c|c|c|}
$L_y \ \downarrow$ \ \ $d \rightarrow$ & 0 & 1 & 2 & 3 & 4 & 5 & 6 & 7 & 8 & 
9 & 10 & $\Delta N_{P,L_y}$ \\
\hline\hline 1  & 1  & 1  &    &    &    &    &   &   &   &   & &
2  \\ \hline 2  & 1  & 0  & 1  &    &    &    &   &   &   &   & &
2  \\ \hline 3  & 2  & 2  & 1  & 1  &    &    &   &   &   &   & &
6  \\ \hline 4  & 2  & 1  & 2  & 0  & 1  &    &   &   &   &   & &
6  \\ \hline 5  & 5  & 5  & 3  & 3  & 1  & 1  &   &   &   &   & &
18 \\ \hline 6  & 5  & 3  & 5  & 1  & 3  & 0  & 1 &   &   &   & &
18 \\ \hline 7  & 13 & 13 & 9  & 9  & 4  & 4  & 1 & 1 &   &   & &
54 \\ \hline 8  & 13 & 9  & 13 & 4  & 9  & 1  & 4 & 0 & 1 &   & &
54 \\ \hline 9  & 35 & 35 & 26 & 26 & 14 & 14 & 5 & 5 & 1 & 1 & &
162 \\ \hline
10 & 35 & 26 & 35 & 14 & 26 & 5 & 14 & 1 & 5 & 0 & 1 & 162 \\
\end{tabular}
\end{center}
\label{nppmtable}
\end{table}

\begin{table}
\caption{\footnotesize{Table of numbers $n_P(sq,L_y,d,\pm)$ for strips of the
square lattice.  For each $L_y$ value, the entries in the first and second
lines are $n_P(sq,L_y,d,+)$ and $n_P(sq,L_y,d,-)$, respectively. Blank entries
are zero. The last entry for each value of $L_y$ is the total
$N_{P,L_y,\lambda}$.}}
\begin{center}
\begin{tabular}{|c|c|c|c|c|c|c|c|c|c|c|c|c|}
$L_y \ \  (d,+) $
   & $0,+$ & $1,+$ & $2,+$ & $3,+$ & $4,+$ & $5,+$ & $6,+$ & $7,+$ & $8,+$ &
   $9,+$ & $10,+$ & \\
\quad \ $(d,-) $
   & $0,-$ & $1,-$ & $2,-$ & $3,-$ & $4,-$ & $5,-$ & $6,-$ & $7,-$ & $8,-$ &
   $9,-$ & $10,-$
& $N_{P,L_y,\lambda}$ \\ \hline\hline
2  & 1   & 1    & 1    &      &     &     &     &    &    &   &   &       \\
   &     & 1    &      &      &     &     &     &    &    &   &   & 4     \\
   \hline
3  & 2   & 3    & 2    & 1    &     &     &     &    &    &   &   &       \\
   &     & 1    & 1    &      &     &     &     &    &    &   &   & 10    \\
   \hline
4  & 3   & 5    & 5    & 2    & 1   &     &     &    &    &   &   &       \\
   & 1   & 4    & 3    & 2    &     &     &     &    &    &   &   & 26    \\
   \hline
5  & 7   & 13   & 12   & 8    & 3   & 1   &     &    &    &   &   &       \\
   & 2   & 8    & 9    & 5    & 2   &     &     &    &    &   &   & 70    \\
   \hline
6  & 13  & 27   & 30   & 20   & 11  & 3   & 1   &    &    &   &   &       \\
   & 8   & 24   & 25   & 19   & 8   & 3   &     &    &    &   &   & 192   \\
   \hline
7  & 32  & 70   & 77   & 61   & 34  & 15  & 4   & 1  &    &   &   &       \\
   & 19  & 57   & 68   & 52   & 30  & 11  & 3   &    &    &   &   & 534   \\
   \hline
8  & 70  & 166  & 199  & 163  & 106 & 49  & 19  & 4  & 1  &   &   &       \\
   & 57  & 157  & 186  & 159  & 97  & 48  & 15  & 4  &    &   &   & 1500  \\
   \hline
9  & 179 & 435  & 528  & 468  & 318 & 174 & 72  & 24 & 5  & 1 &   &       \\
   & 144 & 400  & 502  & 442  & 304 & 160 & 67  & 19 & 4  &   &   & 4246  \\
   \hline
10 & 435 & 1107 & 1405 & 1288 & 946 & 550 & 265 & 96 & 29 & 5 & 1 &       \\
   & 400 & 1081 & 1307 & 1274 & 920 & 545 & 251 & 95 & 24 & 5 &   & 12092 \\
\end{tabular}
\end{center}
\label{npctable}
\end{table}

\begin{table}
\caption{\footnotesize{Table of $\Delta n_P(hc,L_y,d)$ for strips
of the honeycomb lattice with even $L_y$. Blank entries are zero.
The last entry for each value of $L_y$ is the total number of
partitions with self-reflection symmetry.}}
\begin{center}
\begin{tabular}{|c|c|c|c|c|c|c|c|c|c|c|c|c|}
$L_y \ \ \downarrow$ \ \  $d \rightarrow$ 
& 0 & 1 & 2 & 3 & 4 & 5 & 6 & 7 & 8 & 9 & 10 & $\Delta
N_{P,hc,L_y}$ \\
\hline\hline
2  & 1  & 0  & 1  &    &    &    &   &   &   &   & & 2  \\ \hline
4  & 4  & 3  & 3  & 1  & 1  &    &   &   &   &   & & 12 \\ \hline
6  & 7  & 4  & 7  & 1  & 4  & 0  & 1 &   &   &   & & 24 \\ \hline
8  & 36 & 30 & 30 & 17 & 17 & 6  & 6 & 1 & 1 &   & & 144 \\ \hline
10 & 66 & 47 & 66 & 23 & 47 & 7 & 23 & 1 & 7 & 0 & 1 & 288 \\
\end{tabular}
\end{center}
\label{nphcpmtable}
\end{table}

\begin{table}
\caption{\footnotesize{Table of numbers $n_P(hc,L_y,d,\pm)$ for strips of the
honeycomb lattice with even $L_y$.  For each $L_y$ value, the entries in the
first and second lines are $n_P(hc,L_y,d,+)$ and $n_P(hc,L_y,d,-)$,
respectively. Blank entries are zero. The last entry for each value of $L_y$ is
the total $N_{P,hc,L_y,\lambda}$.}}
\begin{center}
\begin{tabular}{|c|c|c|c|c|c|c|c|c|c|c|c|c|}
$L_y \ (d,+) $
   & $0,+$ & $1,+$ & $2,+$ & $3,+$ & $4,+$ & $5,+$ & $6,+$ & $7,+$ & $8,+$ &
   $9,+$ & $10,+$ & \\
\quad \ $(d,-) $
   & $0,-$ & $1,-$ & $2,-$ & $3,-$ & $4,-$ & $5,-$ & $6,-$ & $7,-$ & $8,-$ &
   $9,-$ & $10,-$
& $N_{P,hc,L_y,\lambda}$ \\ \hline\hline
2  & 1   & 1    & 1    &      &     &     &     &    &    &   &   &       \\
   &     & 1    &      &      &     &     &     &    &    &   &   & 4     \\
   \hline
4  & 5   & 8    & 7    & 3    & 1   &     &     &    &    &   &   &       \\
   & 1   & 5    & 4    & 2    &     &     &     &    &    &   &   & 36    \\
   \hline
6  & 25  & 53   & 56   & 35   & 17  & 4   & 1   &    &    &   &   &       \\
   & 18  & 49   & 49   & 34   & 13  & 4   &     &    &    &   &   & 358   \\
   \hline
8  & 194 & 454  & 518  & 404  & 237 & 100 & 32  & 6  & 1  &   &   &       \\
   & 158 & 424  & 488  & 387  & 220 & 94  & 26  & 5  &    &   &   & 3748  \\
   \hline
10 & 1590& 4036 & 4953 & 4324 & 2947& 1565& 664 & 208& 51 & 7 & 1 &       \\
   & 1524& 3989 & 4887 & 4301 & 2900& 1558& 641 & 207& 44 & 7 &   & 40404 \\
\end{tabular}
\end{center}
\label{npchctable}
\end{table}

\begin{table}
\caption{\footnotesize{Factorization properties of characteristic polynomials
$CP(T_{sq,L_y,d},z)$ for cyclic strips of the square lattices. The notation
$(m,n)$ means that $CP(T_{sq,L_y,d},z)$ factorizes into polynomials in $z$ of
degree $m$ and $n$ with integer coefficients.  The notation $j^k$ indicates
that there are $k$ factors of degree $j$.}}
\begin{center}
\begin{tabular}{|c|c|c|c|c|c|c|c|}
$L_y \ \downarrow$ \ \ $d \ \rightarrow$
   & 0       &   1    &   2        & 3        & 4  & 5 & $N_{P,sq,L_y,\lambda}$
\\ \hline\hline
1  & (1)     & (1)    &            &          &       &    & 2    \\ \hline
2  & (1)     & $(1^2)$&  (1)       &          &       &    & 4    \\ \hline
3  & (2)     & (1,3)  & $(1^3)$    & (1)      &       &    & 10   \\ \hline
4  & (1,3)   & (4,5)  & $(3,5)$    & $(1^2,2)$& (1)   &    & 26   \\ \hline
5  & (2,7)   & (8,13) & (9,12)     & (5,8)    & $(2,3)$ &(1) & 70   \\
\end{tabular}
\end{center}
\label{sqfactors}
\end{table}

\begin{table}
\caption{\footnotesize{Factorization properties of characteristic polynomials
$CP(T_{tri,L_y,d},z)$ for cyclic strips of the triangular lattices. The notation
$(m,n)$ means that $CP(T_{tri,L_y,d},z)$ factorizes into polynomials in $z$ of
degree $m$ and $n$ with integer coefficients. The notation $j^k$ indicates that
there are $k$ factors of degree $j$.}}
\begin{center}
\begin{tabular}{|c|c|c|c|c|c|c|c|}
$L_y \ \downarrow$ \ \ $d \ \rightarrow$
   & 0       &   1    &   2     & 3      & 4     & 5  & $N_{P,tri,L_y,\lambda}$
\\ \hline\hline
1  & (1)     & (1)    &         &        &       &     &   2   \\ \hline
2  & (1)     & (2)    & (1)     &        &       &     &   4   \\ \hline
3  & (2)     & (1,3)  & (1,2)   & (1)    &       &     &  10   \\ \hline
4  & (4)     & (9)    & (8)     & (4)    & (1)   &     &  26   \\ \hline
5  & (2,7)   & (8,13) & (9,12) & (5,8)  & (2,3)  & (1) &  70   \\
\end{tabular}
\end{center}
\label{trifactors}
\end{table}

\begin{table}
\caption{\footnotesize{Factorization properties of characteristic polynomials
$CP(T_{hc,L_y,d},z)$ for cyclic strips of the honeycomb lattices. The notation
$(m,n)$ means that the characteristic polynomial factorizes into polynomials in
$z$ of degree $m$ and $n$ with integer coefficients. The notation $j^k$
indicates that there are $k$ factors of degree $j$.}}
\begin{center}
\begin{tabular}{|c|c|c|c|c|c|c|c|}
$L_y \ \downarrow$ \ \ $d \ \rightarrow$
   & 0       &   1    &   2     & 3      & 4    & 5  & $N_{P,hc,L_y,\lambda}$
\\ \hline\hline
2  & (1)     &$(1^2)$ & (1)     &        &       &     &   4    \\ \hline
3  & (3)     & (6  )  & (1,3 )  & (1)    &       &     &  14    \\ \hline
4  & (1,5)   & (5,8)  & (4,7)  &$(1^2,3)$& (1)   &     &  36    \\ \hline
5  & (19)    & (43)   & (40)   & (22)    & (1,6) &  1  &  132   \\ 
\end{tabular}
\end{center}
\label{hcfactors}
\end{table}

\begin{table}
\caption{\footnotesize{Properties of the locus ${\cal B}$ for infinite-length
cyclic/M\"obius strips with width $L_y$ of regular lattices $\Lambda$,
including square (sq), triangular (tri),and honeycomb (hc) lattices.  The
notation BCR denotes ${\cal B}$ crossings on the real $q$ axis, the greatest of
which is $q_c$.  The notation SN indicates whether ${\cal B}$ has some support
for values of $q$ with $Re(q) < 0$, marked as y (yes) or n (no).  The behavior 
of ${\cal B}$ in the neighborhood of the origin is indicated as convex 
or concave to the left, with the convention taken such that, for
example, the circle $|q-1|=1$ is convex (to the left) at $q=0$.  The
final column gives the reference.}}
\begin{center}
\begin{tabular}{|c|c|c|c|c|c|}
$\Lambda$ & $L_y$ & BCR & SN & ${\cal B}$ at $q=0$ & Ref. \\
\hline\hline
sq  & 1 & 2,          \ 0  & n & convex  & $-$      \\ \hline
sq  & 2 & 2,          \ 0  & n & convex  & \protect{\cite{w}}     \\ \hline
sq  & 3 & 2.34,  \ 2, \ 0  & y & concave & \protect{\cite{wcyl}}  \\ \hline
sq  & 4 & 2.49,  \ 2, \ 0  & y & concave & \protect{\cite{s4}}    \\ \hline
sq  & 5 & 2.58,  \ 2, \ 0  & y & concave & \protect{\cite{s5}}  \\ \hline\hline
tri & 2 & 3,     \ 2, \ 0  & n & convex  & \protect{\cite{wcy}}   \\ \hline
tri & 3 & 3,     \ 2, \ 0  & n & convex  & \protect{\cite{t}}     \\ \hline
tri & 4 & 3.23, \ 3,\ 2,\ 0 &y & concave &  \protect{\cite{t}}    \\ \hline
tri & 5 & 3.33, \ 3,\ 2,\ 0 &y & concave &   here              \\ \hline\hline
hc  & 2 & 2,    \ 0        & y & concave & \protect{\cite{pg}}    \\ \hline
hc  & 3 & 2,    \ 0        & y & concave & \protect{\cite{hca}}   \\ \hline
hc  & 4 & 2.155,\ 2,\ 0    & y & concave & here                  \\ \hline
hc  & 5 & 2.26, \ 2.25, \ 2.20, \ 2, \ 0 & y & concave & here    \\ 

\end{tabular}
\end{center}
\label{qctable}
\end{table}

\newpage

\section{Appendix 3}

Here we list polynomials that occur in our explicit expressions for elements of
transfer matrices.  It is convenient to define
\beq
F_{m,n} = D_m - D_n
\label{fmn}
\eeq
One then sees that $q^2-4q+5$ is $F_{4,3}$.  Note that
\beq
F_{m+2,m}=(q-1)^{m-1}(q-2)
\eeq
\beq
F_{m+4,m}=(q-1)^{m-1}(q-2)(q^2-2q+2)
\eeq
\beqs
F_{m+6,m} & = & (q-1)^{m-1}(q-2)(q^2-q+1)D_4 \cr\cr
& = & (q-1)^{m-1}(q-2)(q^2-q+1)(q^2-3q+3)
\eeqs
etc., so e.g., $F_{4,2}=(q-1)(q-2)$.  One can also define
\beq
G_{m,n} = D_m - 2D_n
\label{gmn}
\eeq
so that another frequently occurring entry, $q^2-5q+7$ is seen to be
$G_{4,3}$.  We have not tried to carry out such constructions in an exhaustive
manner and often just give shorthand notation to commonly occurring
polynomials. Because of the various equivalences, there are also several ways
of writing a given polynomial.

We also list a set of polynomials that recur frequently and are given shorthand
names, $p_n$, $r_n$, and $s_n$ for quadratic, cubic, and quartic terms: 
\beq
p_4=q^2-3q+4, \quad p_6=q^2-4q+6, \quad p_8=q^2-5q+8
\eeq
\beq
p_{10}=q^2-6q+10, \quad p_{13}=q^2-7q+13, \quad p_{14}=q^2-7q+14
\eeq
\beq
p_{15}=q^2-7q+15, \quad p_{17}=q^2-8q+17
\eeq

\bigskip

\beq
r_{11}=q^3-5q^2+11q-11, \quad r_{13}=q^3-6q^2+14q-13
\eeq
\beq
r_{19}=q^3-7q^2+19q-19, \quad r_{20}=q^3-7q^2+19q-20
\eeq
\beq
r_{34}=q^3-9q^2+29q-34, \quad r_{46}=q^3-11q^2+39q-46
\eeq
\beq
r_{47}=q^3-10q^2+36q-47, \quad r_{49}=q^3-10q^2+37q-49, 
r_{58}=q^3-11q^2+43q-58
\eeq

\bigskip

\beq
s_7=q^4-5q^3+11q^2-12q+7, \quad s_{21}=q^4-7q^3+21q^2-32q+21
\eeq
\beq
s_{31}=q^4-8q^3+27q^2-45q+31, s_{47}=q^4-9q^3+34q^2-63q+47
\eeq
\beq
s_{61}=q^4-9q^3+35q^2-70q+61, s_{82}=q^4-11q^3+48q^2-99q+82
\eeq
\beq
s_{148}=q^4-13q^3+67q^2-160q+148, s_{173}=q^4-13q^3+68q^2-171q+173
\eeq
\beq
s_{3,5}=(q^2-2q+3)(q^2-4q+5), \quad t_{5,5}=(q^2-4q+5)(q^3-4q^2+6q-5)
\eeq

\vfill
\eject

\newpage

\begin{thebibliography}{99}

\bibitem{potts}
R. B. Potts, Proc. Camb. Phil. Soc. {\bf 48} (1952) 106.

\bibitem{wurev}
F. Y. Wu, Rev. Mod. Phys. {\bf 54} (1982) 235.

\bibitem{wup}
F. Y. Wu, J. Stat. Phys. {\bf 52} (1988) 99.

\bibitem{rrev}
R. C. Read, J. Combin. Theory {\bf 4} (1968) 52.

\bibitem{rtrev}
R. C. Read and W. T. Tutte, ``Chromatic Polynomials'',
in {\it Selected Topics in Graph Theory, 3}, (Academic Press, New York, 1988),
p. 15.

\bibitem{bbook}
N. L. Biggs, {\it Algebraic Graph Theory} (Cambridge
Univ. Press, Cambridge, 1st ed. 1974, 2nd ed. 1993).

\bibitem{saleur}
H. Saleur, Nucl. Phys. B {\bf 360} (1991) 219.  

\bibitem{saleurcmp}
H. Saleur, Commun. Math. Phys. {\bf 132} (1990) 657. 

\bibitem{cf}
S.-C. Chang and R. Shrock, Physica A {\bf 296} (2001) 131.

\bibitem{hca}
S.-C. Chang and R. Shrock, Physica A {\bf 296} (2001) 183.

\bibitem{dg}
S.-C. Chang and R. Shrock, Physica A {\bf 301} (2001) 301.

\bibitem{kf}
P. W. Kasteleyn and C. M. Fortuin, J. Phys. Soc. Jpn. (Suppl.) {\bf 26} (1969)
11. 

\bibitem{fk}
C. M. Fortuin and P. W. Kasteleyn, Physica {\bf 57} (1972) 536.

\bibitem{bkw1}
S. Beraha, J. Kahane, and N. Weiss, J. Combin. Theory B {\bf 27} (1979) 1.

\bibitem{bkw2}
S. Beraha, J. Kahane, and N. Weiss, J. Combin. Theory B {\bf 28} (1980) 52.

\bibitem{read81}
R. C. Read, {\it Proc. 3rd Caribbean Conference on
Combinatorics and Computing} (1981) 23.

\bibitem{read91}
R. C. Read and G. F. Royle, in {\it Graph Theory,
Combinatorics, and Applications}, Y. Alavi et al., eds. (Wiley, NY, 1991),
vol. 2, pp. 1009-1029.

\bibitem{al}
M. Aizenman and E. H. Lieb, J. Stat. Phys. {\bf 24} (1981) 279.

\bibitem{cw}
Y. Chow and F. Y. Wu, Phys. Rev. B {\bf 36} (1987) 285.

\bibitem{w}
R. Shrock and S.-H. Tsai, Phys. Rev. E {\bf 55} (1997) 5165.

\bibitem{wc}
R. Shrock and S.-H. Tsai, Phys. Rev. E {\bf 56} (1997) 1342.

\bibitem{wa}
R. Shrock and S.-H. Tsai, Phys. Rev. E {\bf 56} (1997) 3935.

\bibitem{strip}
M. Ro\v{c}ek, R. Shrock, and S.-H. Tsai, Physica A {\bf 252} (1998) 505.

\bibitem{strip2}
M. Ro\v{c}ek, R. Shrock, and S.-H. Tsai, Physica A {\bf 259} (1998) 367.

\bibitem{hs}
R. Shrock and S.-H. Tsai, Physica A {\bf 259} (1998) 315.

\bibitem{wa2}
R. Shrock and S.-H. Tsai,  Physica A {\bf 265} (1999) 186. 

\bibitem{pg}
R. Shrock and S.-H. Tsai, J. Phys. A Lett. {\bf 32} (1999) L195.

\bibitem{nec}
R. Shrock and S.-H. Tsai, J. Phys. A {\bf 32} (1999) 5053.

\bibitem{wcyl}
R. Shrock and S.-H. Tsai, Phys. Rev. E {\bf 60} (1999) 3512.

\bibitem{wcy}
R. Shrock and S.-H. Tsai, Physica A {\bf 275} (2000) 429.

\bibitem{pm}
R. Shrock, Phys. Lett. A {\bf 261} (1999) 57.

\bibitem{tk}
N. L. Biggs and R. Shrock, J. Phys. A (Letts.) {\bf 32} (1999) L489.

\bibitem{k}
S.-C. Chang and R. Shrock, Phys. Rev. E {\bf 62} (2000) 4650

\bibitem{bcc}
R. Shrock, in {\it Proceedings of the 1999 British
Combinatorial Conference, BCC99}, Discrete Math. {\bf 231} (2001) 421.

\bibitem{s4}
S.-C. Chang and R. Shrock, Physica A {\bf 290} (2001) 402.

\bibitem{t}
S.-C. Chang and R. Shrock, Ann. Phys. {\bf 290} (2001) 124.

\bibitem{tor4}
S.-C. Chang and R. Shrock, Physica A {\bf 292} (2001) 307.

\bibitem{hd}
J. Salas and R. Shrock, Phys. Rev. E {\bf 64} (2001) 011111.

\bibitem{s5}
S.-C. Chang and R. Shrock, Physica A {\bf 316} (2002) 335.

\bibitem{lenard}
A. Lenard (unpublished), cited in E. H. Lieb, Phys. Rev. {\bf 162} (1967) 162.

\bibitem{baxter86}
R. J. Baxter, J. Phys. A {\bf 19} (1986) 2821.

\bibitem{baxter87}
R. J. Baxter, J. Phys. A {\bf 20} (1987) 5241.

\bibitem{p3afhc}
R. Shrock and S.-H. Tsai, J. Phys. A {\bf 30} (1997) 495.

\bibitem{ww}
R. Shrock and S.-H. Tsai, Phys. Rev. E {\bf 55} (1997) 6791.

\bibitem{w3}
R. Shrock and S.-H. Tsai, Phys. Rev. E {\bf 56} (1997) 2733.

\bibitem{wn}
R. Shrock and S.-H. Tsai, Phys. Rev. E {\bf 56} (1997) 4111.

\bibitem{w2d}
R. Shrock and S.-H. Tsai, Phys. Rev. E {\bf E58} (1998) 4332 
(cond-mat/9808057). 

\bibitem{ssbounds}
J. Salas and A. Sokal, J. Stat. Phys. {\bf 86} (1997) 551.

\bibitem{a}
R. Shrock, Physica A {\bf 283} (2000) 388.

\bibitem{bds}
N. L. Biggs, R. M. Damerell, and D. A. Sands, J. Combin. Theory
B {\bf 12} (1972) 123.

\bibitem{bm}
N. L. Biggs and G. H. Meredith, J. Combin. Theory B {\bf 20} (1976) 5.

\bibitem{b}
N. L. Biggs, Bull. London Math. Soc. {\bf 9} (1977) 54.

\bibitem{matmeth}
N. L. Biggs, J. Combin. Theory B {\bf 82} (2001) 19.

\bibitem{matmeth2}
N. L. Biggs, Bull. London Math. Soc. {\bf 34} (2002) 129.

\bibitem{matmeth3}
N. L. Biggs, London School of Economics Centre for Discrete and Applicable
Mathematics, report LSE-CDAM-2000-04. 

\bibitem{klein} 
D. Klein and W. Seitz, Studies in Physical and
Theoretical Chemistry, vol. 63 (1988), pp. 155-166.

\bibitem{sqtran}
J. Salas and A. Sokal, J. Stat. Phys. {\bf 104} (2001) 609. 

\bibitem{cyltran}
J.Jacobsen and J. Salas, J. Stat. Phys. {\bf 104} (2001) 701.

\bibitem{tritran}
J. Jacobsen, J. Salas, and A.D. Sokal, J. Stat. Phys. {\bf 112} (2003) 921.

\bibitem{ts}
S.-C. Chang, J. Salas, and R. Shrock, J. Stat. Phys. {\bf 107} (2002) 1207.

\bibitem{tt}
S.-C. Chang, J. Jacobsen, J. Salas, and R. Shrock,
J. Stat. Phys. {\bf 114} (2004) 763.

\bibitem{dn}
S.-C. Chang, Physica A {\bf 296} (2001) 495.

\bibitem{cprg}
N. L. Biggs, M. Klin, and P. Reinfeld, in {\it Proc. Algebraic Graph Theory 
Workshop}, European Journal of Combinatorics {\bf 25} (2004) 147. 

\bibitem{ka3}
S.-C. Chang and R. Shrock, in {\it Proc. of the CRM Workshop on Tutte
Polynomials, 2001}, Advances in Applied. Math. {\bf 32} (2004) 44.

\bibitem{js} J. Salas has kindly informed that he, A. Sokal, and J.  Jespersen
have also generalized their transfer matrix method used for chromatic
polynomials on free and cylindrical strips of the square and triangular lattice
to cyclic strips of these lattices.

\bibitem{zt}
S.-C. Chang and R. Shrock, cond-mat/0404524. 

\bibitem{jacobson}
N. Jacobson, {\it Basic Algebra} (Freeman, San Francisco, 1974). 

\bibitem{sokalzeros}
A. Sokal, Combin. Probab. Comput. {\bf 10} (2001) 41. 

\bibitem{jz}
S.-C. Chang and R. Shrock,  Physica A {\bf 301} (2001) 196. 

\bibitem{sdg}
S.-C. Chang and R. Shrock, Phys. Rev. E {\bf 64} (2001) 066116. 

\end{thebibliography}
\end{document}